\documentclass[aps,prl,twocolumn,showpacs,superscriptaddress,floatfix]{revtex4-2}

\usepackage{enumitem}
\setlist[itemize]{noitemsep, topsep=-\parskip}
\usepackage{amsmath}          
\usepackage{amssymb}          
\usepackage{graphicx}         
\usepackage{dcolumn}          
\usepackage{bm}               
\usepackage{hyperref}         
\usepackage{xcolor}           
\usepackage{empheq}
\usepackage{latexsym}
\usepackage[capitalise]{cleveref}
\usepackage{fontawesome5}
\newcommand{\dd}{\text{d}}

\begin{document}

\title{Hydrodynamic flow in the gut hinders the fixation of beneficial mutants but accelerates their invasion}

\author{Arthur Alexandre}
\affiliation{Institute of Bioengineering, School of Life Sciences, École Polytechnique Fédérale de Lausanne (EPFL), Lausanne, CH-1015, Switzerland}
\affiliation{SIB Swiss Institute of Bioinformatics, Lausanne, CH-1015, Switzerland}

\author{Claude Loverdo}
\affiliation{Sorbonne Université, CNRS, Institut de Biologie Paris-Seine (IBPS), Laboratoire Jean Perrin (LJP), Paris, France}

\author{Anne-Florence Bitbol}
\email{anne-florence.bitbol@epfl.ch}
\affiliation{Institute of Bioengineering, School of Life Sciences, École Polytechnique Fédérale de Lausanne (EPFL), Lausanne, CH-1015, Switzerland}
\affiliation{SIB Swiss Institute of Bioinformatics, Lausanne, CH-1015, Switzerland}

\date{\today}

\begin{abstract}
  Gut bacteria experience a spatially structured environment with hydrodynamic flow. Using a branching-process approach, we show that beneficial mutants arising randomly upon division have a reduced fixation probability compared to the well-mixed case. The suppression factor is set by the spatial heterogeneity of mutant fate across the gut. In contrast, the invasion of mutant lineages that avoid early stochastic extinction is accelerated in the gut compared to a matched well-mixed case. Stronger advective flow decreases fixation probabilities while accelerating mutant invasion.
\end{abstract}

\maketitle

The gut microbiota is very diverse \cite{qin2010human,li2014integrated}, and its diversity is shaped by evolutionary processes occurring within the gut \cite{walter2011human,zhao2019adaptive,garud2019evolutionary}. Understanding these processes is of substantial public-health importance, as the gut can act as a reservoir for antibiotic-resistance genes in both humans and livestock \cite{rolain2013food,ifedinezi2024environmental}. The gut has a specific spatial structure, featuring hydrodynamic transport and nutrient and bacterial concentration gradients \cite{cremer2016effect,Cremer2017,tropini2017gut,salari2025diurnal}. Effective one-dimensional models have been developed to describe the key ingredients of this spatial structure within the ascending colon~\cite{cremer2016effect,Cremer2017}. As transport makes bacterial growth strongly heterogeneous across the gut, a key question is how this impacts the evolution of gut bacteria. In discrete models on graphs, spatial structure can strongly impact mutant fixation probability~\cite{lieberman2005evolutionary,Kaveh15,moller_exploring_2019,Yagoobi23,sharma_graph-structured_2025}, especially with asymmetric migration flows~\cite{marrec2021toward,abbara2023frequent,fruet2026environment}. In continuous systems, fluid flow can strongly affect mutant fate~\cite{Pigolotti12,gralka2017convection,plummer_fixation_2019,herrerias2019motion}. In the gut, the fixation probability of neutral mutants was predicted to be increased~\cite{labavic2022hydrodynamic}. An effective stochastic theory was then developed for beneficial mutants~\cite{ghosh2022emergent}. Here, we ask how the spatial structure of the gut impacts the fate of beneficial mutants relative to the well-mixed case, and how this depends on hydrodynamic flow.

Using a branching-process approach, we derive a reaction-diffusion equation for the fixation probability of beneficial mutants. 
We demonstrate that the fixation probability of a beneficial mutant appearing randomly upon division is suppressed compared to the well-mixed case, by a factor that reflects the heterogeneity of mutant fate in the gut. We show that mutant invasion is faster than in a matched well-mixed population. Both results hold for any positive relative mutant fitness advantage $s$. Finally, we find that higher advective flow in the gut leads to lower fixation probabilities but faster invasion of mutant lineages that survive initial stochastic extinction.

\textit{Effective one-dimensional model.} We consider the minimal one-dimensional model of transport and bacterial growth in the ascending colon introduced
in Refs. \cite{cremer2016effect, Cremer2017}. In a segment of length $L$, the linear concentrations of bacteria $B(x,t)$ and food $F(x,t)$ are subject to effective diffusion with coefficient $D$, modeling mixing due to contractions of the gut, and to advection with velocity $v$. Food, but no bacteria (because bacterial concentrations in the small intestine are very low), enter the system at $x=0$, and both food and bacteria exit to the rest of the colon at $x = L$. Bacteria consume food to divide, and division is limited by food availability. Specifically, wild-type bacteria divide at rate $r_w\rho(x,t)$ with $r_w$ their maximal growth rate and $\rho(x,t) = F(x,t)/[k+F(x,t)]$ the division rate profile, where $k$ is a Monod constant. This yields coupled reaction-diffusion partial differential equations on $B(x,t)$ and $F(x,t)$ (see SI Sec.~\ref{sec:effective_1D_model}).  

\begin{figure}[htbp]
    \centering
    \includegraphics[width=\linewidth]{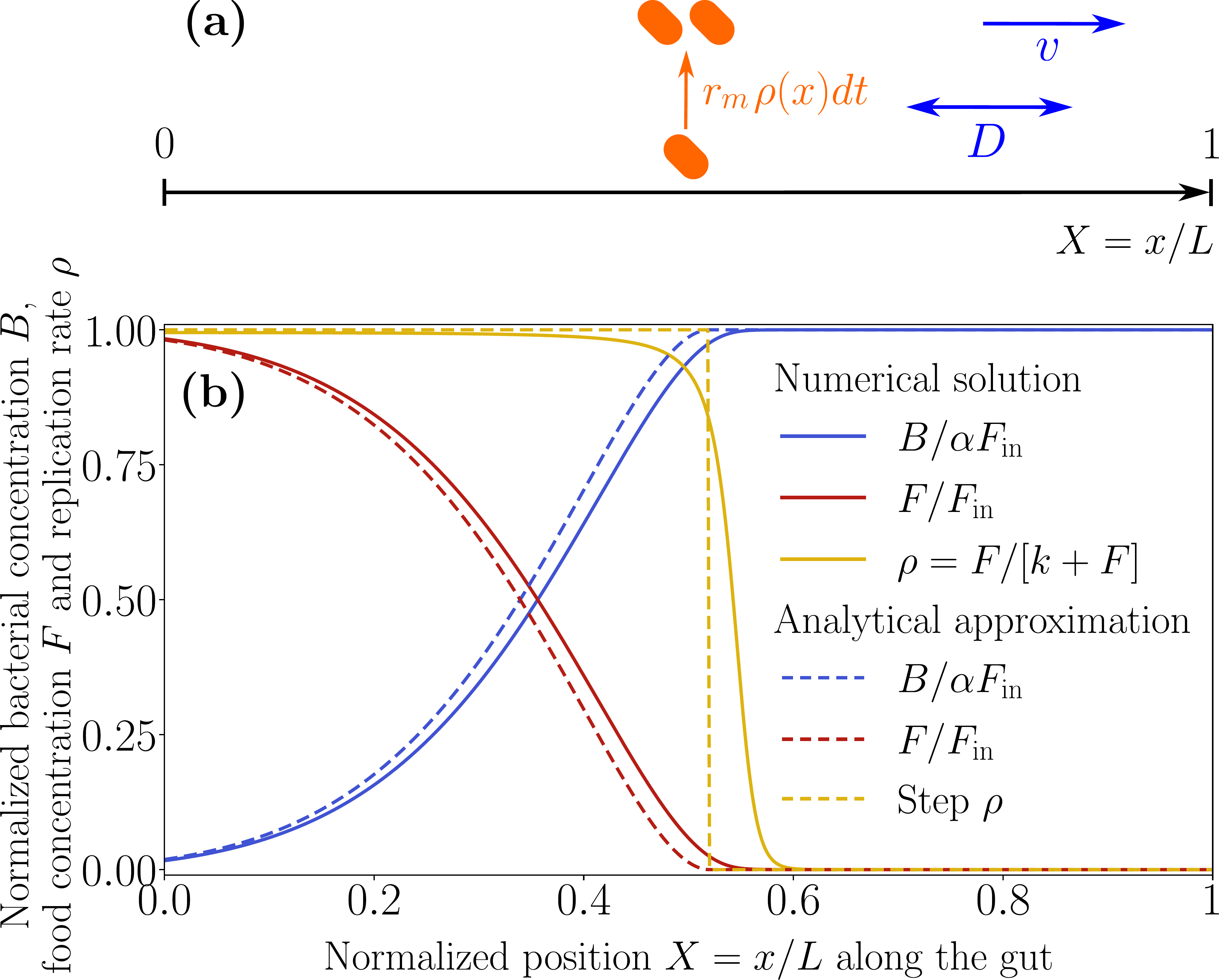}
    \caption{(a) Schematic of the one-dimensional model of the gut~\cite{cremer2016effect,Cremer2017,labavic2022hydrodynamic,ghosh2022emergent}: each mutant at position $x\in [0,L]$ divides with rate $r_m\rho(x)$, and is transported with an advective velocity $v$ and an effective diffusion coefficient $D$. (b) Steady-state wild-type bacteria concentration $B$, food concentration $F$ and division rate $\rho$ along the gut. Numerical solutions of Eqs.~\eqref{eq:eq_F},~\eqref{eq:eq_B} are compared to analytical solutions assuming a step function $\rho$~\cite{ghosh2022emergent}, see SI Sec.~\ref{steadyBstep}. Parameter values: $\lambda_w = r_w L^2 / D= 75.6$, $k/ F_{\text{in}} =5 \times 10^{-3}$ with $F_\mathrm{in}$ the incoming food concentration at $x=0$, food-bacteria conversion factor $\alpha = 200$, and $\text{Pe} =v L / D = 15.3=0.92 \,\mathrm{Pe}_\mathrm{max}$, with $\mathrm{Pe}_\mathrm{max}=16.6$ corresponding to washout.  }
    \label{fig:fig1}
\end{figure}

Our goal is to investigate the fate of a beneficial mutant in this system. 
We focus on mutants with a uniform relative fitness advantage $s$ with respect to wild-types throughout the system. Their maximum division rate is $r_m = r_w(1+s)$, but they follow the same transport dynamics as wild-types. To investigate the early dynamics of the mutant population, after a single mutant has appeared in the system at steady state, we approximate the division rate profile $\rho(x,t)$ with
\begin{equation}
    \rho(x)=\frac{F(x)}{k+F(x)}\,, \label{eq:rhostat}
\end{equation}
where $F(x)$ denotes the steady-state food concentration in the absence of mutants. Hence, mutants divide at rate $r_m\rho(x)$, and are transported, see Fig.~\ref{fig:fig1}(a). They experience a steady-state environment illustrated in Fig.~\ref{fig:fig1}(b). 

\textit{Fixation probability. } Let $\phi(x)$ be the mutant fixation probability, starting from a mutant at initial position $x$ in the gut. Reasoning on what happens to the mutant just after it appears (transport and division events, see Fig.~\ref{fig:fig1}(a)) allows us to establish the differential equation satisfied by $\phi(x)$, see SI Sec.~\ref{sec:eq-fix}: 
\begin{align}
    &D\phi''(x) + v\phi'(x) + r_m\rho(x) \phi(x)\left[1-\phi(x)\right] = 0. \label{eq:phiMT}
\end{align}
The associated boundary conditions are
\begin{equation}
     \phi'(0) = 0, \quad D\phi'(L) + v\phi(L) = 0, \label{eq:BCphi}
\end{equation}
see SI Sec.~\ref{sec:BBM}. 
To obtain Eqs.~\eqref{eq:phiMT} and~\eqref{eq:BCphi}, the mutant environment is assumed to be the steady-state wild-type one (Eq.~\eqref{eq:rhostat}), and different mutant lineages are assumed to behave independently (branching process approximation). These assumptions are appropriate if the fate of the mutant lineage is determined early, when the mutant population is still very small compared to the wild-type one, which is assumed to be large.
We further assume that if the mutant population does not go extinct while it is rare, then it fixes, which holds for beneficial mutations~\cite{patwa_fixation_2008}. 
Such branching-process descriptions are often used to describe the evolution of well-mixed populations~\cite{Haldane27,Ewens79}, and have been generalized to spatially structured ones on graphs~\cite{abbara2023frequent,fruet2026environment}. Furthermore, continuous equations analogous to \cref{eq:phiMT} describe the survival of branching Brownian motion \cite{derrida2007survival}, and gene surfing in expanding populations \cite{Hallatschek08,lehe2012rate}. 

\begin{figure}[htbp]
    \centering\includegraphics[width=\linewidth]{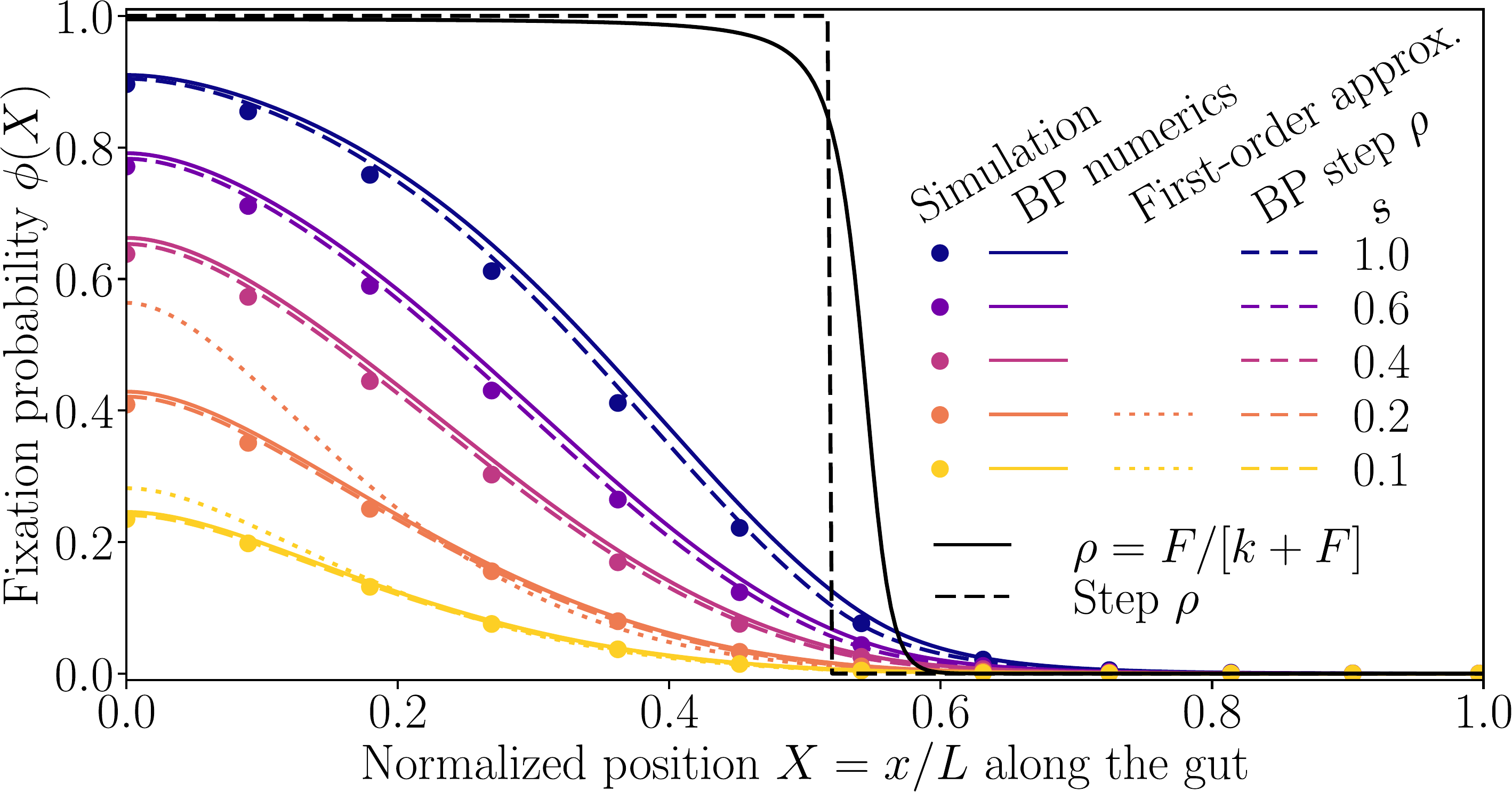}
    \caption{Mutant fixation probability $\phi(X)$ versus normalized initial mutant position $X=x/L$ along the gut, for different values of the mutant relative fitness advantage $s$. Stochastic simulations are described in SI Sec.~\ref{sec:stochsim}. Numerical solutions of the branching process Eq.~\eqref{eq:phiMT} use either the steady-state mutant-free profile of the division rate $\rho$ (``BP numerics'') or its step approximation (``BP step $\rho$''). The first-order approximation Eq.~\eqref{phi_order1_MT} in the first case is shown for small $s$. The two profiles of $\rho$ are shown for reference. Other parameter values: same as in Fig.~\ref{fig:fig1} }
    \label{fig:fig2}
\end{figure}

Fig.~\ref{fig:fig2} shows that the fixation probability $\phi(x)$ of a beneficial mutant is much larger if the mutant appears upstream (small $x$) than downstream. We also observe that a mutant appearing close to $x=0$ has an enhanced fixation probability compared to the well-mixed case: for $s=0.1$, we have $\phi(0)\approx 0.25$ with the parameters considered, while the Moran process for large populations would give a fixation probability $s/(1+s)\approx 0.09$. Meanwhile, the fixation probability is strongly suppressed downstream. Fig.~\ref{fig:fig2} also shows that a numerical solution of Eqs.~\eqref{eq:phiMT} and~\eqref{eq:BCphi} is in good agreement with stochastic simulations of the model, and that Eq.~\eqref{phi_order1_MT} gives a good approximation for $s\ll 1$. Note that the small discrepancy between simulation and numerical solution can be attributed to fluctuations of the division rate profile $\rho(x)$ in simulations due to finite-size effects, see SI Sec.~\ref{compMoranGal}. We also observe in Fig.~\ref{fig:fig2} that approximating $\rho$ with a step function gives a good approximation with the parameters used (satisfying $k\ll F_\mathrm{in}$). 

To get analytical insight into $\phi(x)$, we introduce the linear differential operator $\mathcal{A}_s$ such that
\begin{equation}
    \mathcal{A}_s \phi(x)=D\phi''(x) + v\phi'(x) +(1+s)r_w\rho(x)\phi(x), \label{def_As_MT}
\end{equation}
with boundary conditions in Eq.~\eqref{eq:BCphi}. 
Then, Eq.~\eqref{eq:phiMT} can be written as
\begin{equation}
    \mathcal{A}_s\phi(x)=(1+s)r_w\rho(x)\phi^2(x). \label{eqphi_A_MT}
\end{equation}
Linearizing this equation around $\phi=0$ gives $\mathcal{A}_s\phi(x)=0$, which leads to the eigenvalue problem for $\mathcal{A}_s$. Transforming it into a self-adjoint Sturm-Liouville problem~\cite{teschl_book} allows us to show that the principal eigenvalue of $\mathcal{A}_s$ is zero at $s=0$ and positive if and only if $s>0$ (see SI Sec.~\ref{critpt}). This entails that Eq.~\eqref{eq:phiMT} admits a non-zero solution if and only if $s>0$~\cite{englander_local_2004}: the critical point is $s=0$. This result is consistent with branching process results in well-mixed~\cite{Haldane27} and deme-structured~\cite{abbara2023frequent} populations. Performing a perturbative expansion of $\phi$ for $s\ll 1$ yields
\begin{equation}
    \phi(x)=s\frac{\left<f\right>}{\left<f^2\right>}f(x)+O(s^2), \label{phi_order1_MT}
\end{equation}
where we introduced 
\begin{equation}
f(x)=B(x)e^{-vx/D},
\end{equation}
with $B(x)$ the steady-state wild-type concentration in the absence of mutants (see SI Sec.~\ref{phi-expansion}), and the division-event-weighted average~\cite{labavic2022hydrodynamic,ghosh2022emergent},
\begin{equation}
    \left<f\right>=\frac{\int_0^L \dd x \,f(x) \rho(x) B(x) }{\int_0^L \dd x\, \rho(x) B(x)}.\label{dwa}
\end{equation}
Note that the first-order result in Eq.~\eqref{phi_order1_MT} is consistent with Ref.~\cite{ghosh2022emergent} in the limit $L\to\infty$, but that higher orders differ (see SI Sec.~\ref{comp_GG}). 

Studying the mean number of $n$-th generation descendants of a neutral mutant appearing at $x$ allows us to show that in the limit $n\to\infty$, its long-term contribution is proportional to $f(x)$, see SI Sec.~\ref{sec-Rn}. This long-term contribution to future generations can be interpreted as reproductive value~\cite{Taylor90,Maciejewski14}. Eq.~\eqref{phi_order1_MT} therefore shows that the fixation probability of a beneficial mutant appearing at $x$ is proportional to reproductive value there, to first order in $s$. Close relationships between fixation probability and reproductive value exist in other contexts~\cite{barton_relation_2011}.

Let us now average the fixation probability $\phi(x)$ over the initial position $x$ of the mutant. Assuming that mutants appear randomly upon cell division, which is realistic for mutations resulting from DNA replication errors, we focus on the division-event-weighted fixation probability $\langle \phi\rangle$ (see Eq.~\eqref{dwa}). To first order in $s$, Eq.~\eqref{phi_order1_MT} gives
\begin{equation}
    \left<\phi\right>=s\frac{\left<f\right>^2}{\left<f^2\right>}+O(s^2)=s\frac{1}{1+c_v^2(f)}+O(s^2) \leq s, \label{phibar_order1_MT}
\end{equation}
where $c_v(f)=(\left<f^2\right>-\left<f\right>^2)^{1/2}/\left<f\right>$ is the coefficient of variation of reproductive value. While mutations are produced proportionally to $\rho  B$, their long-term contribution is controlled by $f$, showing that mutant production and fate are distinct in a spatial system with advection. In addition, for a step-function $\rho$ and at first order in $s$, $\langle\phi\rangle/s$ depends only on $v/v_F$, where $v_F=2\sqrt{r_wD}$ is the Fisher speed (see SI Sec.~\ref{step_fix_mod}). 

Beyond the perturbative regime, we obtain
\begin{equation}
    \langle\phi\rangle=\frac{s}{1+s}\frac{1}{1+c_v^2(\phi)}\leq \frac{s}{1+s}, \quad \forall s>0, \label{compMoran_MT}
\end{equation}
see SI Sec.~\ref{compMoranGal}. 
Therefore, $\langle\phi\rangle$ cannot exceed the mutant fixation probability of the Moran model in the large-population limit, $\phi=s/(1+s)$~\cite{Moran58,Ewens79}, which constitutes the well-mixed reference for our model where birth and death events happen at the individual level.

\begin{figure}[htbp]
    \centering
    \includegraphics[width=\linewidth]{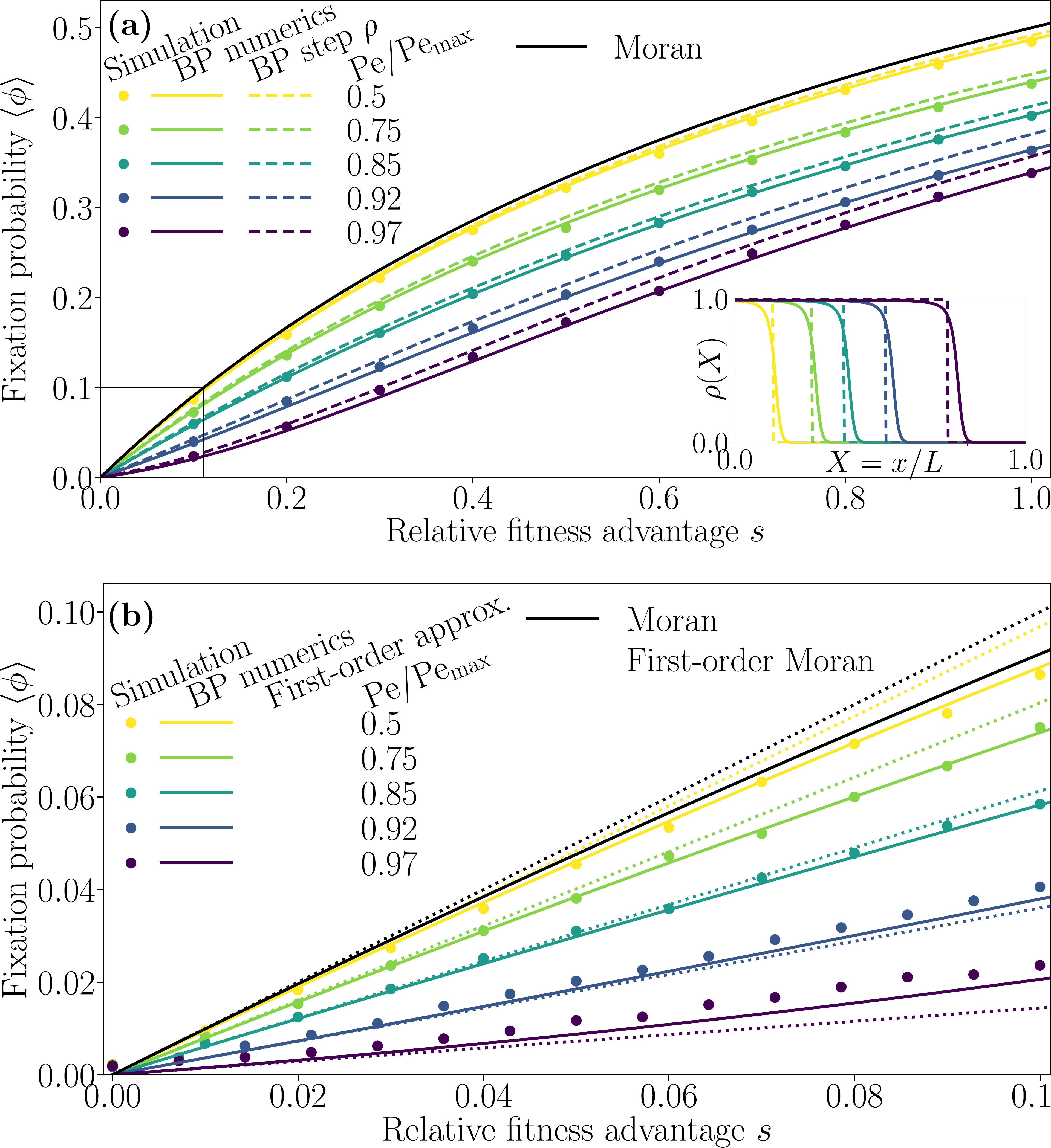}
    \caption{Division-event-weighted mutant fixation probability $\langle\phi\rangle$ versus mutant relative fitness advantage $s$, for different values of the Péclet number Pe. Panel (b) reports results for smaller $s$ than (a), shown by a rectangle in (a). Stochastic simulations are described in SI Sec.~\ref{sec:stochsim}. Numerical solutions of the branching process Eq.~\eqref{eq:phiMT} use either the steady-state mutant-free profile of the division rate $\rho$ (``BP numerics'') or its step approximation (``BP step $\rho$'' in (a)). The first-order approximation Eq.~\eqref{phibar_order1_MT} is shown in (b). The Moran model for large populations, $\phi=s/(1+s)$, gives the well-mixed reference. Inset in (a): profiles of $\rho$ along the gut. Parameters: same $\lambda_w$ (and $\mathrm{Pe}_\mathrm{max}$), $k/F_\mathrm{in}$ and $\alpha$ as in Fig.~\ref{fig:fig1}.}
    \label{fig:fig3}
\end{figure}

Fig.~\ref{fig:fig3} shows that the division-event-weighted fixation probability $\langle \phi\rangle$ is strongly impacted by the Péclet number $\mathrm{Pe}=vL/D$, which quantifies the intensity of advective flow versus diffusive flow. Larger Péclet numbers lead to stronger suppression of selection compared to the well-mixed Moran model reference. Importantly, Péclet numbers close to washout are biologically relevant, given the strong spatial gradients of food and bacterial concentration in the colon~\cite{cremer2016effect,Cremer2017,ghosh2022emergent} (see SI Sec.~\ref{humanparams}). Fig.~\ref{fig:fig3}(b) confirms that Eq.~\eqref{phibar_order1_MT} gives a good approximation for $s\ll 1$. Analytical predictions slightly underestimate simulation results for $s=0$ and for the largest Péclet numbers shown, because finite population effects yield a nonzero fixation probability for neutral mutants (see SI Sec.~\ref{sec:neutral})~\cite{labavic2022hydrodynamic,ghosh2022emergent}.

\textit{Mutant invasion growth rate.} Once the lineage arising from a single beneficial mutant with $r_m=(1+s)r_w>r_w$ has passed the initial stochastic phase where it may go extinct (described by the branching process), the concentration of beneficial mutants grows deterministically~\cite{patwa_fixation_2008,desai_beneficial_2007}. Let us focus on the initial growth phase, assuming that the division rate profile remains unperturbed and given by Eq.~\eqref{eq:rhostat}. Looking for a separable solution of the form $M(x,t)=g(x)h(t)$ for the partial differential equation on mutant concentration $M(x,t)$, we find that mutant number exponentially grows with a rate given by the principal (largest) eigenvalue $\gamma_\mathrm{max}$ of the operator $\mathcal{A}_s^\dagger$, defined by
\begin{equation}
    \mathcal{A}_s^\dagger g(x)=Dg''(x)-vg'(x)+(1+s)r_w\rho(x)g(x),
\end{equation}
with boundary conditions
\begin{equation}
     Dg'(0) -vg(0)= 0, \quad Dg'(L) = 0, \label{eq:BCg}
\end{equation}
see SI Sec.~\ref{detgrowth-general}. This operator is the adjoint of $\mathcal{A}_s$ in Eq.~\eqref{def_As_MT}. It comprises a forward transport operator, while $\mathcal{A}_s$, which acts on the fixation probability, comprises a backward one. The eigenvalue problem of operator $\mathcal{A}_s^\dagger$ can be mapped onto the same Sturm-Liouville problem as that of $\mathcal{A}_s$, yielding the principal eigenvalue
\begin{equation}
    \gamma_\mathrm{max}=s\,r_w\frac{\int_0^L\rho(x)f(x)B(x)\,\dd x}{\int_0^L f(x)B(x)\,\dd x}+O(s^2). \label{gammamax-pert_MT}
\end{equation}
which gives the invasion growth rate of mutants. Note that the first-order term is consistent with Ref.~\cite{ghosh2022emergent} (see SI Sec.~\ref{detgrowth-pert}). Eq.~\eqref{gammamax-pert_MT} shows that, to first order in $s$, the invasion growth rate is obtained by spatially averaging the mutant division rate advantage $s\, r_w \rho(x)$, weighing each position $x$ by $f(x)B(x)$, which is proportional to its long-term contribution to the future population. Beyond the perturbative regime, denoting by $\gamma_{\mathrm{WM}}$ the mutant invasion rate in the reference well-mixed population with the same average per capita division rate, we obtain
\begin{equation}
    \gamma_\mathrm{max}\geq\gamma_{\mathrm{WM}}, \quad \forall s>0,
\end{equation}
see SI Sec.~\ref{gamma-general}.
Hence, mutant invasion is accelerated in the gut compared to a matched well-mixed population. 

\begin{figure}[h!]
    \centering
    \includegraphics[width=\linewidth]{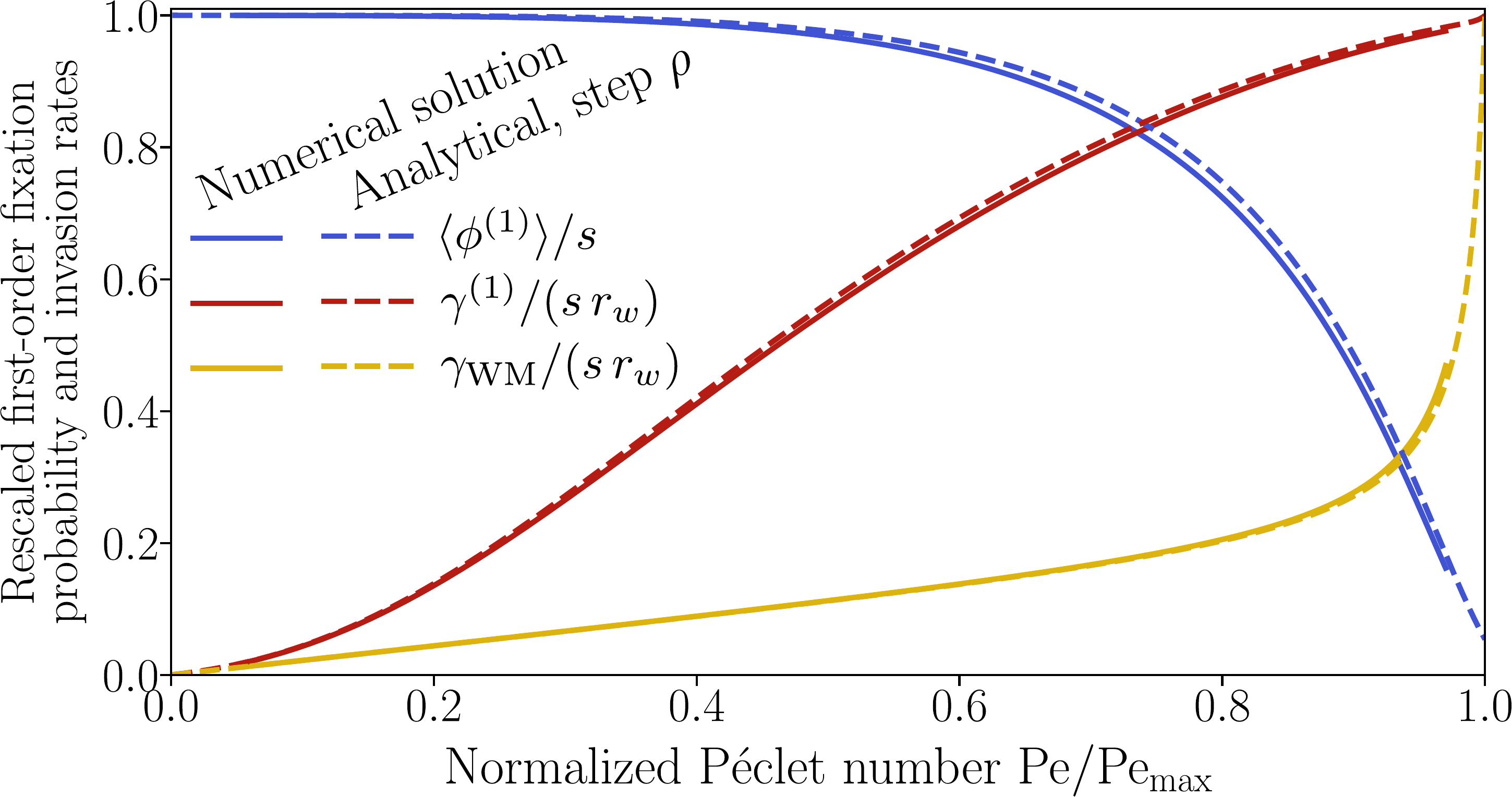}
    \caption{The rescaled first-order approximations of the division-event-weighted fixation probability $\langle \phi\rangle$ and of the mutant invasion growth rate $\gamma_\mathrm{max}$ are shown versus the normalized Péclet number Pe. Specifically, we plot $\langle \phi^{(1)} \rangle/s$ where $\langle \phi\rangle = \langle \phi^{(1)} \rangle + O(s^2)$ and $\gamma^{(1)}/(s\,r_w)$ where $\gamma_\mathrm{max} = \gamma^{(1)} + O(s^2)$ versus $\mathrm{Pe}/\mathrm{Pe}_\mathrm{max}$. For reference, we also show the rescaled mutant invasion growth rate $\gamma_\mathrm{WM}/(s\,r_w)$ in a well-mixed population with the same mean per-capita division rate. Note that $s\,r_w$ is the maximum possible invasion growth rate (see SI Sec.~\ref{detgrowth-general}). Parameter values: same $\lambda_w$ (and hence $\mathrm{Pe}_\mathrm{max}$), $k/F_\mathrm{in}$ and $\alpha$ as in Fig.~\ref{fig:fig1}. }
    \label{fig:fig4}
\end{figure}

Fig.~\ref{fig:fig4} shows that, to first order in $s$, the division-event-weighted fixation probability $\langle \phi \rangle$ decreases  with the Péclet number, while the mutant invasion growth rate $\gamma_\mathrm{max}$ grows with it. Increasing the importance of advection with respect to diffusion makes a mutant's fate strongly depend on where it appears, thus decreasing $\langle \phi \rangle$. However, it also increases the importance of fast-dividing regions in terms of long-term contributions to the population, thus increasing $\gamma_\mathrm{max}$.

\textit{Discussion.}  
Using a branching-process approach, we showed that spatial structure and transport in the gut reduce the fixation probability of beneficial mutants arising randomly upon division compared to the well-mixed case. The suppression factor is set by the spatial heterogeneity of mutant fate across the gut. We also showed that mutant invasion is faster in the gut than in a matched well-mixed population. These results hold for all positive relative mutant fitness advantages $s$, including large ones, which are relevant for adaptation to the gut~\cite{Barroso2014}. Furthermore, increased hydrodynamic flow reduces fixation probability but accelerates invasion. This combination is reminiscent of models on graphs with directional migration flows, where mutants with uniform $s$ feature suppression of natural selection and accelerated dynamics in the branching process regime for frequent migrations~\cite{abbara2023frequent,fruet2026environment}. It emphasizes that conditions yielding faster mutant invasion do not always reduce stochastic extinction~\cite{Alexander08}. 
While our fixation probability is consistent with Ref.~\cite{ghosh2022emergent} to leading order in $s$, it differs beyond that, and better agrees with simulation results (see SI Sec.~\ref{comp_GG}).
 
In the future, it will be interesting to investigate the fate of beneficial mutants in a more detailed gut model~\cite{salari2025diurnal} than the one-dimensional one~\cite{cremer2016effect,Cremer2017} used here for tractability. Incorporating clonal interference will be important as well~\cite{desai_beneficial_2007,Barroso2014}. Including ecological interactions between different bacterial strains would also be highly relevant, given the rich coupling between ecological and evolutionary dynamics~\cite{li2026emergent}.
More generally, the framework developed here can give insight in populations growing in systems with net flow, including microfluidic channels~\cite{tesser2017finite}, flow reactors~\cite{ballyk1998effects} and streams~\cite{speirs_population_2001,plummer_fixation_2019}.

\textit{Acknowledgments. } We thank Antoine Aragon for a preliminary exploration of this question by simulations. This research was partly funded by the Swiss National Science Foundation (SNSF, grant No.~315230 208196, to A.-F.~B.) and by the Chan-Zuckerberg Initiative (CZI, to A.-F.~B.). C.~L. acknowledges funding by the Agence Nationale de la Recherche (grants ANR-20-CE30-0001, ANR-21-CE45-0015 and ANR-25-CE44-4822, to C.~L.). AI tools (Claude Opus 5, 5.5, GPT~5.6 Sol, 6 Astra) were used for literature analysis, mathematical assistance on Sturm-Liouville problems, and proofreading. We checked all AI suggestions and take responsibility for the content.

\textit{Data availability. } Our stochastic simulation code is freely available at \url{https://github.com/Bitbol-Lab/Gut_evolutionary_dynamics}

\newpage

\onecolumngrid

\begin{center}
 {\LARGE \bf Supporting Information}   
\end{center}

\renewcommand{\thefigure}{S\arabic{figure}}
\renewcommand{\theHfigure}{S\arabic{figure}}
\setcounter{figure}{0}
\renewcommand{\thetable}{S\arabic{table}}
\renewcommand{\theHtable}{S\arabic{table}}
\setcounter{table}{0}
\renewcommand{\theequation}{S\arabic{equation}}
\renewcommand{\theHequation}{S\arabic{equation}}
\setcounter{equation}{0}

\tableofcontents

\newpage

\section{General results for the one-dimensional gut model}

\subsection{Equations of the model}
\label{sec:effective_1D_model}

We start from a description of transport and division in the gut from Ref. \cite{cremer2016effect}, which was extended in Ref. \cite{labavic2022hydrodynamic} to include neutral mutants, and in Ref.~\cite{ghosh2022emergent} to include beneficial mutants. We consider three fields: the food concentration $F(x,t)$, the wild type bacteria concentration $B(x,t)$, and the mutant concentration $M(x,t)$. They satisfy the following coupled partial differential equations:
\begin{empheq}[left=\empheqlbrace]{align}
\frac{\partial F}{\partial t}
&= D \frac{\partial^2 F}{\partial x^2}
 - v \frac{\partial F}{\partial x}
 - \frac{1}{\alpha}
 \frac{(r_w B+r_m M)F}{k+F}, \label{eq:eq_F}
 \\
\frac{\partial B}{\partial t}
&= D \frac{\partial^2 B}{\partial x^2}
 - v \frac{\partial B}{\partial x}
 + r_w \frac{BF}{k+F}, \label{eq:eq_B}
\\
\frac{\partial M}{\partial t}
&= D \frac{\partial^2 M}{\partial x^2}
 - v \frac{\partial M}{\partial x}
 + r_m \frac{MF}{k+F}, \label{eq:eq_M}
\end{empheq}
where $D$ is the effective diffusion constant and $v$ is the effective flow velocity. The wild-type and mutant maximal division rates are respectively $r_w$ and $r_m = r_w(1+s)$, where we introduced the mutant relative fitness advantage $s$. Given the definition of $s$, we will always assume that $s>-1$, and we will mainly focus on beneficial mutants with $s>0$. Moreover, $k$ is the Monod constant and $\alpha$ is a conversion factor. For simplicity, we assume that $F$, $B$ and $M$ are linear concentrations, i.e.\ they are integrated over the gut cross section, and $\int_0^L B(x,t)\dd x$ is the total number of wild-type bacteria in the system at time $t$.

The boundary conditions are 
\begin{empheq}[left=\empheqlbrace]{align}
\left[D \frac{\partial}{\partial x} (F;B;M) - v (F;B;M)\right] (x = 0) 
&=  (- v\,F_{\text{in}};0;0) ,\label{bc_fwd1}
\\
D \frac{\partial}{\partial x} (F;B;M) (x = L) 
&=  (0;0;0) ,\label{bc_fwd2}
\end{empheq}
where $F_{\text{in}}$ is the incoming food concentration at the upstream boundary, corresponding to a food inflow $v F_{\text{in}}$. 

\subsection{Ordinary differential equation for the fixation probability of a beneficial mutant} \label{sec:eq-fix}

\subsubsection{Discrete derivation} \label{discrete1}

\begin{figure}[htbp]
    \centering
    \includegraphics[width=0.5\linewidth]{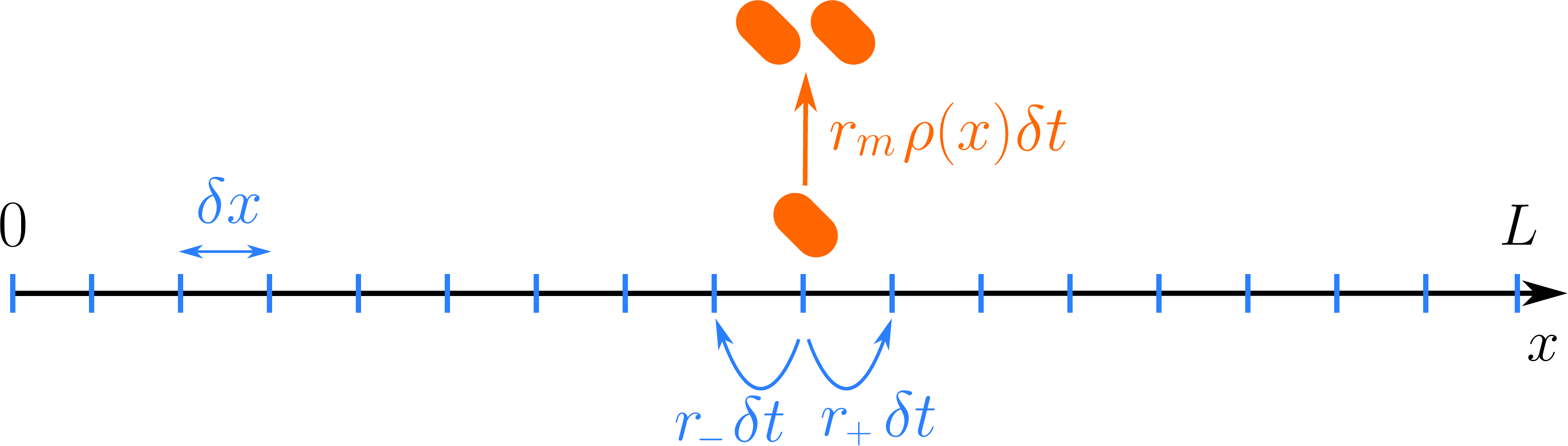}
    \caption{Schematic of the discretized one-dimensional model of the gut. The segment $[0,L]$ is discretized into $K+1$ sites located at $x_i=i\delta x$ with $0\leq i\leq K$, with $\delta x=L/K$ (light blue). Each mutant (resp.\ wild-type) at site $x_i$ divides with rate $r_m\rho(x_i)$ (resp.\ $r_w\rho(x_i)$) (shown in orange in the mutant case). Bacteria perform discrete jumps from their site of origin to its left (resp.\ right) neighbor with rate $r_-=D/(\delta x)^2$ (resp.\ $r_+=D/(\delta x)^2+v/\delta x$). There is no upstream jump from $x_0$, and there are only advective downstream jumps from $x_K$. Recall that in the continuous model~\cite{cremer2016effect,Cremer2017}, bacteria are transported with an advective velocity $v$ and an effective diffusion coefficient $D$ (see main text Fig.~1(a)). }
    \label{fig:fig1SI}
\end{figure}

To obtain an equation on the extinction probability $u(x)$, we reason on what happens to the mutant just after it appears. Here, we present an intuitive derivation in discrete space, where the $x$ axis is divided into $K+1$ sites located at $x_i=i\delta x$ with $0\leq i\leq K$, with $\delta x=L/K$, see Fig.~\ref{fig:fig1SI}. Afterwards, we will present a more rigorous derivation based on a Langevin equation describing the stochastic motion of mutant bacteria. 

In the discrete framework, a given bacterium at $x_i$ with $1\leq i\leq K-1$ (i.e.\ at a non-boundary site) may move to the neighboring site on the right (resp.\ left) with a rate $r_+=D/(\delta x)^2+v/\delta x$ (resp.\ $r_-=D/(\delta x)^2$), see Fig.~\ref{fig:fig1SI}. Boundary conditions will be addressed in Section~\ref{sec:discrete}. 
Let us consider a short time $\delta t$ such that, to first order in $\delta t$, at most one event may happen. During this time, either:
\begin{itemize}
    \item the mutant moves right with probability $r_+ \delta t$, and the extinction probability becomes $u(x+\delta x)$;
    \item the mutant moves left with probability $r_- \delta t$, and the extinction probability becomes $u(x-\delta x)$;
    \item the mutant divides, with probability $r_m\rho(x)\delta t$, and the extinction probability becomes $u^2(x)$ under the branching process approximation, where different mutant lineages do not interact;
    \item nothing happens, with probability $1-[r_++r_-+r_m\rho(x)]\delta t$, and the extinction probability remains $u(x)$.
\end{itemize}
This leads to 
\begin{align}
u(x)= r_+ \delta t u(x+\delta x) + r_- \delta t u(x-\delta x) + r_m\rho(x) \delta t u^2(x)  +\left[1-r_-\delta t - r_+\delta t - r_m\rho(x)\delta t \right] u(x) + o(\delta t) \,,
\end{align}
which gives
\begin{align}
    0= \frac{D}{\delta x^2}\left[u(x+\delta x)-2u(x) + u(x-\delta x)\right] +\frac{v}{\delta x} \left[u(x+\delta x)-u(x)\right]-r_m\rho(x)u(x)[1-u(x)]\,.
\end{align}
In the continuous limit $\delta x \rightarrow 0$, we obtain the differential equation satisfied by the extinction probability,
\begin{align}
    &Du''(x) + vu'(x) - r_m\rho(x) u(x)\left[1-u(x)\right] = 0. 
\end{align}
The fixation probability $\phi(x)=1-u(x)$ thus satisfies
\begin{align}
    &D\phi''(x) + v\phi'(x) + r_m\rho(x) \phi(x)\left[1-\phi(x)\right] = 0. \label{eq:phi1}
\end{align}

\subsubsection{Continuous derivation}

In our effective model of transport in the gut, with advection and effective diffusion, the Langevin equation for bacterial motion is
\begin{align}
    \dd X_t =  v \dd t + \sqrt{2D} \, \dd  W_t ,\label{eq:stoch_process_1D}
\end{align}
with $\dd t$ the infinitesimal time step, and $\dd  W_t$ the Wiener increment such that $\overline{ \dd  W_t} = 0$ and  $\overline{ \dd  {W_t}^2} = \dd t$. This is the Langevin equation associated to the transport part of the Fokker-Planck equation for the probability density of bacteria, or equivalently, for their concentration $B(x,t)$, i.e.\ to the transport part of Eq.~\eqref{eq:eq_B}.

We assume that a beneficial mutant arises in the bacterial population upon division at a position $x$ along the gut. We are interested in the early dynamics of the new mutant population, where its lineage is still rare. Thus, we make the approximation that the division rate profile 
\begin{equation}
    \rho(x,t)=\frac{F(x,t)}{k+F(x,t)}
\end{equation} 
is time-independent and matches the mutant-free stationary profile $ \rho(x)$. This assumes that the mutant is rare enough not to perturb $F$ and $B$, and that the food and wild-type bacteria profiles are initially at steady state.

Let $u(x)$ be the extinction probability of the mutant lineage for one mutant starting from $x$. Let us focus on the infinitesimal time interval $\dd t$ just after the mutant has appeared. This time interval is so short that at most one division event happens during it. During $\dd t$, either the mutant does not replicate, in which case it may be transported by advection or diffusion, or the mutant replicates. In the latter case, the two offspring mutant lineages are assumed to be independent within the branching process approximation, so that the mutant extinction probability becomes $u^2(x)$, and the spatial displacement only contributes at higher order. Summing over these two possibilities yields
\begin{align}
    u(x)  =    \left[1-r_m\rho(x)\dd t\right]\, \overline{u(x+ \dd X_t)} + r_m\rho(x)\dd t \,u^2(x) \label{ext_p_Eq_1_SI} + o(\dd t),
\end{align}
where the bar denotes the average over the stochastic increment $\dd X_t$ introduced in Eq.~\eqref{eq:stoch_process_1D}.
Expanding to second order in the stochastic increment and using $\overline{ \dd  X_t} = v \dd t$ and  $\overline{ (\dd  {X_t})^2} = 2D\dd t+o(\dd t)$, we obtain
\begin{align}
    \overline{u(x+ \dd X_t)} =u(x)  + [Du''(x) + vu'(x)] \dd t  + o(\dd t).\label{uxpdx}
\end{align}
Let us introduce the differential operator $\mathcal{L}_x$ such that 
\begin{equation}
\mathcal{L}_xu(x)=Du''(x) + vu'(x).
\label{bwd}
\end{equation}
We observe that it is the backward operator adjoint to the forward transport operator $\mathcal{L}_x^\dagger$ that enters Eq.~\eqref{eq:eq_B}, 
\begin{equation}
\mathcal{L}_x^\dagger B(x)=DB''(x) - vB'(x).
\label{fwd}
\end{equation}
Hence, $\mathcal{L}_x^\dagger$ acts on bacteria concentration $B$, while $\mathcal{L}_x$ acts on extinction probability $u$, which is a function of the initial mutant position. 
Combining Eq. (\ref{ext_p_Eq_1_SI}) with Eq.~\eqref{uxpdx} yields
\begin{align}
    \mathcal{L}_x u(x) - r_m\rho(x) u(x)[1-u(x)] = 0.
    \label{eq:u}
\end{align}
Assuming that once a mutant lineage has escaped early extinctions described within the branching process regime, it fixes in the whole population~\cite{patwa_fixation_2008}, the fixation probability $\phi(x)$ can be expressed as $\phi(x)=1-u(x)$. Hence, it satisfies the differential equation 
\begin{align}
    \mathcal{L}_x\phi(x) + r_m\rho(x) \phi(x)\left[1-\phi(x)\right] = 0. \label{eq:phi}
\end{align}

\subsection{Boundary conditions for the fixation probability profile of a beneficial mutant}
\label{sec:BBM}

In this section, we derive the boundary conditions for $\phi$. As for the bulk differential equation, it is possible to derive them using two methods, either reasoning on discrete jumps or directly in the continuous limit. Here, we present both, which allows us to check consistency.

\subsubsection{Discrete derivation} \label{sec:discrete}

For this derivation, we will use the discretization of the system introduced in Section~\ref{discrete1}, with $K+1$ sites $x_i=i\delta x$ for $0\leq i\leq K$.\\

\noindent\textbf{Boundary condition at $x=0$.} Let us assume that a mutant appears in the upstream boundary site $x_0=0$. 
Recall that the total transport flux vanishes at $x=0$, both for wild-types and for mutants: $DB'(0)-vB(0)=0$ and $DM'(0)-vM(0)=0$. We implement this zero-flux condition by allowing no upstream jumps from site $x_0$. 
Let us consider a short time $\delta t$ such that, to first order in $\delta t$, at most one event may happen. During this time, either:
\begin{itemize}
    \item the mutant moves right with probability $r_+\delta t$, where $r_+=D/(\delta x)^2+v/\delta x$, and the extinction probability becomes $u(x_1)$;
    \item the mutant divides, with probability $r_m\rho(x_0)\delta t$, and the extinction probability becomes $u^2(x_0)$ under the branching process approximation;
    \item nothing happens, with probability $1-[r_+ +r_m\rho(x_0)]\delta t$, and the extinction probability remains $u(x_0)$.
\end{itemize}
This leads to 
\begin{align}
u(x_0) = \left[\frac{v\,\delta t}{\delta x} + \frac{D\,\delta t}{\delta x^2} \right]u(x_1)+ r_m\rho(x_0) \delta t \,u^2(x_0)  +\left[1-\frac{v\,\delta t}{\delta x} - \frac{D\,\delta t}{\delta x^2} - r_m\rho(x_0)\delta t \right] u(x_0) + o(\delta t) \,.
\end{align}
Identifying the first-order terms in $\delta t$ yields
\begin{align}
    0&= \frac{D}{(\delta x)^2}\left[u(x_1)-u(x_0) \right]  +\frac{v}{\delta x} \left[u(x_1)-u(x_0)\right]-r_m\rho(x_0)u(x_0)[1-u(x_0)]\nonumber\\
    &= \frac{D}{\delta x}\left[u'(x_0)+ O(\delta x)\right]  +\frac{v}{\delta x} \left[u'(x_0)\delta x+ O((\delta x)^2)\right]-r_m\rho(x_0)u(x_0)[1-u(x_0)]\,.
\end{align}
This yields $Du'(x_0)+O(\delta x)=0$. Since $x_0=0$, we obtain
\begin{align}
    u'(0) =0\,.
\end{align}
Finally, using $\phi(x)=1-u(x)$ yields the boundary condition on $\phi$ at $x=0$:
\begin{align}
    \phi'(0) = 0\,.
\end{align}

\noindent\textbf{Boundary condition at $x=L$.} Let us assume that a mutant appears in the downstream boundary site $x_K=L$. 
Recall that we assume that diffusive flux is zero at $x=L$, both for wild-type and for mutant bacteria: $M'(L)=B'(L)=0$. We implement this condition by suppressing downstream diffusive jumps from site $x_K$. 
Let us consider a short time $\delta t$ such that, to first order in $\delta t$, at most one event may happen. During this time, either:
\begin{itemize}
    \item the mutant moves right due to advection with probability $v \delta t/\delta x$, and exits the system: the extinction probability becomes 1;
    \item the mutant moves left due to diffusion with probability $D \delta t/(\delta x)^2$, and the extinction probability becomes $u(x_{K-1})$;
    \item the mutant divides, with probability $r_m\rho(x_K)\delta t$, and the extinction probability becomes $u^2(x_K)$ under the branching process approximation;
    \item nothing happens, with probability $1-[v/\delta x+D /(\delta x)^2+r_m\rho(x_K)]\delta t$, and the extinction probability remains $u(x_K)$.
\end{itemize}
This leads to 
\begin{align}
u(x_K) = \frac{v\,\delta t}{\delta x}   + \frac{D\,\delta t}{\delta x^2}  u(x_{K-1})+ r_m\rho(x_K) \delta t \,u^2(x_K)  +\left[1-\frac{v\,\delta t}{\delta x} - \frac{D\,\delta t}{\delta x^2} - r_m\rho(x_K)\delta t \right] u(x_K) + o(\delta t) \,.
\end{align}
Identifying the first-order terms in $\delta t$ yields
\begin{align}
    0&= -\frac{D}{\delta x^2}\left[u(x_K)-u(x_{K-1}) \right] +\frac{v}{\delta x} \left[1-u(x_K)\right]-r_m\rho(x_K)u(x_K)[1-u(x_K)]\nonumber\\
    &= -\frac{D}{\delta x}\left[u'(x_K)+ O(\delta x) \right] +\frac{v}{\delta x} \left[1-u(x_K)\right]-r_m\rho(x_K)u(x_K)[1-u(x_K)]\,.
\end{align}
This yields $-Du'(x_K) +v \left[1-u(x_K)\right]+O(\delta x)=0$. Since $x_K=L$, we obtain
\begin{align}
    -Du'(L) +v \left[1-u(L)\right]=0\,.
\end{align}
Finally, using $\phi(x)=1-u(x)$ yields the boundary condition on $\phi$ at $x=L$:
\begin{align}
    D\phi'(L) + v\phi(L) = 0\,.
\end{align}

\subsubsection{Continuous derivation}

Let us consider the two transport operators $\mathcal{L}_x$ and $\mathcal{L}_x^\dagger$. In the bulk, they are defined respectively by Eqs.~\eqref{bwd} and~\eqref{fwd}. The forward operator $\mathcal{L}_x^\dagger$ describes the evolution of probability densities associated with the stochastic transport process considered, while the backward one $\mathcal{L}_x$ acts on quantities that depend on the initial position of the stochastic process. Since both operators describe the same stochastic process, the boundary conditions on $\phi$ should be chosen so that they are adjoint on the whole interval $[0,L]$. Hence, we impose the condition
\begin{align}
    \int_0^L \psi\, \mathcal{L}_x \phi \,\dd x = \int_0^L \phi\, \mathcal{L}_x^{\dagger} \psi \,\dd x , \label{cstrt}
\end{align}
where $\psi$ is any function on which the operator $\mathcal{L}_x^\dagger$ is allowed to act. Thus, $\psi$ should be sufficiently smooth and satisfy 
the same boundary conditions as $B$ in Eqs.~\eqref{bc_fwd1} and~\eqref{bc_fwd2}, namely
\begin{align}
    D \psi'(0) - v \psi(0) &= 0, \label{bc1_bwd}\\
    \psi'(L) &= 0  .\label{bc2_bwd}
\end{align}
Let us consider 
\begin{align} 
  \int_0^L \phi \mathcal{L}_x^{\dagger} \psi \,\dd x = \int_0^L \phi(x) \left[ D \psi''(x) - v \psi'(x) \right]\, \dd x.
\end{align}
After integration by parts (IBP), it yields
\begin{align}
    \int_0^L \phi \mathcal{L}_x^{\dagger} \psi \,\dd x  =  - \int_0^L \phi'(x) \left[ D\psi'(x) - v\psi(x) \right]\, \dd x +\left[ \phi(x) \left(D \psi'(x) - v \psi(x) \right) \right]_0^L.
\end{align}
Using Eqs.~\eqref{bc1_bwd} and~\eqref{bc2_bwd}, the expression simplifies and a second IBP gives
\begin{align}
    \int_0^L \phi \mathcal{L}_x^{\dagger} \psi \,\dd x  & = \int_0^L \psi(x) \left[ D\phi''(x) + v\phi'(x) \right] \dd x -\left[ \phi'(x) D \psi(x) \right]_0^L  - v \psi(L) \phi(L). 
\end{align}
Imposing Eq.~\eqref{cstrt} then yields
\begin{align}
    - \psi(L) [  D\phi'(L) + v \phi(L)]   + \psi(0)D \phi'(0) =0.
\end{align}
As this identity must be satisfied for all admissible forward functions $\psi$, whose values at $x=0$ and $x=L$ can be varied independently, we obtain 
\begin{align}
 \phi'(0) & = 0, \label{bcphi1}\\
    D\phi'(L) + v \phi(L) & = 0 ,\label{bcphi2}
\end{align}
which yields the boundary conditions associated to the fixation probability $\phi$. Note that the adjointness condition on the transport operator fixes the boundary conditions of the full equation for $\phi$, because the reaction term involves no derivatives (and thus, it produces no boundary terms).

\subsection{Fixation probability in the branching process approximation} \label{secfixgal}

\subsubsection{Critical point} \label{critpt}

Let us first identify the critical point where a nonzero fixation probability appears. The differential equation for the fixation probability, Eq.~\eqref{eq:phi}, can be written as
\begin{equation}
    \mathcal{A}_s\phi(x)-(1+s)r_w\rho(x)\phi^2(x)=0, \label{eqphi_A}
\end{equation}
where we introduced the backward operator
\begin{equation}
    \mathcal{A}_s=\mathcal{L}_x+(1+s)r_w\rho(x), \label{def_As}
\end{equation}
with $\mathcal{L}_x$ denoting the backward transport operator defined in Eq.~\eqref{bwd}. 
We observe that $\phi=0$ is always a solution of Eq.~\eqref{eqphi_A}. We are interested in the onset of a positive solution of Eq.~\eqref{eqphi_A}. To determine it, we linearize Eq.~\eqref{eqphi_A} around $\phi=0$, yielding $\mathcal{A}_s\phi(x)=0$, and we search for the point where the largest eigenvalue of the linearized operator $\mathcal{A}_s$ crosses zero~\cite{englander_local_2004}. 

Although $\mathcal{A}_s$ is not self-adjoint because of the advective term, its eigenvalue problem
\begin{equation}
    \mathcal{A}_s q(x)=\gamma\, q(x),
    \label{eqAs_bwd_eig}
\end{equation}
with the backward boundary conditions Eqs.~\eqref{bcphi1} and~\eqref{bcphi2}, can be transformed into a self-adjoint Sturm-Liouville problem~\cite{teschl_book} by introducing $y(x)=e^{vx/(2D)} q(x)$. Indeed, Eq.~\eqref{eqAs_bwd_eig} then becomes
\begin{equation}
    -D y''(x)+\left[\frac{v^2}{4D}-(1+s)r_w\rho(x)\right]y(x)=-\gamma\, y(x),
    \label{SL-bulk}
\end{equation}
with boundary conditions
\begin{equation}
    y'(0)=\frac{v}{2D}y(0),\qquad y'(L)=-\frac{v}{2D}y(L).
    \label{SL-bc}
\end{equation}
In such a regular Sturm-Liouville problem with separated boundary conditions, all eigenvalues are real and simple. Furthermore, the smallest eigenvalue $-\gamma_\mathrm{max}$ is associated with an eigenfunction that can be chosen positive, while all other eigenfunctions change sign~\cite{teschl_book}. This entails that the largest (or principal) eigenvalue $\gamma_\mathrm{max}$ of $\mathcal{A}_s$ is simple, and is associated with an eigenfunction $q(x)=e^{-vx/(2D)}y(x)$ that can be chosen positive (i.e.\ $q(x)>0$ for $x\in(0,L)$), while all other eigenfunctions of $\mathcal{A}_s$ change sign.

Let us consider 
\begin{equation}
f(x)=B(x) e^{-vx/D}, \label{deff}
\end{equation}
where $B(x)$ is the steady-state wild-type bacteria concentration profile in the absence of mutants, i.e.\ the solution to Eqs.~\eqref{eq:eq_B},~\eqref{bc_fwd1} and~\eqref{bc_fwd2} for $M=0$. 
In the case of a neutral mutant, $s=0$, it satisfies
\begin{equation}
    \mathcal{A}_0f(x)=\mathcal{L}_xf(x)+r_w\rho(x)f(x)=Df''(x) + vf'(x)+r_w\rho(x)f(x)=0, \label{eq:A0f}
\end{equation}
as well as the backward boundary conditions Eqs.~\eqref{bcphi1} and~\eqref{bcphi2}. Since $B(x)>0$ for $x\in[0,L]$, we have $f(x)>0$ for $x\in[0,L]$. The corresponding eigenvalue, $0$, is therefore the principal eigenvalue of $\mathcal{A}_0$. Moreover, since $\mathcal A_s=\mathcal A_0+s\,r_w\rho(x)$ (see Eq.~\eqref{def_As}), $\mathcal A_s$ differs from $\mathcal A_0$ by a growth term whose sign is that of $s$. Hence, the principal eigenvalue of $\mathcal{A}_s$ is larger than that of $\mathcal{A}_0$ for $s>0$ and smaller for $s<0$: it crosses zero at
$s=0$. This entails that the critical point where a nonzero fixation probability appears is $s=0$. Note that this result is consistent with the fixation probability expressions within the branching process approximation in well-mixed populations~\cite{Haldane27} and in deme-structured populations~\cite{abbara2023frequent}. Note also that the function introduced in Eq.~\eqref{deff} coincides with the one in Eq.~6 of Ref.~\cite{ghosh2022emergent} up to normalization when $D$ and $v$ are constant.

\subsubsection{Moderate selection regime}
\label{phi-expansion}

Let us perform a perturbative calculation of the fixation probability in the regime where the relative selective advantage $s$ of the mutant, defined via $r_m=(1+s)r_w$, is small: $s\ll 1$. Within the branching-process approximation, the fixation probability vanishes at the critical point $s=0$. Thus, we write 
\begin{equation}
    \phi(x)=s\phi_1(x)+s^2\phi_2(x)+O(s^3). \label{phiexpansion}
\end{equation}
Note that both the present expansion and the branching process approximation neglect the fact that neutral mutants have a nonzero fixation probability in finite populations. In particular, this expansion requires $1/N_\mathrm{active}\ll s\ll 1$, where $N_\mathrm{active}$ is the active population size which controls the fixation of a neutral mutant appearing randomly upon division in the system~\cite{labavic2022hydrodynamic}. Let us call this regime the moderate selection regime. We will give a more precise definition of the moderate selection regime in Section~\ref{sec:neutral}.

Substituting Eq.~\eqref{phiexpansion} into the differential equation for $\phi$ Eq.~\eqref{eq:phi} and identifying terms of first and second order in $s$ yields:
\begin{align}
    \mathcal{A}_0\phi_1(x)=D\phi_1''(x)+v\phi'_1(x)+r_w\rho(x)\phi_1(x)&=0, \label{eqphi1}\\
    \mathcal{A}_0\phi_2(x)=D\phi_2''(x)+v\phi'_2(x)+r_w\rho(x)\phi_2(x)&=r_w\rho(x)\phi_1(x) [\phi_1(x)-1]. \label{eqphi2}
\end{align}
In addition, the boundary conditions on $\phi$, namely Eqs.~\eqref{bcphi1} and~\eqref{bcphi2}, also apply to $\phi_1$ and $\phi_2$. 

We notice that any multiple of $f$ defined in Eq.~\eqref{deff} satisfies Eqs.~\eqref{eqphi1},~\eqref{bcphi1} and~\eqref{bcphi2}. As shown above, zero is the principal eigenvalue of $\mathcal A_0$. Since this eigenvalue is simple~\cite{teschl_book}, we have
\begin{equation}
    \phi_1(x)=A f(x)=A B(x) e^{-vx/D}, \label{phi1}
\end{equation}
where $A$ is a constant. 

Next, let us consider Eq.~\eqref{eqphi2}: $\mathcal{A}_0\phi_2(x)=r_w\rho(x)\phi_1(x) [\phi_1(x)-1]$. Multiplying both sides by $B(x)$ and integrating between $0$ and $L$ gives:
\begin{equation}
\int_0^L B(x)\mathcal{A}_0 \phi_2(x) \,\dd x=r_w\int_0^L\rho(x)B(x)\phi_1(x) [\phi_1(x)-1]\,\dd x.
\end{equation}
Let us introduce the adjoint $\mathcal{A}_0^\dagger$ of $\mathcal{A}_0$ as
\begin{equation}
    \mathcal{A}_0^\dagger=\mathcal{L}_x^\dagger+r_w\rho(x), \label{def_A0dag}
\end{equation}
where $\mathcal{L}_x^\dagger$ denotes the forward transport operator defined in Eq.~\eqref{fwd}. Performing integrations by parts, and using the adjointness of $\mathcal A_0$ and $\mathcal A_0^\dagger$ together with the backward and forward boundary conditions, we obtain
\begin{equation}
\int_0^L \phi_2(x)\mathcal{A}_0^\dagger B(x) \,\dd x=r_w\int_0^L\rho(x)B(x)\phi_1(x) [\phi_1(x)-1]\,\dd x.
\end{equation}
As $B$ satisfies the stationary Eq.~\eqref{eq:eq_B}, i.e.\ $\mathcal A_0^\dagger B(x)=0$, the left-hand side of this equation vanishes, leading to
\begin{equation}
\int_0^L\rho(x)B(x)\phi_1(x) [\phi_1(x)-1]\,\dd x=0. \label{integrphi1}
\end{equation}
Combining Eq.~\eqref{phi1} with Eq.~\eqref{integrphi1} and assuming $A\neq 0$ gives
\begin{equation}
    A=\frac{\int_0^L \rho(x)B(x)f(x)\dd x}{\int_0^L \rho(x)B(x)f^2(x)\dd x}.
\end{equation}
Let us define the division-event-weighted average $\left<q\right>$ of a quantity $q(x)$~\cite{labavic2022hydrodynamic,ghosh2022emergent} as:
\begin{align}
     \left<q\right>  = \frac{\int_0^L \dd x \,\rho(x) B(x)q(x) }{\int_0^L \dd x\, \rho(x) B(x)}.\label{rwa}
\end{align}
We can then write $A=\left<f\right>/\,\left<f^2\right>$.

Assembling our results gives the mutant fixation probability $\phi$ in the moderate selection regime:
\begin{equation}
    \phi(x)=s\frac{\left<f\right>}{\left<f^2\right>}f(x)+O(s^2)=s\frac{\left<f\right>}{\left<f^2\right>}B(x)e^{-vx/D}+O(s^2), \label{phi_order1}
\end{equation}
where $B(x)$ is the steady-state wild-type bacteria concentration profile in the absence of mutants, i.e.\ the solution to Eqs.~\eqref{eq:eq_B},~\eqref{bc_fwd1} and~\eqref{bc_fwd2} for $M=0$. Note that, although it was obtained using a different approach (see Section~\ref{comp_GG}) and formulated differently, the first-order result in Eq.~\eqref{phi_order1} is consistent with Ref.~\cite{ghosh2022emergent} in the limit $L\to\infty$ (see their Eq.~S112, with $s_e=s$ from Eq.~S100 without cell death, $N_e$ from Eq.~10 and $u(x)$ from Eq.~6 when $D$ and $v$ are constant). 

The division-event-weighted average of $\phi$ reads
\begin{equation}
    \left<\phi\right>=s\frac{\left<f\right>^2}{\left<f^2\right>}+O(s^2)=s\frac{1}{1+c_v^2(f)}+O(s^2), \label{phibar_order1}
\end{equation}
where we introduced the coefficient of variation $c_v(f)=\text{std}(f)/\left<f\right>$, with std denoting the standard deviation using the division-event-weighted average: $\text{std}^2(f)=\left<f^2\right>-\left<f\right>^2$. 
Eq.~\eqref{phibar_order1} demonstrates that, to leading order in $s$, we have $\left<\phi\right>\leq s$ in the gut. In addition, the inequality is strict whenever $f(x)$ varies spatially over the region where bacteria reproduce. Hence, to leading order in the moderate selection regime for beneficial mutants, the specific spatial structure and environmental gradients in the gut suppress natural selection compared to a well-mixed Moran process~\cite{Moran58,Ewens79}, which gives $\phi=s+O(s^2)$ in the limit of large population sizes.

From a formal point of view, we note that the derivation of the first-order approximation of the fixation probability in the moderate selection regime is similar to that presented in Ref.~\cite{abbara2023frequent} in the case of populations on graphs. In particular, the second-order equation allows us to determine the amplitude of the first-order term. Both calculations are branching process calculations, but the present one involves dependencies on a continuous variable $x$, while the one in Ref.~\cite{abbara2023frequent} involves dependencies on a discrete variable $i$ that labels graph nodes. 

\subsubsection{Beyond moderate selection} \label{compMoranGal}

The ordinary differential equation on $\phi$, Eq.~\eqref{eq:phi1}, can be written as
\begin{equation}
    \mathcal{A}_0\phi(x)=(1+s)r_w\rho(x)\phi^2(x)-sr_w\rho(x)\phi(x).
\end{equation}
Inspired by the above reasoning on $\phi_2$, let us multiply both sides by $B(x)$ and integrate between $0$ and $L$. It gives:
\begin{equation}
\int_0^L B(x)\mathcal{A}_0 \phi(x) \,\dd x=(1+s)r_w\int_0^L\rho(x)B(x)\phi^2(x)\,\dd x-sr_w\int_0^L\rho(x)B(x)\phi(x)\,\dd x.
\end{equation}
Performing integrations by parts, and using the adjointness of $\mathcal A_0$ and $\mathcal A_0^\dagger$ together with the backward and forward boundary conditions, we obtain
\begin{equation}
\int_0^L \phi(x)\mathcal{A}_0^\dagger B(x) \,\dd x=(1+s)r_w\int_0^L\rho(x)B(x)\phi^2(x)\,\dd x-sr_w\int_0^L\rho(x)B(x)\phi(x)\,\dd x.
\end{equation}
As $B$ satisfies $\mathcal A_0^\dagger B(x)=0$, the left-hand side of this equation vanishes, leading to
\begin{equation}
(1+s)\int_0^L\rho(x)B(x)\phi^2(x)\,\dd x=s\int_0^L\rho(x)B(x)\phi(x)\,\dd x
\end{equation}
Dividing both sides by $\int_0^L\rho(x)B(x)\,\dd x$, and recognizing division-event-weighted averages, we obtain:
\begin{equation}
    (1+s)\langle\phi^2\rangle=s\langle\phi\rangle,
\end{equation}
which gives
\begin{equation}
    \langle\phi^2\rangle=\frac{s}{1+s}\langle\phi\rangle. \label{moyPhi2-1}
\end{equation}

Let us consider the coefficient of variation of the fixation probability $c_v(\phi)=(\left<\phi^2\right>-\left<\phi\right>^2)^{1/2}/\left<\phi\right>$. We have
\begin{equation}
    \langle\phi^2\rangle=\left[c_v^2(\phi)+1\right]\langle\phi\rangle^2. \label{moyPhi2-2}
\end{equation}
Combining Eqs.~\eqref{moyPhi2-1} and~\eqref{moyPhi2-2} and using $\langle\phi\rangle>0$ (which holds for $s>0$) yields
\begin{equation}
    \langle\phi\rangle=\frac{s}{1+s}\frac{1}{1+c_v^2(\phi)}\leq \frac{s}{1+s}. \label{compMoran}
\end{equation}
Hence, the division-event-weighted average $\langle\phi\rangle$ of the fixation probability cannot exceed the large-population limit of the Moran process, which constitutes the well-mixed reference. This generalizes our first-order result above, and shows that the spatial structure of the gut can only decrease the division-event-weighted average of the fixation probability of beneficial mutants. Note that Eq.~\eqref{compMoran} reduces to Eq.~\eqref{phibar_order1} to first order in $s$ (as it should), as we then have $\phi(x)=sA f(x)+O(s^2)$ (see Eq.~\eqref{phi1}), which yields $c_v^2(\phi)=c_v^2(f)+O(s)$.

\begin{figure}[htbp]
    \centering
    \includegraphics[width=0.55\linewidth]{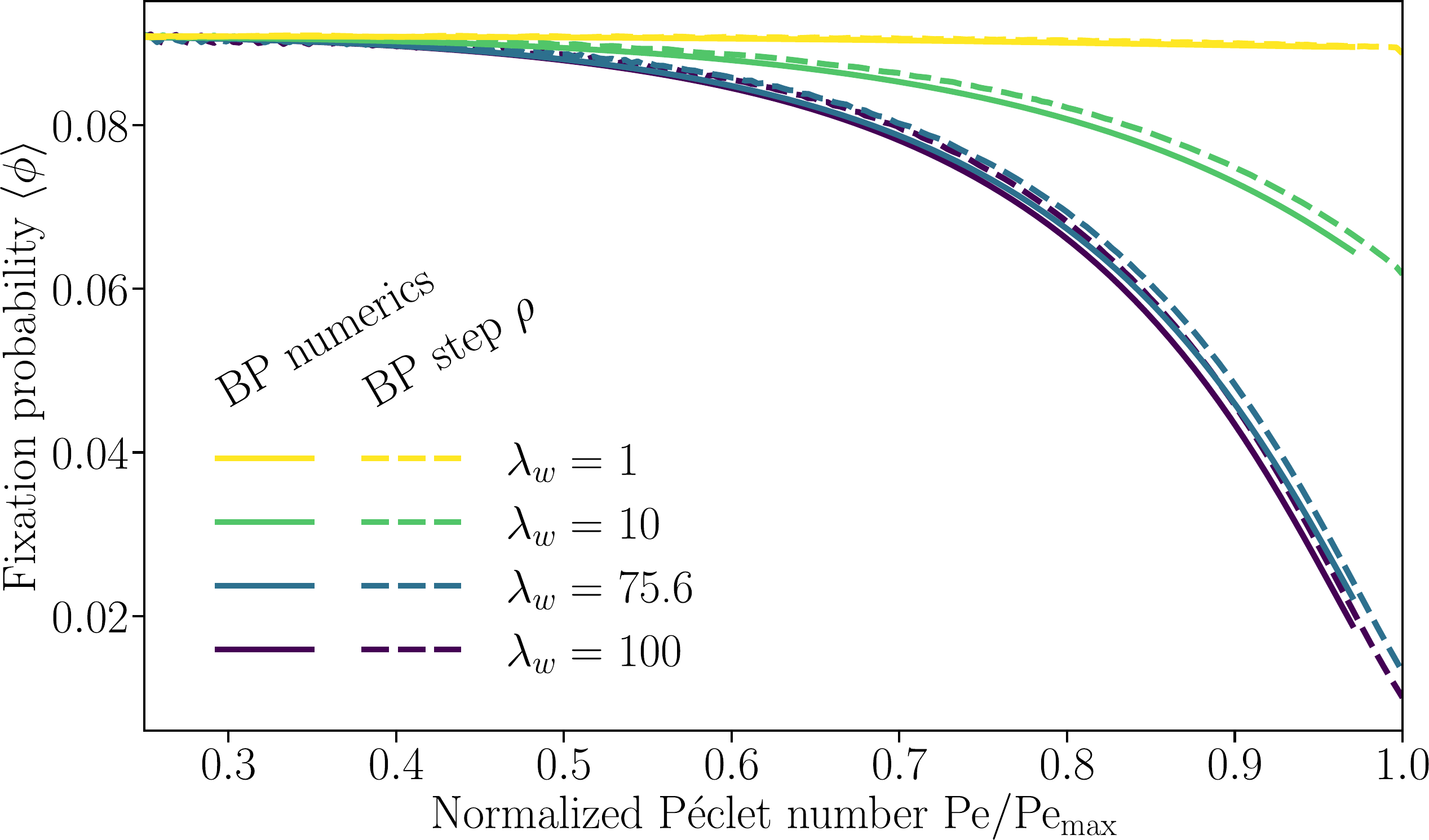}
    \caption{Division-event-weighted mutant fixation probability $\langle\phi\rangle$ versus normalized Péclet number $\mathrm{Pe}/\mathrm{Pe}_\mathrm{max}$, for a mutant relative fitness advantage $s=0.1$. Numerical solutions of the branching process Eq.~\eqref{eq:phi1} use either the steady-state mutant-free profile of the division rate $\rho$ (``BP numerics'') or its step approximation (``BP step $\rho$''). Note that the Moran model for large populations, which is the well-mixed reference, gives $\langle\phi\rangle=s/(1+s)=0.09$ for $s=0.1$, which matches with the result obtained for small $\mathrm{Pe}$ here. Parameters: same $k/F_\mathrm{in}$ and $\alpha$ as in main text Figs.~1, 2 and 3.
    }
    \label{fig:phi_Pe}
\end{figure}

Fig.~\ref{fig:phi_Pe} shows that the division-event-weighted mutant fixation probability $\langle\phi\rangle$ decreases as the Péclet number increases, and that this decrease is more pronounced for larger values of $\lambda_w= r_w L^2 / D$. Indeed, spatial heterogeneities in the system, and in particular in mutant fate, increase as advection becomes dominant over diffusion (larger $\mathrm{Pe}$) and as bacteria divide faster (larger $\lambda_w$).

\begin{figure}[htbp]
    \centering
    \includegraphics[width=0.55\linewidth]{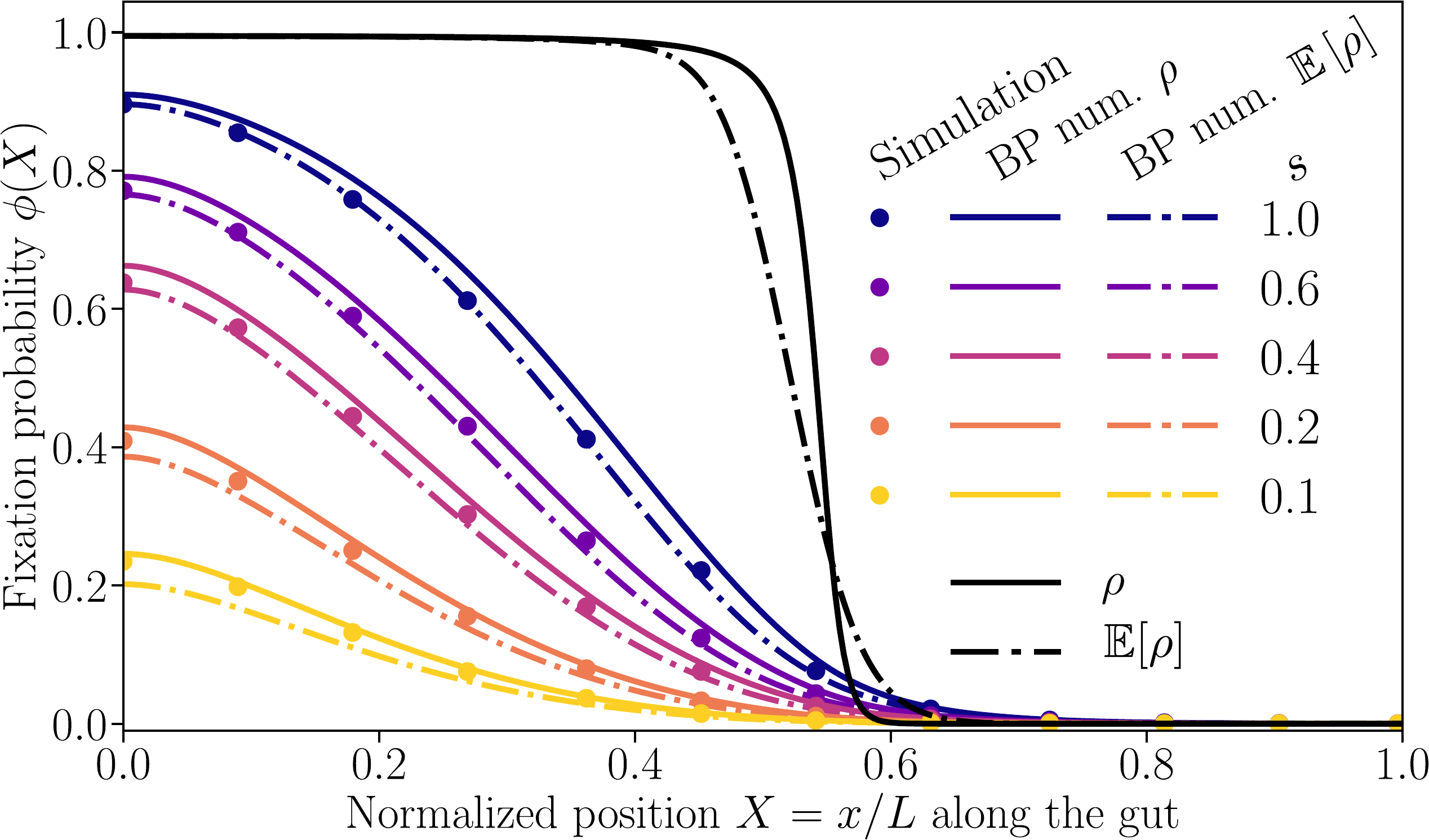}
    \caption{Mutant fixation probability $\phi(X)$ versus normalized initial mutant position $X=x/L$ along the gut, for different values of the mutant relative fitness advantage $s$. Stochastic simulations are described in Section~\ref{sec:stochsim}. Numerical solutions of the branching process Eq.~\eqref{eq:phi1} use either the steady-state mutant-free profile of the division rate $\rho(x)$ from numerical integrations (``BP num. $\rho$'') or the expectation value $\mathbb{E}\left[\rho(x)\right]=\mathbb{E}\left[F(x)/(k+F(x))\right]$ of $\rho(x)$ from stochastic simulations (``BP num. $\mathbb{E}[\rho]$''). The profiles of $\rho$ and of $\mathbb{E}[\rho]$ are shown as well. Parameter values: same as in main text Figs.~1 and 2. Simulation results and pure numerical solutions (``BP num. $\rho$'') are the same as in main text Fig.~2.
    }
    \label{fig:phi_2profiles}
\end{figure}

Fig.~\ref{fig:phi_2profiles} shows a comparison between our predictions and the results of stochastic simulations for the mutant fixation probability $\phi(x)$ of a mutant appearing at position $x$ along the gut, as in main text Fig.~2, but including an additional variant for the calculation of the predictions. Specifically, our predictions are plotted using a numerical solution of Eq.~\eqref{eq:phi1}. This requires first computing the steady-state division rate profile $\rho(x)=F(x)/[k+F(x)]$ in the absence of mutants. Here, we compute it two ways. The first one uses the food concentration profile $F(x)$ determined by numerical integration of Eqs.~\eqref{eq:eq_F} and \eqref{eq:eq_B} at steady state (see Section~\ref{numerics} for details), and is the same as in Fig.~2 of the main text. The second one uses the expectation value $\mathbb{E}\left[\rho(x)\right]=\mathbb{E}\left[F(x)/(k+F(x))\right]$ of $\rho(x)$ directly obtained from stochastic simulations. We observe in Fig.~\ref{fig:phi_2profiles} that these two estimates of the division rate profiles differ. This is due to the fluctuations from finite-size effects which exist in stochastic simulations but are not taken into account in the deterministic partial differential equation description. Note that in our stochastic simulations, the total number of bacteria is in the $1200-2400$ range, see Section~\ref{sec:stochsim}. The discretization employed in stochastic simulations could also play a role in the observed discrepancy, but in practice, decreasing spatial and temporal discretization steps resulted in negligible changes here. We observe in Fig.~\ref{fig:phi_2profiles} that the predictions of fixation probability $\phi(x)$ from Eq.~\eqref{eq:phi1} differ as a result of these differences in $\rho(x)$. Stochastic simulation results lie between these two cases. This suggests that the small but systematic discrepancies observed in main text Fig.~2 between our predictions and the results of stochastic simulations can be attributed to finite-size effect fluctuations.

\subsection{Mean number of descendant branches across successive generations}
\label{sec-Rn}

\subsubsection{Mean number of first-generation offspring branches}

\noindent\textbf{Definition.} 
Let us consider a single mutant bacterium starting at $x$, and track its divisions~\cite{barril_reproduction_2017}. At each division, one daughter bacterium is considered as continuing the parental branch, while the other is considered as a new offspring branch. In this convention (reminiscent of a ``spine decomposition'' viewpoint in branching process theory~\cite{englander_local_2004}), the number of division events along the parental branch is equal to the number of first-generation offspring branches it produces. Let us define the mean number $R_1(x)$ of division events along the parental branch, before the tracked parental branch exits the system, as: 
\begin{align}
    R_1(x) = r_m\int_0^{+\infty} \dd t\int_0^{L}\dd y\,\rho(y)  p(y,t|x),
    \label{eq:R1def0}
\end{align}
where $p(y,t|x)$ is the probability density for the tracked parental branch, initially located at $x$ at time $t=0$, to be at position $y$ at time $t$. In other words, the number $R_1(x)$ is the expected integrated division rate along the tracked parental branch trajectory before it exits the system. 
As division does not change the position or transport dynamics of the bacterium continuing the parental branch, its position changes by transport alone. Hence, the probability density $p(y,t|x)$ satisfies both the backward and forward Kolmogorov equations
\begin{equation}
    \frac{\partial p}{\partial t}(y,t|x) = \mathcal{L}_x p(y,t|x),  \;\;\textrm{and}\;\;  \frac{\partial p}{\partial t}(y,t|x) = \mathcal{L}_y^\dagger p(y,t|x),
\end{equation}
with the associated backward and forward boundary conditions and the initial condition
\begin{equation}
    p(y,0|x)=\delta(y-x).
\end{equation}
Note that trajectories that have already exited the system no longer contribute to $p$. Assuming that the parental branch eventually exits the system, we have $p(y,t|x)\to 0$ for $t\to\infty$. 
Let us introduce the Green function $G$ of the backward transport operator $\mathcal{L}_x$ defined in Eq.~\eqref{bwd} and acting on $x$. It satisfies
\begin{align}
    \mathcal{L}_x G(y|x) = - \delta(x-y).
\end{align}
Since $G(y|x)$, seen as a function of $x$, is a backward observable of the same stochastic transport process as $\phi$, it satisfies the same boundary conditions, namely
\begin{align}
    \partial_x G(y|0) = 0,\\
    D\partial_x G(y|L) + v G(y|L) = 0.
\end{align}
It can be expressed as
\begin{align}
G(y|x) = \int_0^{+\infty} \dd t\,p(y,t|x).
\end{align}
Indeed, we have
\begin{align}
    \mathcal{L}_x\int_0^{+\infty} \dd t\, p(y,t|x) = \int_0^{+\infty} \dd t\,\mathcal{L}_x p(y,t|x)  = \int_0^{+\infty} \dd t\,\frac{\partial p}{\partial t }(y,t|x) = \left[ p(y,t|x) \right]_0^{+\infty} = - \delta(x-y).
\end{align}
Hence, Eq.~\eqref{eq:R1def0} can be expressed as
\begin{align}
    R_1(x) = r_m\int_0^L \dd y\,\rho(y) G(y|x).
    \label{eq:R1def}
\end{align}
Furthermore, the Green function can be expressed explicitly as
\begin{equation}
    G(y|x)=
    \begin{cases}
        \dfrac{1}{v}\exp\left[-\dfrac{v}{D}(x-y)\right], & x>y,\\[7pt]
        \dfrac{1}{v}, & x<y.
    \end{cases}
    \label{Gexplicit}
\end{equation}
This yields
\begin{equation}
    R_1(x)=\frac{r_m}{v}\left[\int_0^x \dd y\,\rho(y)e^{-v(x-y)/D} +\int_x^L \dd y\,\rho(y)\right]. \label{eq:R_eq}
\end{equation}

\noindent\textbf{Differential equation for $R_1$.} 
We observe that
\begin{align}
    \mathcal{L}_x R_1(x) &= r_m\int_0^L \dd y\,\rho(y) \mathcal{L}_x G(y|x)  = -r_m\int_0^L \dd y\,\rho(y) \delta(x-y) = - r_m \rho(x).
\end{align}
Hence, $R_1$ satisfies the ordinary differential equation
\begin{align}
    \mathcal{L}_x R_1(x) = DR_1''(x) + vR_1'(x) =- r_m\rho(x) .\label{eq:eq_R1}
\end{align}
In addition, the boundary conditions on $G$ lead to
\begin{align}
    R_1'(0) = 0, \label{bc_bwd1_mt}\\
    DR_1'(L) + v R_1(L) = 0. \label{bc_bwd2_mt}
\end{align}
Note that the solution can be expressed as
\begin{align}
    R_1(x) = \frac{r_m}{D} \int_{x}^{L} \dd x' e^{-v x' /D}  \int_{0}^{x'}  \dd y\, e^{v y /D} \rho(y) + \frac{r_m\,e^{-v L /D}}{v} \int_{0}^{L}  \dd x' e^{v x' /D} \rho(x'),
    \label{R1sol}
\end{align}
which gives back Eq.~\eqref{eq:R_eq} after integration by parts.\\

\noindent\textbf{Alternative derivation of the differential equation for $R_1$.} One can also derive the bulk differential equation for $R_1(x)$ using a similar first-step decomposition as for the differential equation for the fixation probability, see Section~\ref{sec:eq-fix}. We focus on the infinitesimal time interval $\dd t$ just after the tracked parental branch starts at position $x$, with $0<x<L$. During $\dd t$, either no division occurs, in which case its position changes according to the advective and diffusive transport in the gut, or a division event occurs. In the latter case, one daughter bacterium continues the parental branch, which still has an expected $R_1(x)$ further division events after the one that just occurred, giving a contribution $R_1(x)+1$, and the spatial displacement only contributes at higher order. Note that subsequent divisions along the offspring branch and its descendants do not contribute to $R_1$, since we specifically track the parental branch. Summing over these two possibilities yields
\begin{align}
 R_1(x) &= \left[1-r_m\rho(x) \dd t\right] \,\overline{ R_1(x+ \dd X_t) }+ r_m\rho(x) \dd t  [ R_1(x)+1]+o(\dd t) ,
\end{align}
where the bar denotes the average over the stochastic increment $\dd X_t$ introduced in Eq.~\eqref{eq:stoch_process_1D}.
Expanding to second order in $\dd X_t$ and using $\overline{ \dd  X_t} = v \dd t$ and  $\overline{ (\dd  {X_t})^2} = 2D\dd t+o(\dd t)$, we obtain
\begin{align}
    \overline{ R_1(x+ \dd X_t) } = R_1(x)  + \mathcal{L}_x  R_1(x) \dd t +o(\dd t),
\end{align}
and we recover \cref{eq:eq_R1}:
\begin{align}
\mathcal{L}_x  R_1(x) = -r_m\rho(x).
\end{align}

\noindent\textbf{Comparison with the wild-type case.} 
The mean number $R_1^0$ of division events along the tracked parental branch of a wild-type bacterium starting at $x$ is given by Eq.~\eqref{eq:R1def} where $r_m$ is replaced with $r_w$. 
Recall that $r_m=(1+s)r_w$ where $s$ is the relative mutant fitness advantage. 
Hence, we have
\begin{equation}
    R_1(x)=\frac{r_m}{r_w}R_1^0(x)=(1+s)R_1^0(x).
\end{equation}
This relation between $R_1^0$ and $R_1$ shows that the parental branch of a mutant with relative fitness advantage $s$ undergoes, on average, $1+s$ times as many division events as that of a wild-type starting from the same initial position $x$. This result is in good agreement with intuition, as the mutant and wild type experience the same transport process, but the mutant division rate is $1+s$ times larger than that of the wild-type everywhere.

\subsubsection{Mean number of descendant branches across successive generations and reproductive value}
\label{sec:Rn-rv}

\noindent\textbf{Second generation.} So far, we determined the mean number of first-generation offspring branches a tracked parental branch produces before it exits the system. Similarly, the mean number of second-generation descendant branches it produces can be expressed as
\begin{align}
    R_2(x) = {r_m}^2\int_0^L \dd y_1  \int_0^L \dd y_2  \rho(y_1) \rho(y_2) G(y_2|y_1) G(y_1|x)  ,
\end{align} 
where we integrate over the position $y_1$ where a first-generation offspring branch is produced and over the subsequent trajectory of this branch and the positions $y_2$ of its division events, each of which produces a second-generation offspring branch. Here, each first-generation offspring branch is treated according to the parental-branch convention used for the calculation of $R_1$. 
We have
\begin{align}
    \mathcal{L}_x R_2(x) &= {r_m}^2\int_0^L \dd y_1  \int_0^L \dd y_2  \rho(y_1) \rho(y_2) G(y_2|y_1) \mathcal{L}_xG(y_1|x) \nonumber \\& = - {r_m}^2 \int_0^L \dd y_1  \int_0^L \dd y_2  \rho(y_1) \rho(y_2) G(y_2|y_1) \delta(y_1-x) \nonumber\\& = - {r_m}^2\rho(x)  \int_0^L \dd y_2   \rho(y_2) G(y_2|x)\nonumber \\& = - r_m\rho(x)  R_1(x).
\end{align}
Hence, $R_2$ satisfies
\begin{align}
    \mathcal{L}_x R_2(x) =D R_2''(x) + v R_2'(x) =- r_m\rho(x)  R_1(x),
\end{align}
and the boundary conditions in Eqs.~\eqref{bc_bwd1_mt} and~\eqref{bc_bwd2_mt}. In addition, we have
\begin{equation}
    R_2(x)=(1+s)^2R_2^0(x),
\end{equation}
where $R_2^0$ denotes the mean number of second-generation descendant branches of a wild-type. \\

\noindent\textbf{Subsequent generations.} The same calculation can be generalized to $n\geq 2$, yielding
\begin{align}
    \mathcal{L}_x R_n(x) =  D R_n''(x) + v R_n'(x) = - r_m\rho(x)  R_{n-1}(x). \label{edRn}
\end{align}
Each $R_n$, being a backward observable of the same transport process, satisfies the same boundary conditions as $R_1$, given in Eqs.~\eqref{bc_bwd1_mt} and~\eqref{bc_bwd2_mt}. In addition, we have
\begin{equation}
    R_n(x)=(1+s)^nR_n^0(x), \label{Rn_Rn0}
\end{equation}
where $R_n^0$ denotes the mean number of $n$-th-generation descendant branches of a wild-type. Indeed, each of the $n$ successive generations of offspring branches contributes one factor $r_m/r_w=1+s$, while mutant and wild type follow the same stochastic transport process.
\\

\noindent\textbf{Long-term limit.} Let us first focus on the wild-type. Assuming convergence of $R_n^0$ to a finite nonzero limit $R_\infty^0$, we obtain in the limit $n\rightarrow \infty$
\begin{align}
    \mathcal{L}_x R_{\infty}^0(x) =D [R_{\infty}^0]''(x) + v [R_{\infty}^0]'(x) =- r_w\rho(x)  R_{\infty}^0(x) ,\label{eqinf_wt}
\end{align}
and $R_{\infty}^0$ satisfies the same boundary conditions as $R_1^0$ and $R_1$. We recognize Eqs.~\eqref{eqphi1}, the differential equation for the first-order term $\phi_1$ of the fixation probability, as well as the associated boundary conditions. Hence, we have
\begin{equation}
    R_\infty^0(x)=A_\infty f(x) =A_\infty B(x) e^{-vx/D}, \label{Rinf}
\end{equation}
where $A_\infty$ is a constant. 
Combining Eqs.~\eqref{Rn_Rn0} and~\eqref{Rinf}, we obtain for $n\to\infty$
\begin{equation}
    R_n(x)\sim (1+s)^n A_\infty B(x) e^{-vx/D}. \label{Rinf2}
\end{equation}
Hence, for $n\to\infty$:
\begin{itemize}
    \item If $s<0$, $R_n(x)\to 0$;
    \item If $s=0$, $R_n(x)\to A_\infty B(x) e^{-vx/D}$;
    \item If $s>0$, $R_n(x)\to \infty$.
\end{itemize}
Note that we determine the value of the constant $A_\infty$ in the next section, see Eq.~\eqref{Ainf}. 

These results are consistent with our conclusion from Section~\ref{critpt}: the critical point is at $s=0$. Furthermore, the asymptotic long-term contribution $R_\infty^0(x)$ of a neutral individual starting at position $x$ to future generations can be interpreted as its reproductive value, and can be viewed as a continuous-space analog of the discrete-class reproductive value~\cite{Taylor90}. Note that the reproductive value notion is also used in evolutionary graph theory~\cite{Maciejewski14}. We showed that $R_\infty^0(x)\propto f(x)$, but also that, to first order in relative mutant selective advantage $s$, the fixation probability satisfies $\phi(x)\propto f(x)$, see Eq.~\eqref{phi_order1}. Hence, the fixation probability of a mutant appearing at a given position along the gut is proportional to reproductive value. Such close relationships between fixation probability and reproductive value were evidenced in other contexts~\cite{barton_relation_2011}.

\begin{figure}[htbp]
    \centering
    \includegraphics[width=0.55\linewidth]{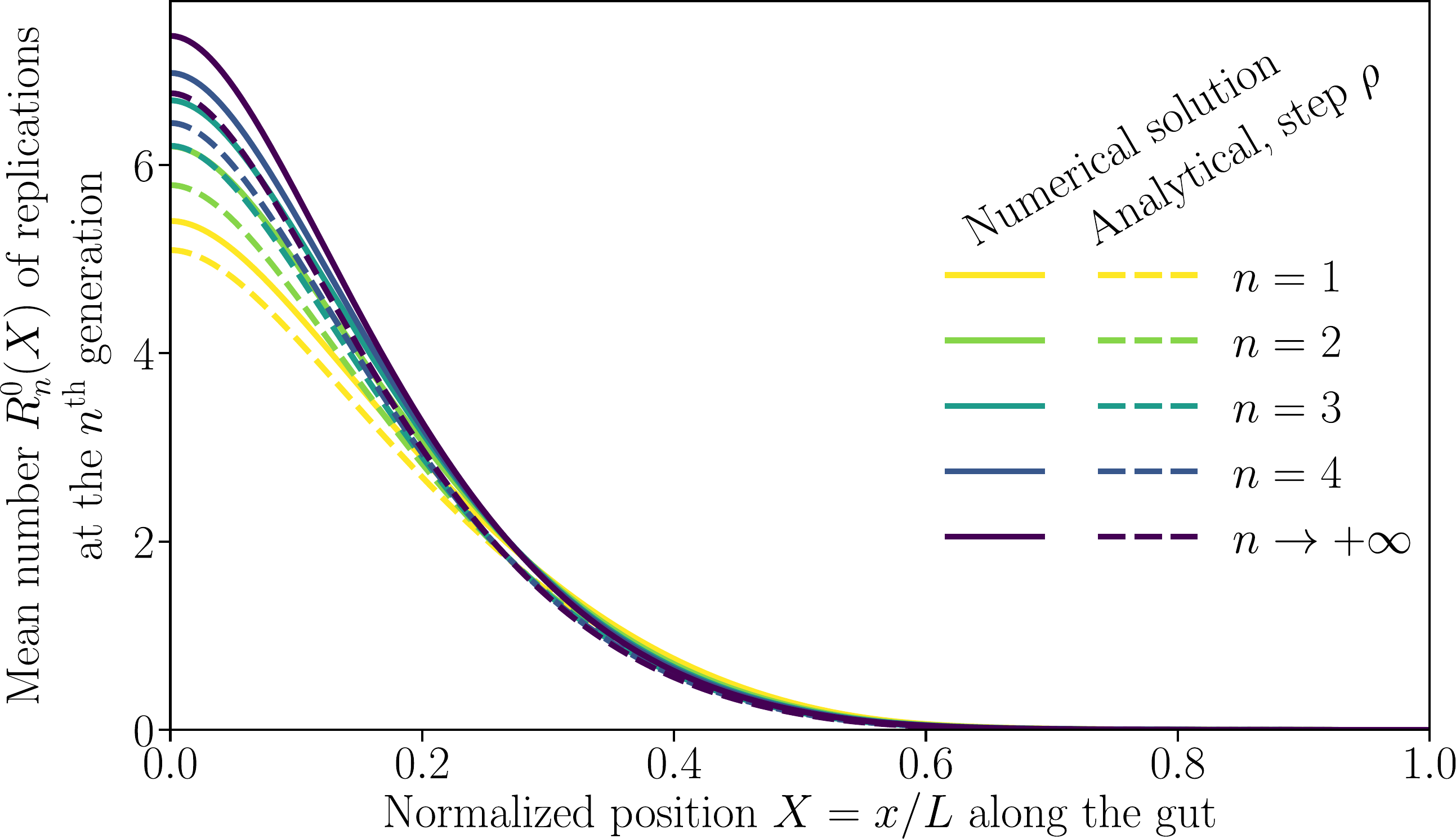}
    \caption{Mean number $R_n^0$ of descendant branches of a tracked wild-type (or neutral mutant) parental branch at the $n$-th generation versus normalized initial mutant position $X=x/L$ along the gut, for different values of $n$. For finite $n$, we plot numerical solutions of Eq.~\eqref{edRn} using either the steady-state mutant-free profile of the division rate $\rho$ (``Numerical solution'') or its step approximation (``Analytical, step $\rho$''). For $n\to+\infty$, we plot Eq.~\eqref{Rinf} using the steady-state wild-type profile $B$ either from numerical integrations (``Numerical solution'') or from the analytical formula in Section~\ref{steadyBstep} in the step division rate case (``Analytical, step $\rho$''). Parameter values: same as in main text Figs.~1 and 2. 
    }
    \label{fig:Rn_X}
\end{figure}

The spatial profiles of $R_n^0$ for $n=1,2,3,4$ and of $R_\infty^0$ are shown in Fig.~\ref{fig:Rn_X} using the same parameter values as in the main text. We observe that $R_n^0$ is strongly space-dependent, with large values upstream and very small ones downstream, illustrating the strong impact of initial position on the fate of a bacterium in the gut.

\newpage
\subsubsection{Division-event-weighted average}

\noindent\textbf{Wild-type case.} Let us consider the division-event-weighted average of the number $R_n^0$ of $n$-th-generation descendant branches of a wild-type parental branch (see Eq.~\eqref{rwa}):
\begin{align}
     \left< R_n^0\right>  = \frac{\int_0^L \dd x \,\rho(x) B(x)R_n^0(x) }{\int_0^L \dd x\, \rho(x) B(x)}.
\end{align}
This quantity is the mean number of $n$-th-generation descendant branches of a wild-type parental branch that appeared upon a random division event occurring in the stationary wild-type population. The starting position $x$ which we integrate over is the position where the tracked wild-type branch was produced by a division event. 
We have $\mathcal{L}_xR_n^0(x)=-r_w\rho(x)R_{n-1}^0(x)$ (see Eq.~\eqref{edRn} in the wild-type case), while the stationary wild-type concentration $B$ satisfies $\mathcal{L}_x^\dagger B(x)=-r_w\rho(x)B(x)$ (see Eq.~\eqref{eq:eq_B}). 
Since $\mathcal{L}_x$ and $\mathcal{L}_x^\dagger$ are adjoint (including the corresponding boundary conditions), we have
\begin{equation}
    \int_0^L \dd x\,B(x)\mathcal{L}_xR_n^0(x)=\int_0^L \dd x\,R_n^0(x)\mathcal{L}_x^\dagger B(x).
\end{equation}
Using the two differential equations mentioned above yields
\begin{align}
    \int_0^L \dd x\,\rho(x)B(x)R_{n-1}^0(x)=\int_0^L \dd x\,\rho(x)B(x)R_n^0(x).
\end{align}
Iterating this identity gives
\begin{equation}
    \int_0^L \dd x\,\rho(x)B(x)R_n^0(x)=\int_0^L \dd x\,\rho(x)B(x)R_1^0(x). \label{part1}
\end{equation}
Furthermore, since $\mathcal{L}_xR_1^0(x)=-r_w\rho(x)$ (see Eq.~\eqref{eq:eq_R1} in the wild-type case), and since
$\int_0^L \dd x\,B(x)\mathcal{L}_xR_1^0(x)=\int_0^L \dd x\,R_1^0(x)\mathcal{L}_x^\dagger B(x)$, we have
\begin{equation}
    \int_0^L \dd x\,\rho(x)B(x)=\int_0^L \dd x\,\rho(x)B(x)R_1^0(x). \label{part2}
\end{equation}
Combining Eq.~\eqref{part1} and Eq.~\eqref{part2} yields
\begin{equation}
    \int_0^L \dd x\,\rho(x)B(x)R_n^0(x)=\int_0^L \dd x\,\rho(x)B(x),
\end{equation}
and hence
\begin{equation}
    \left<R_n^0\right>=\frac{\int_0^L \dd x \,\rho(x) B(x)R_n^0(x) }{\int_0^L \dd x\, \rho(x) B(x)}=1, \label{R0bar}
\end{equation}
for all $n\geq 1$. This result is consistent with intuition: at steady-state, a wild-type branch sampled at a random division event has, on average, one descendant branch at each subsequent generation, as the population does not grow or shrink.

Note that Eq.~\eqref{R0bar} allows us to determine the value of the constant $A_\infty$ involved in Eqs.~\eqref{Rinf} and~\eqref{Rinf2}. Indeed, taking the limit of Eq.~\eqref{R0bar} when $n\to\infty$ gives $\left<R_\infty^0\right>=1$. Hence, $A_\infty \left<f\right>=1$, and
\begin{align}
     A_\infty=\frac{1}{\left<f\right>}, \label{Ainf}
\end{align}
where $f$ is defined in Eq.~\eqref{deff} as $f(x)=B(x) e^{-vx/D}$. \\

\noindent\textbf{Mutant case.} In the case of a mutant parental branch, since $R_n(x)=(1+s)^nR_n^0(x)$ (see Eq.~\eqref{Rn_Rn0}), we obtain
\begin{equation}
    \left<R_n\right>=(1+s)^n \label{Rnbar}
\end{equation}
for all $n\geq 1$.

\subsection{Deterministic growth of the mutant concentration}
\label{detgrowth-general}

\subsubsection{Eigenvalue problem}
\label{detgrowth-general-eigval}

Let us focus on the deterministic growth of the concentration of beneficial mutants with $r_m>r_w$ in the system. We start from the bulk partial differential equation which describes the time evolution of the mutant concentration $M(x,t)$, namely Eq.~\eqref{eq:eq_M}. We assume that mutants are sufficiently rare not to perturb the wild-type and food concentration profiles. Consequently, $F(x,t)$, and thus $\rho(x,t)$, are assumed to be time-independent and equal to their mutant-free stationary profiles, $\rho(x,t)=\rho(x)$, as in Sections~\ref{sec:eq-fix},~\ref{secfixgal} and~\ref{sec-Rn}. 
Looking for a separable solution of the form $M(x,t)=g(x)h(t)$ for this equation gives
\begin{equation}
    \frac{1}{h(t)}\frac{\dd h}{\dd t}(t)=D\frac{g''(x)}{g(x)}-v\frac{g'(x)}{g(x)}+r_m\rho(x)=\gamma,
\end{equation}
where $\gamma$ is a constant. Solving the differential equation for $h$ yields
\begin{equation}
    h(t)=h_0e^{\gamma t}, \label{thisisgrowth}
\end{equation}
where $h_0$ is a constant. 

Let us now consider the corresponding spatial differential equation, describing $g$:
\begin{equation}
    Dg''(x)-vg'(x)+\left[r_m\rho(x)-\gamma\right]g(x)=0, \label{eqspatial}
\end{equation}
with the forward boundary conditions given by Eqs.~\eqref{bc_fwd1} and~\eqref{bc_fwd2}. Note that this is similar to Eqs.~S122-S123 in Ref.~\cite{ghosh2022emergent} for constant $D$ and $v$ and zero death rate. Introducing the adjoint operator $\mathcal{A}_s^\dagger$ of $\mathcal{A}_s$ from Eq.~\eqref{def_As}, namely
\begin{equation}
    \mathcal{A}_s^\dagger=\mathcal{L}_x^\dagger+(1+s)r_w\rho(x),
\end{equation}
where we wrote $r_m=(1+s)r_w$, and where $\mathcal{L}_x^\dagger$ is the forward transport operator defined in Eq.~\eqref{fwd}, Eq.~\eqref{eqspatial} can be written as
\begin{equation}
    \mathcal{A}^\dagger_s g=\gamma g. \label{eqAs}
\end{equation}
Our goal is to solve this eigenvalue problem, i.e.\ to determine the eigenvalues $\gamma$ of the operator $\mathcal{A}_s^\dagger$. Indeed, each eigenvalue corresponds to a spatial mode whose amplitude evolves as $e^{\gamma t}$ (see Eq.~\eqref{thisisgrowth}). In particular, the long-time deterministic behavior of the mutant concentration is controlled by the largest eigenvalue $\gamma_\mathrm{max}$.

Introducing $y(x)=e^{-vx/(2D)}g(x)$ maps the eigenvalue problem in Eq.~\eqref{eqAs} with the forward boundary conditions in Eqs.~\eqref{bc_fwd1} and~\eqref{bc_fwd2} to the self-adjoint Sturm-Liouville problem introduced in Section~\ref{critpt}, namely Eqs.~\eqref{SL-bulk} and~\eqref{SL-bc}. Therefore, $\mathcal A_s$ and $\mathcal A_s^\dagger$ have the same spectrum, and in particular the same principal (largest) eigenvalue $\gamma_\mathrm{max}$. In addition, the principal eigenfunction of $\mathcal A_s^\dagger$ can be chosen positive, while every higher eigenfunction changes sign~\cite{teschl_book}. The transformation introduced here further shows that their eigenfunctions are related by $q(x)\propto g(x)e^{-vx/D}$, where $q$ and $g$ are the respective eigenfunctions of $\mathcal{A}_s$ and $\mathcal{A}^\dagger_s$ associated with the same eigenvalue. Note that this is consistent with our earlier observation that $\mathcal A_0 f(x)=0$ while $\mathcal A_0^\dagger B(x)=0$ with $f(x)=B(x)e^{-vx/D}$, see Eq.~\eqref{deff}.

Here, the largest admissible eigenvalue $\gamma_\mathrm{max}$ will determine the mutant invasion growth rate (see Eq.~\eqref{thisisgrowth}). The value of $\gamma_\mathrm{max}$ can be determined numerically, e.g.\ by discretizing the associated self-adjoint Sturm-Liouville operator and computing its lowest eigenvalue. Below, we further perform a perturbative expansion in the weak selection regime $s\ll 1$.

\subsubsection{Weak selection regime} \label{detgrowth-pert}

\noindent\textbf{Neutral mutant case.} For $s=0$, the relevant forward operator is $\mathcal{A}^\dagger_0=\mathcal{L}_x^\dagger+r_w\rho(x)$. Since the stationary wild-type concentration profile $B(x)$ satisfies $\mathcal{A}_0^\dagger B=0$ and the boundary conditions in Eqs.~\eqref{bc_fwd1} and~\eqref{bc_fwd2}, it is an eigenfunction of the system in the neutral mutant case. Furthermore, since $B(x)>0$ for all $x\in[0,L]$, the corresponding transformed eigenfunction $B(x)e^{-vx/(2D)}$ is positive as well. Thus, it is the principal Sturm-Liouville eigenfunction~\cite{teschl_book}. Hence, the principal eigenvalue is $\gamma_\mathrm{max}=0$, with right eigenfunction $B(x)$. In the neutral case, the mutant concentration has zero invasion exponential growth rate, consistent with the result $\gamma_\mathrm{max}=0$.

The adjoint operator is $\mathcal{A}_0=\mathcal{L}_x+r_w\rho(x)$, where $\mathcal{L}_x$ is the backward transport operator defined in Eq.~\eqref{bwd}. The associated boundary conditions are given by Eqs.~\eqref{bcphi1} and~\eqref{bcphi2}. 
The corresponding left eigenfunction is $f(x)=B(x)e^{-vx/D}$, see Eq.~\eqref{deff}, as it satisfies $\mathcal{A}_0 f=0$ and the boundary conditions. \\

\noindent\textbf{Perturbative expansion.} 
Let us now determine $\gamma_\mathrm{max}$ perturbatively for $s\ll1$. Since
\begin{equation}
    \mathcal{A}^\dagger_s=\mathcal{A}^\dagger_0+s\,r_w\rho(x),
\end{equation}
we expand the principal eigenfunction as
\begin{equation}
    g(x)=B(x)+s g_1(x)+O(s^2),
\end{equation}
and the associated principal eigenvalue as
\begin{equation}
    \gamma_\mathrm{max}=s\gamma_1+O(s^2).
\end{equation}
Substituting these expansions into Eq.~\eqref{eqAs} and identifying first-order terms in $s$ gives
\begin{equation}
    \mathcal{A}^\dagger_0g_1(x)+r_w\rho(x)B(x)=\gamma_1B(x).
\end{equation}
Multiplying both sides by $f(x)$ and integrating between $0$ and $L$ gives
\begin{equation}
    \int_0^L f(x)\mathcal{A}^\dagger_0g_1(x)\,\dd x+r_w\int_0^L\rho(x)f(x)B(x)\,\dd x=\gamma_1\int_0^L f(x)B(x)\,\dd x.
\end{equation}
Since $\mathcal{A}_0$ and $\mathcal{A}_0^\dagger$ are adjoint, this can be rewritten as
\begin{equation}
    \int_0^L g_1(x)\mathcal{A}_0 f(x)\,\dd x+r_w\int_0^L\rho(x)f(x)B(x)\,\dd x=\gamma_1\int_0^L f(x)B(x)\,\dd x.
\end{equation}
Using $\mathcal A_0 f=0$ (as shown in the neutral case above), we obtain
\begin{equation}
    \gamma_1=r_w\frac{\int_0^L\rho(x)f(x)B(x)\,\dd x}{\int_0^L f(x)B(x)\,\dd x}.
\end{equation}
This provides the first-order expansion of the principal eigenvalue $\gamma_\mathrm{max}$ for $s\ll 1$ (including the moderate-selection regime discussed above):
\begin{equation}
    \gamma_\mathrm{max}=s\,r_w\frac{\int_0^L\rho(x)f(x)B(x)\,\dd x}{\int_0^L f(x)B(x)\,\dd x}+O(s^2). \label{gammamax-pert}
\end{equation}
We observe that $\gamma_\mathrm{max}>0$ to leading order in the present case of a beneficial mutant ($s>0$). This principal eigenvalue $\gamma_\mathrm{max}$ gives the deterministic invasion growth rate of the mutant concentration in the system. Note that an equivalent expression was obtained for the mutant invasion growth rate in Ref.~\cite{ghosh2022emergent} to first order in $s$, see their Eq.~S126 (see Section~\ref{comp_GG} for a comparison between our study and Ref.~\cite{ghosh2022emergent}). 

Eq.~\eqref{gammamax-pert} can be interpreted as follows: The absolute mutant division-rate advantage over wild-types is $s\, r_w \rho(x)$ at position $x$. To first order in $s$, the exponential growth rate of the overall mutant population is obtained by averaging this quantity spatially, weighting each position $x$ by its long-term contribution to the future population, which is proportional to $f(x)B(x)$, i.e.\ to the reproductive value at $x$ times the bacterial concentration at $x$. Qualitatively, although bacteria located downstream are numerous, they have a larger probability of exiting the system before contributing to the future population, hence the need to weight with reproductive value and not only with bacterial concentration.

\textbf{Comparison with the well-mixed case.} In the case of a well-mixed population where wild-types divide at rate $r_w$ and mutants at rate $r_m=(1+s)r_w$, corresponding to the maximal division rates in the gut, the mutant to wild type ratio grows at rate $r_m-r_w=s\,r_w$ (whether the well-mixed population exponentially grows or undergoes dilutions maintaining it at a steady state size). In other words, the invasion growth rate is $s\,r_w$. Since $\rho(x)\leq 1$ for all $x\in[0,L]$, we have $\int_0^L\rho(x)f(x)B(x)\dd x\leq \int_0^Lf(x)B(x)\dd x$. Hence, Eq.~\eqref{gammamax-pert} shows that, to first order in $s$, $\gamma_\mathrm{max}\leq s\,r_w$: invasion is slower in the gut than in this maximally dividing well-mixed population. Note that the limit  $\gamma_\mathrm{max}\to s\,r_w$ is approached as $\mathrm{Pe}\to\mathrm{Pe}_{\max}$ for $k\ll F_\mathrm{in}$, since all bacteria divide at a rate close to the maximal one at washout.

However, this comparison is not the most relevant one, since it compares the population in the gut to a well-mixed population where all bacteria divide at the maximal rate. Let us consider a well-mixed population with the same mean per-capita division rate as in the gut. In this reference well-mixed population, the wild-type division rate is
\begin{equation}
    \tilde{r}_w=r_w\frac{\int_0^L\rho(x)B(x)\dd x}{\int_0^LB(x)\dd x}\,,
\end{equation}
and the mutant division rate is $\tilde{r}_m=(1+s)\tilde{r}_w$. Hence, the mutant invasion growth rate is
\begin{equation}
    \gamma_\text{WM}=s\,\tilde{r}_w=s\,r_w\frac{\int_0^L\rho(x)B(x)\dd x}{\int_0^LB(x)\dd x}\,. \label{gammaWM}
\end{equation}
Let us now compare $\gamma_\mathrm{max}$ in Eq.~\eqref{gammamax-pert} with $\gamma_\text{WM}$ in Eq.~\eqref{gammaWM}. We remark that in our system, food concentration decreases along the gut, leading to a decreasing division rate profile $\rho$ with $x$. Furthermore, $f$ also decreases with $x$. Indeed, recalling that $f(x)=B(x)e^{-vx/D}$, we have
\begin{equation}
    f'(x)=-\frac{1}{D}\left[vB(x)-DB'(x)\right]e^{-vx/D}\,.
\end{equation}
In this expression, we recognize the net downstream flux of wild-type bacteria between the square brackets. This quantity is nonnegative, thus showing that $f'(x)\leq 0$ for $x\in[0,L]$. Since both $\rho$ and $f$ decrease with $x$, their covariance is nonnegative. Defining the average over wild-type bacteria of a function $g$ as
\begin{equation}
    \langle g\rangle_B =\frac{\int_0^Lg(x)B(x)\dd x}{\int_0^LB(x)\dd x}\,,
\end{equation}
this leads to
\begin{equation}
    \langle\rho f\rangle_B\geq \langle\rho \rangle_B\langle f\rangle_B \;\;\;\text{i.e.}\;\;\; \frac{\int_0^L\rho(x)f(x)B(x)\,\dd x}{\int_0^L f(x)B(x)\,\dd x}\geq \frac{\int_0^L\rho(x)B(x)\dd x}{\int_0^LB(x)\dd x}\,, \label{compInt}
\end{equation}
which shows that $\gamma_\mathrm{max}\geq\gamma_\text{WM}$ to first order in $s$. Hence, to first order in $s$, mutant invasion is accelerated in the gut compared to the reference well-mixed population with the same average division rate.

\subsubsection{Beyond weak selection}
\label{gamma-general}

Let us now compare the mutant invasion growth rate in the gut and in the well-mixed population with same mean per-capita division rate, beyond weak selection. Recall that the eigenvalue problem $\mathcal{A}_s^\dagger g=\gamma g$ (i.e., Eq.~\eqref{eqAs}) can be mapped, by introducing $y(x)=e^{-vx/(2D)}g(x)$, onto the self-adjoint Sturm-Liouville problem
\begin{equation}
H_sy(x)=-\gamma y(x),\quad\mathrm{with}\quad H_s y(x)=-Dy''(x)+\left[\frac{v^2}{4D}-(1+s)r_w\rho(x)\right]y(x),
\end{equation}
and with boundary conditions
\begin{equation}
y'(0)=\frac{v}{2D}y(0), \quad y'(L)=-\frac{v}{2D}y(L),\label{bcs}
\end{equation}
see Section~\ref{detgrowth-general-eigval}. 
For any nonzero sufficiently regular function $y$ on $[0,L]$ satisfying Eq.~\eqref{bcs}, let us consider the Rayleigh quotient
\begin{equation}
\mathcal{R}_s[y]=\frac{\int_0^L y(x)\,H_s y(x)\,\dd x}{\int_0^L y^2(x)\,\dd x}.
\end{equation}
The smallest eigenvalue of this Sturm-Liouville problem, namely $-\gamma_\mathrm{max}$, satisfies~\cite{teschl_book}
\begin{equation}
-\gamma_\mathrm{max}=\min_{y\neq 0}\mathcal{R}_s[y].
\end{equation}
We have
\begin{equation}
\int_0^L y(x)\,H_s y(x)\,\dd x=D\int_0^L [y'(x)]^2\,\dd x+\int_0^L\left[\frac{v^2}{4D}-r_w\rho(x)\right]y^2(x)\,\dd x+\frac{v}{2}\left[y^2(0)+y^2(L)\right]-s\, r_w\int_0^L\rho(x)y^2(x)\,\dd x,
\end{equation}
which allows us to write
\begin{equation}
\mathcal{R}_s[y]=b[y]+s\,a[y],
\end{equation}
where $a$ and $b$ are independent of $s$. Hence, 
\begin{equation}
\gamma_\mathrm{max}=-\min_{y\neq 0} \left\{b[y]+s\,a[y]\right\}.
\end{equation}
Since the pointwise minimum of affine functions is concave, $\gamma_\mathrm{max}$ is a convex function of $s$.

For $s=0$, we have $\gamma_\mathrm{max}=0$ (see Section~\ref{critpt}). Convexity thus implies, for $s>0$,
\begin{equation}
\gamma_\mathrm{max}(s)\geq \gamma_\mathrm{max}(0)+s\,\gamma_\mathrm{max}'(0)=s\,\gamma_\mathrm{max}'(0). \label{ineq1}
\end{equation}
Besides, Eq.~\eqref{gammamax-pert} yields
\begin{equation}
\gamma_\mathrm{max}'(0)=r_w\frac{\int_0^L \rho(x)f(x)B(x)\,dx}{\int_0^L f(x)B(x)\,dx}.
\end{equation}
Eq.~\eqref{compInt} can thus be rewritten as
\begin{equation}
    \gamma_\mathrm{max}'(0)\geq \frac{\gamma_\text{WM}(s)}{s}, \label{ineq2}
\end{equation}
where the well-mixed mutant invasion growth rate is given by Eq.~\eqref{gammaWM}. 
Combining Eqs.~\eqref{ineq1} and~\eqref{ineq2} yields
\begin{equation}
\gamma_\mathrm{max}(s)\geq\gamma_{\mathrm{WM}}(s),
\end{equation}
for all $s>0$.
Hence, for all $s>0$, mutant invasion is accelerated in the gut compared to the reference well-mixed population with the same average division rate.

\begin{figure}[htbp]
    \centering
    \includegraphics[width=0.72\linewidth]{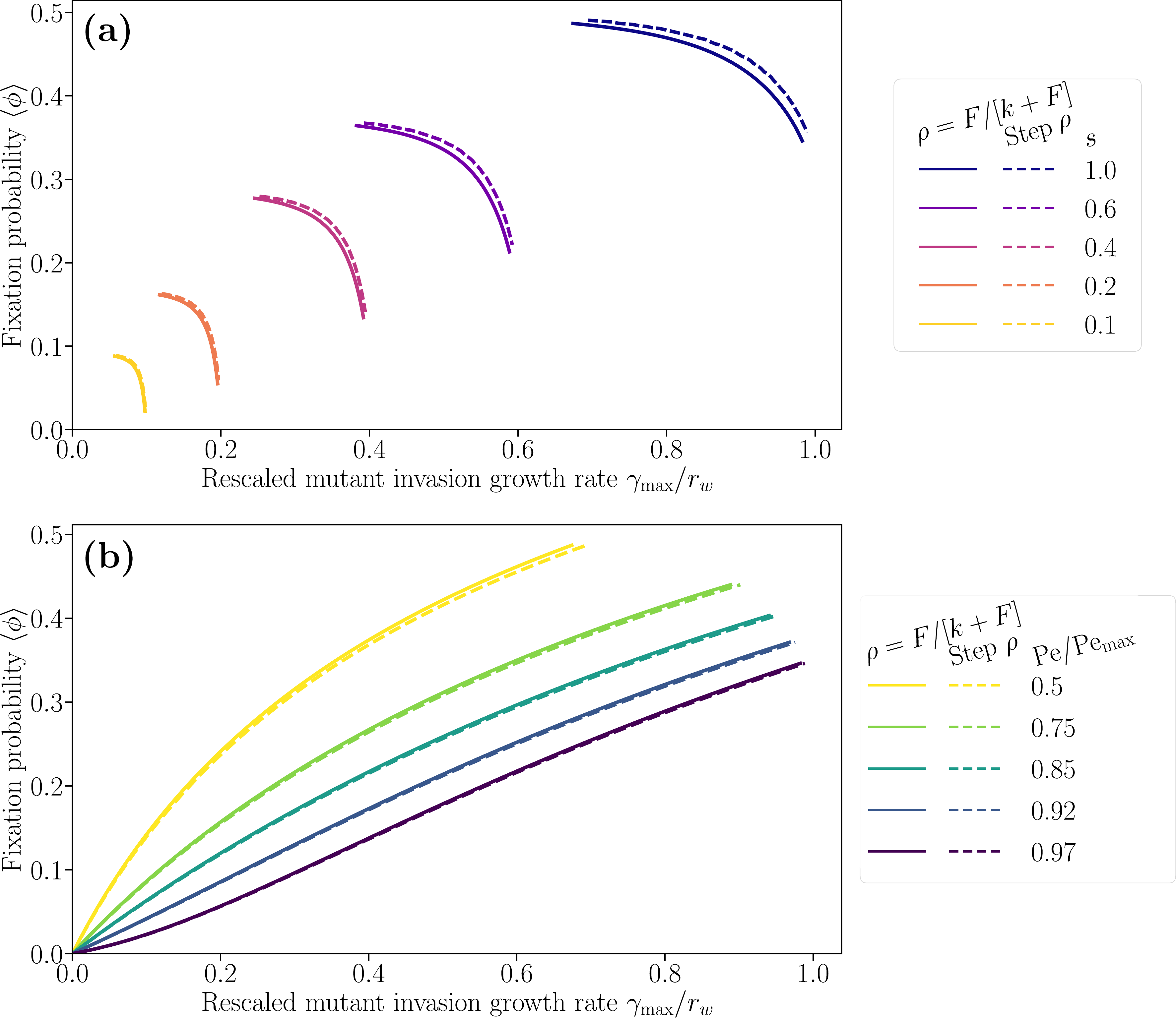}
    \caption{Division-event-weighted mutant fixation probability $\langle\phi\rangle$ versus rescaled mutant invasion growth rate $\gamma_\mathrm{max}/r_w$. In panel (a), the normalized Péclet number $\mathrm{Pe}/\mathrm{Pe}_\mathrm{max}$ is varied between 0.5 and 0.97 while holding fixed the mutant relative fitness advantages $s$. In panel (b), $s$ is varied between 0 and 1 while holding fixed $\mathrm{Pe}/\mathrm{Pe}_\mathrm{max}$. Numerical solutions of the branching process Eq.~\eqref{eq:phi1} using the steady-state mutant-free profile of the division rate $\rho$ (``$\rho=F/[k+F]$'') or its step approximation (``Step $\rho$'') are used to obtain $\langle\phi\rangle$. The principal eigenvalue $\gamma_\mathrm{max}$ of the operator $\mathcal{A}_s^\dagger$ is determined numerically in both cases. Parameters: same $\lambda_w$ (and $\mathrm{Pe}_\mathrm{max}$), $k/F_\mathrm{in}$ and $\alpha$ as in main text Figs.~1, 2 and 3. 
    }
    \label{fig:phi_gamma}
\end{figure}

Fig.~\ref{fig:phi_gamma} shows the division-event-weighted mutant fixation probability $\langle\phi\rangle$ versus the rescaled mutant invasion growth rate $\gamma_\mathrm{max}/r_w$. We observe in Fig.~\ref{fig:phi_gamma}(a) that when varying the Péclet number Pe at fixed mutant relative fitness advantage $s$, $\langle\phi\rangle$ decreases as $\gamma_\mathrm{max}/r_w$ increases. This is due to the fact that $\langle\phi\rangle$ decreases as Pe increases, while $\gamma_\mathrm{max}/r_w$ increases as Pe increases, see main text, in particular Fig.~4 there. \\

\subsection{Fixation probability of a neutral mutant and active population size}
\label{sec:neutral}

Within the branching process approximation, the fixation probability of a neutral mutant vanishes (see Section~\ref{critpt}). Let us now focus specifically on this fixation probability, which is nonzero in finite populations, and express it in the gut model considered here. Ref.~\cite{labavic2022hydrodynamic} showed that it is well-approximated by $1/N_\mathrm{active}$, where $N_\mathrm{active}$ is the number of actively dividing bacteria. In practice, in cases where the profiles of food and bacteria feature strong heterogeneity, with $[F(0)-F(L)]/F_\mathrm{in}\gtrsim 0.9$, and where the division rate profile is close to a step, with $k\ll F_\mathrm{in}$, $N_\mathrm{active}$ can be estimated by counting the bacteria whose division rate is larger than half its maximum value in the system. Then, Ref.~\cite{ghosh2022emergent} proposed a more detailed calculation of the fixation probability of a neutral mutant. Here, our focus is on beneficial mutants, but we nevertheless revisit the case of neutral mutants for completeness.

Let us consider a single neutral mutant (with $s=0$) appearing at $x_0$ in the stationary wild-type population. For simplicity, let us assume that a neutral mutant replaces an existing wild-type bacterium, thus not affecting the population state. In the neutral mutant case, the division rate profile remains equal to its stationary value $\rho(x)$, apart from fluctuations, even when the mutant lineage is no longer rare. Let $m(x,t)$ denote the concentration of descendants of the initial mutant located at $x_0$. Note that it is a linear concentration, like $B(x,t)$, $M(x,t)$ and $F(x,t)$ (see Section~\ref{sec:effective_1D_model}). In the deterministic limit, $m(x,t)$ satisfies (see Eq.~\eqref{eq:eq_M})
\begin{equation}
    \frac{\partial m}{\partial t}(x,t)=\mathcal{A}_0^\dagger m(x,t), \qquad m(x,0)=\delta(x-x_0),
    \label{eq:m_neutral}
\end{equation}
with the forward boundary conditions in Eqs.~\eqref{bc_fwd1} and~\eqref{bc_fwd2}, where $\mathcal{A}_0^\dagger$ is defined in Eq.~\eqref{def_A0dag}. Since $\mathcal{A}_0$ and $\mathcal{A}_0^\dagger$ are adjoint, and since $\mathcal{A}_0 f=0$ (see Eq.~\eqref{eq:A0f}), we have
\begin{equation}
    \frac{\dd}{\dd t}\int_0^L \dd x\, f(x)\,m(x,t)=\int_0^L \dd x\, f(x)\,\frac{\partial m}{\partial t}(x,t)=\int_0^L \dd x\, f(x)\mathcal{A}_0^\dagger m(x,t) =\int_0^L \dd x\, m(x,t)\mathcal{A}_0 f(x)=0 .
\end{equation}
This entails that 
\begin{equation}
\int_0^L f(x)\,m(x,t)\,\dd x=\int_0^L f(x)\,m(x,0)\,\dd x=f(x_0), \label{conservation}
\end{equation}
at all times $t\geq 0$. In addition, we saw in Section~\ref{detgrowth-general-eigval} that the principal (largest) eigenvalue of $\mathcal{A}_0^\dagger$ is $0$, and that it is simple and associated with the eigenfunction $B$. Thus, at long times, the profile of $m$ tends towards that of $B$: there exists a constant $C$ such that $m(x,t\to\infty)\sim C\,B(x)$. The long-time limit of the conservation law in Eq.~\eqref{conservation} yields
\begin{equation}
    m(x,t\to\infty)\sim C\,B(x)\;\;\;\textrm{with}\;\;\;C=\frac{f(x_0)}{\int_0^L f(x)B(x)\,\dd x} .
\end{equation}
This shows that in the long term, the descendants of the initial mutant constitute a fraction $C$ of the population, distributed along the gut as the wild type, and that this fraction is proportional to the reproductive value $f(x_0)$ associated with the mutant initial position $x_0$ (see Section~\ref{sec:Rn-rv}).

In the neutral case, mutants behave exactly like wild-types and do not modify the steady-state of the system (neglecting fluctuations). Thus, the expectation of mutant concentration over stochastic realizations obeys the same linear equation as the deterministic large-size limit $m(x,t)$, namely Eq.~\eqref{eq:m_neutral}, with the same initial and boundary conditions. Hence, they coincide. In addition, in a stochastic description, each realization eventually ends with a mutant frequency of either $0$ (extinction) or $1$ (fixation). Therefore, we can interpret the expected final mutant frequency $C$ as the fixation probability $\phi_0(x_0)$ of a neutral mutant starting at position $x_0$:
\begin{equation}
    \phi_0(x_0)=\frac{f(x_0)}{\int_0^L f(x)B(x)\,\dd x} .
    \label{eq:phi_neutral}
\end{equation}
Hence, the fixation probability of a neutral mutant is proportional to reproductive value, see also Ref.~\cite{ghosh2022emergent}. 
Averaging over the positions of division events, where mutants appear, i.e.\ calculating the division-event-weighted average, gives
\begin{equation}
    \left\langle \phi_0 \right\rangle=\frac{\langle f\rangle}{\int_0^L f(x)B(x)\,\dd x}=\frac{1}{N_\mathrm{active}},
    \;\;\; \textrm{with}\;\;\; N_\mathrm{active}=\frac{\int_0^L f(x)B(x)\,\dd x}{\langle f\rangle} .
    \label{eq:Nactive}
\end{equation}
Note that this result is valid to leading order in $1/N_\mathrm{active}$, since fluctuations of the population and food profiles are neglected.

If $f$ is uniform, the active population size $N_\mathrm{active}$ reduces to the total population size $N=\int_0^L B(x)\,\dd x$. If $f$ is uniform over the region where bacteria divide and negligible elsewhere, $N_\mathrm{active}$ is the number of bacteria in that region, thus matching the active population introduced heuristically in Ref.~\cite{labavic2022hydrodynamic}. 
Furthermore, Eq.~\eqref{eq:phi_neutral} coincides with the neutral fixation probability derived in Ref.~\cite{ghosh2022emergent} (see their Eqs.~6 and~8, for constant $D$ and $v$). Note that in Ref.~\cite{ghosh2022emergent}, $fB$ is interpreted as the distribution of future common ancestors, following Ref.~\cite{Hallatschek08}. Indeed, in the neutral case, the probability density that a randomly chosen bacterium in the distant future descends from one that is now at $x_0$ is given by $B(x_0)\phi_0(x_0)=B(x_0)f(x_0)/\int_0^LB(x)f(x)\dd x$. 

The active population size can be connected to the mutant invasion growth rate:
\begin{equation}
    \frac{N_\mathrm{active}}{N}=\frac{\int_0^L f(x)B(x)\,\dd x}{\langle f\rangle}\,\frac{1}{\int_0^L B(x)\,\dd x}=\frac{\int_0^L f(x)B(x)\,\dd x}{\int_0^L f(x)\rho(x)B(x)\,\dd x}\,\frac{\int_0^L \rho(x)B(x)\,\dd x}{\int_0^L B(x)\,\dd x}=\frac{\gamma_\text{WM}}{\gamma^{(1)}}\,, \label{NactiveOverN}
\end{equation}
where we recognized the mutant invasion growth rate in the well-mixed reference case (Eq.~\eqref{gammaWM}) and the first-order approximation of the mutant invasion growth rate in the gut (Eq.~\eqref{gammamax-pert}, written as $\gamma_\text{max}=\gamma^{(1)}+O(s^2)$). Since we showed above that $\gamma^{(1)}\geq\gamma_\text{WM}$, we have $N_\mathrm{active}\leq N$. Furthermore, compared to a well-mixed population, the reduction of the active population size and the acceleration of mutant invasion in the gut both stem from the positive covariance between $\rho$ and $f$.

\begin{figure}[htbp]
\centering
\includegraphics[width=0.72\textwidth]{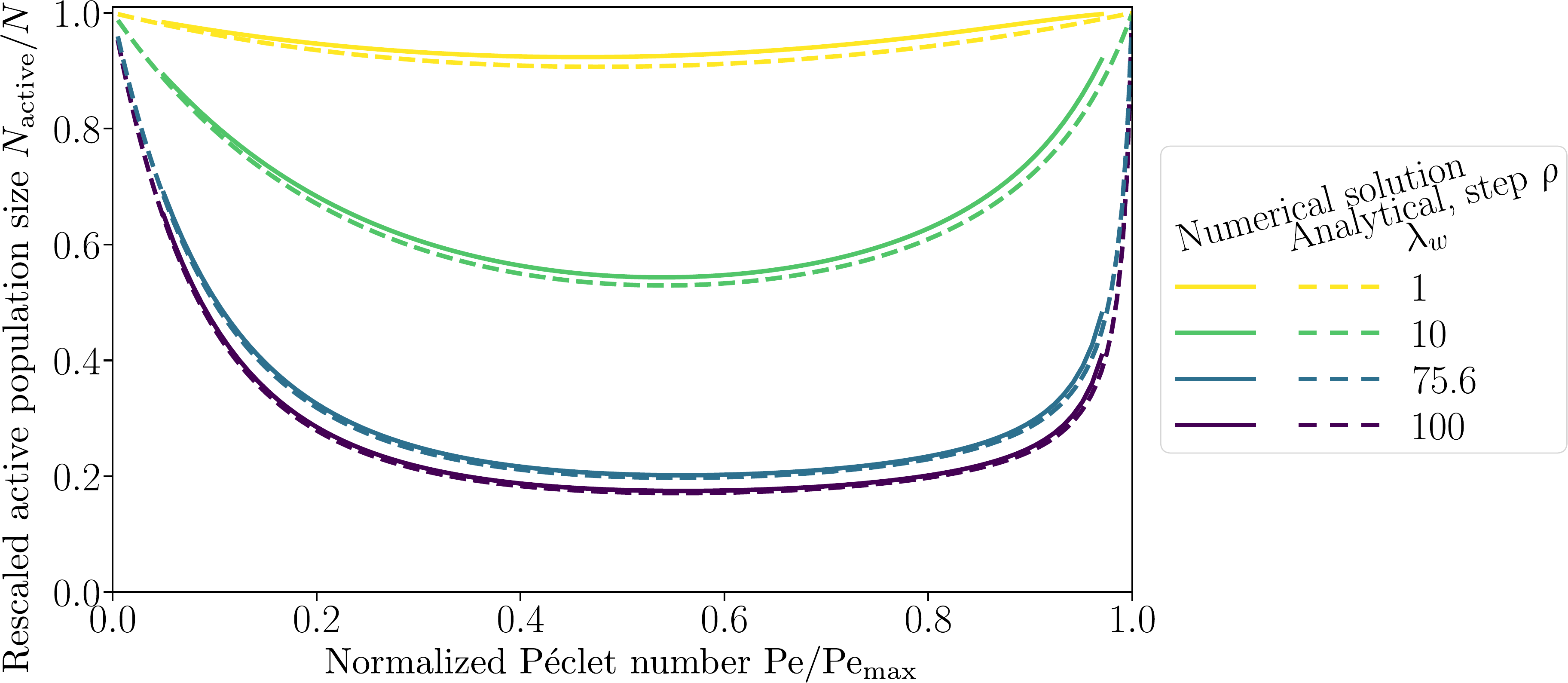}
\caption{Rescaled active population $N_{\text{active}}/ N$  as a function of the normalized Péclet number $\text{Pe}/\text{Pe}_{\text{max}}$ (where $N = \int_0^L \dd x\, B(x)$ corresponds to the total population size). Solid lines: numerical integration of Eq.~\eqref{NactiveOverN} using the numerically-obtained steady-state mutant-free wild-type bacteria concentration profile $B(x)$. Dashed lines: analytical result in Eq.~\eqref{NactiveStep} obtained using the step profile of $\rho$. Parameters: same $k/F_\mathrm{in}$ and $\alpha$ as in main text Figs.~1, 2 and 3.
\label{Nactive_fig}
}
\end{figure}

In Fig.~\ref{Nactive_fig}, we observe that $N_{\text{active}}$ is close to $N$ in the two cases with nearly uniform division rate where Pe is close to zero or close to $\text{Pe}_{\text{max}}$ but that $N_{\text{active}}$ can become significantly smaller than $N$ for intermediate Pe values. This decrease is more pronounced if $\lambda_w$ is larger, as faster-growing bacteria deplete food earlier, resulting in a growth region that occupies a smaller fraction of the system.  

\begin{figure}[htbp]
\centering
\includegraphics[width=0.56\textwidth]{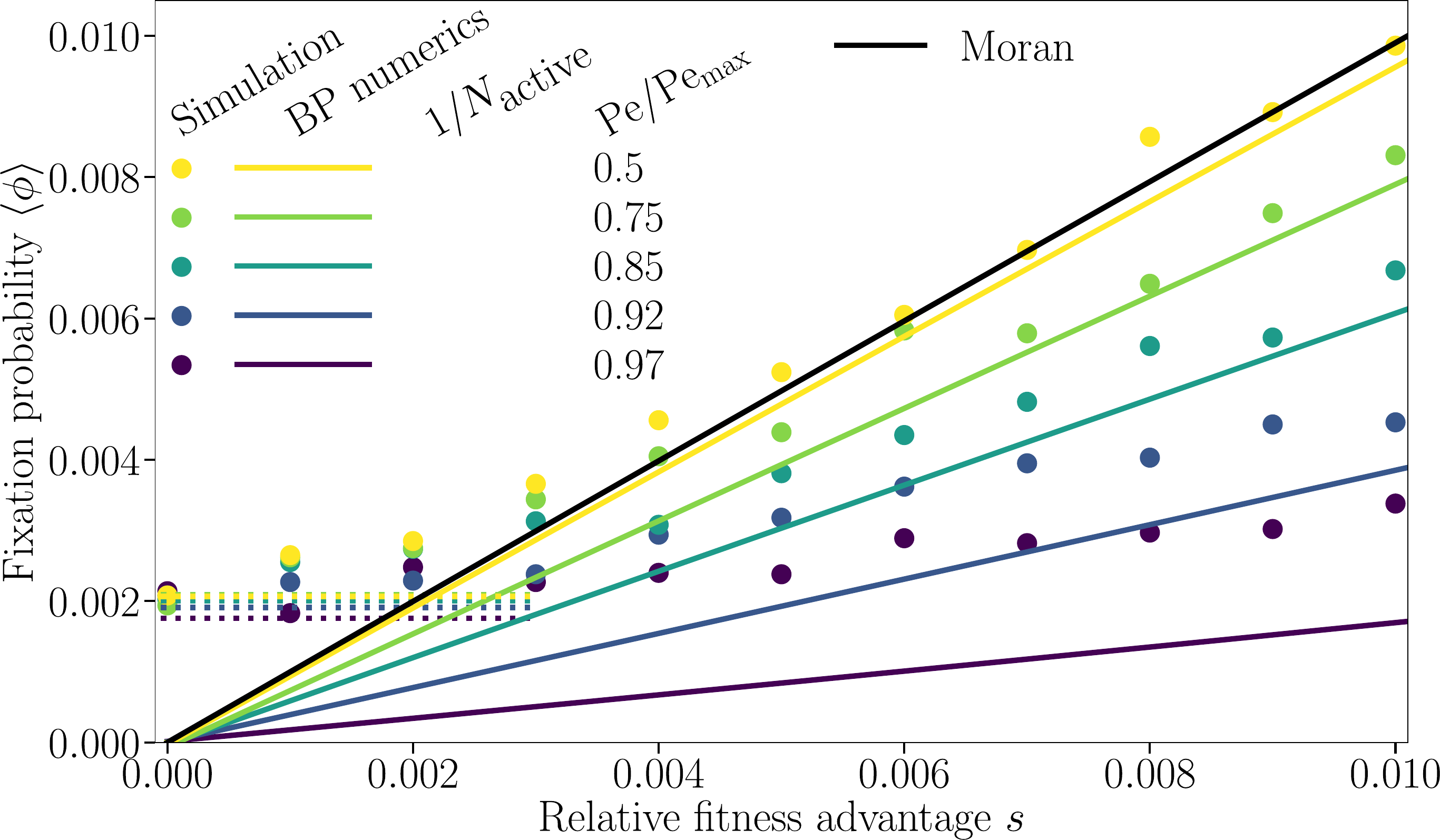}
\caption{Division-event-weighted mutant fixation probability $\langle\phi\rangle$ versus mutant relative fitness advantage $s$, for different values of the Péclet number Pe, for smaller $s$ values than in main text Fig.~3. Stochastic simulations are described in Section~\ref{sec:stochsim}. Numerical solutions of the branching process Eq.~\eqref{eq:phi1} use the steady-state mutant-free profile of the division rate $\rho$ and give our prediction in the large-population limit (``BP numerics''). The prediction for neutral mutants, $1/N_\mathrm{active}$, is obtained by numerical integration of Eq.~\eqref{NactiveOverN} using the numerically-obtained steady-state mutant-free wild-type bacteria concentration profile $B(x)$. The Moran model for large populations, $\langle\phi\rangle=s/(1+s)$, gives the well-mixed reference in the large-population regime. Parameters: same $\lambda_w$ (and $\mathrm{Pe}_\mathrm{max}$), $k/F_\mathrm{in}$ and $\alpha$ as in main text Figs.~1, 2 and 3.
\label{phimoy_smalls}}
\end{figure}

In Fig.~\ref{phimoy_smalls}, we observe that the division-event-weighted mutant fixation probability $\langle\phi\rangle$ is well-approximated by $1/N_\mathrm{active}$ in the neutral limit $s\rightarrow 0$. In Fig.~\ref{phimoy_smalls} and in main text Fig.~3(b), we note that the branching process results underestimate the fixation probability for small values of $s$ and for the largest Péclet numbers shown. This is due to the fact that the branching process calculation neglects neutral fixation. Comparing Eq.~\eqref{eq:Nactive} with Eq.~\eqref{phibar_order1}, the branching-process result dominates over the neutral contribution when $s\,\langle f\rangle^2/\langle f^2\rangle\gg1/N_\mathrm{active}$, i.e.\ when $s\gg\left[1+c_v^2(f)\right]/N_\mathrm{active}$. This allows us to make more precise the lower bound of the moderate selection regime of Section~\ref{phi-expansion}. In our stochastic simulations, $N_\mathrm{active}$ is on the order of 500 bacteria (see Section~\ref{sec:stochsim}), so the moderate selection regime requires $s\gg 0.01$ for $\mathrm{Pe}=0.97\,\mathrm{Pe}_\mathrm{max}$, while it only requires $s\gg 0.002$ for $\mathrm{Pe}=0.5\,\mathrm{Pe}_\mathrm{max}$. This explains the slight underestimate of the fixation probability for the largest Péclet numbers shown in main text Fig.~3(b).

\newpage
\section{Additional analytical insight with a step-function division rate profile}

In this Section, we will make the simplification that the stationary division rate profile in the absence of mutants is a step function:
\begin{align}
\rho(x) = \frac{F(x)}{k+F(x)} \approx \Theta(x_0 -x),
\end{align}
where $x_0$ denotes the position of the step~\cite{ghosh2022emergent}. This means that we make the approximation $k \ll F_\mathrm{in}$. Furthermore, we assume that the stationary food concentration satisfies $F(x)\gg k$ in the upstream food-rich region, while $F(x)=0$ in the downstream food-poor region, with a sharp transition. The position $x_0$ of the step is determined self-consistently, ensuring that the coupling between food and bacterial concentration, which is crucial in our system, is preserved~\cite{ghosh2022emergent}. This approximation facilitates calculations and allows us to obtain some explicit analytical results.

\subsection{Dimensionless variables} \label{dimlessvars}

To streamline our analytical calculations, we will employ dimensionless variables throughout this section. These variables are listed below:
\begin{align}
    \begin{array}{c|c}
    \text {Dimensionless variable}   & \hspace{1cm} \text{Expression} \hspace{1cm} \ \\\hline
    X& x/L \\  \hline
    X_0&x_0/L \\\hline
     \tau & D t / L^2 \\\hline
     1+ s & r_m/ r_w  \\\hline
    \lambda_w & r_w L^2 / D \\\hline 
    \lambda_m & r_m L^2 / D  \\\hline
\text{Pe} & v L / D \\\hline
\text{Pe}_0 & 2\sqrt{\lambda_w}  \\\hline
K_w & \sqrt{4\lambda_w -\text{Pe}^2}/2=\sqrt{\text{Pe}_0^2 -\text{Pe}^2}/2\\\hline 
K_m & \sqrt{4\lambda_m -\text{Pe}^2}/2\\
\end{array}
\end{align}
Note that Pe denotes the Péclet number. This dimensionless number is used in fluid mechanics to compare advective transport to diffusive transport.
 
\subsection{Stationary wild-type concentration profile in the absence of mutants} \label{steadyBstep}

Let us first focus on the stationary wild-type concentration profile $B(x)$ in the absence of mutants, which provides the background when the mutant is introduced in our system. Our goal is to express it analytically in the case where the division rate profile is a step function.

At steady state in the absence of mutants, and using dimensionless variables, Eqs.~\eqref{eq:eq_B} and~\eqref{eq:eq_F} become
\begin{align}
    0 &= B''(X) - \text{Pe}B'(X) +\lambda_w \Theta (X_0-X)B(X) ,\\
    0 &= F''(X) - \text{Pe} F'(X) -\frac{\lambda_w}{\alpha} \Theta (X_0-X)B(X).
\end{align}
with boundary conditions
\begin{align}
B'(0)-\text{Pe}\,B(0) &= 0,
&\;
F'(0)-\text{Pe}\,F(0) &= -\text{Pe}\,F_{\mathrm{in}},\\
B'(1) &= 0,
&\;
F'(1) &= 0.
\end{align}
Following Ref.~\cite{labavic2022hydrodynamic}, we notice that 
\begin{align}
0 &= (B +\alpha F)''(X) - \text{Pe}\,(B +\alpha F)'(X).
\end{align}
Solving this differential equation and using the boundary conditions yields
\begin{align}
    B(X) +\alpha F(X)  = \alpha F_{\text{in}} \label{eq:B_F}.
\end{align} 
This conserved quantity arises from the fact that bacteria need to eat food to divide, which dictates the structure of the reaction term of the model. \\

\noindent\textbf{Upstream food-rich region.} In the upstream part of the gut $X< X_0$, where $\rho(X)=1$, we have
\begin{align}
    B_u''(X) -\text{Pe} B_u'(X) +\lambda_w B_u(X) = 0.
\end{align}
Solving this equation and using the boundary conditions yields 
\begin{align}
    B_u(X) = B_u^0 e^{\text{Pe} X/ 2} \left[ \frac{2K_w}{\text{Pe}} \cos(K_w X) +\sin(K_w X)\right],
\end{align} 
where we assumed $\text{Pe}<2\sqrt{\lambda_w}=\text{Pe}_0$ and we introduced
\begin{align}
   K_w = \frac{\sqrt{4\lambda_w -\text{Pe}^2}}{2}.
\end{align}

\noindent\textbf{Downstream food-poor region.} In the downstream part of the gut $X>X_0$, where $\rho(X)=0$, we further assume that $F(X) = 0$. Furthermore, the bacteria concentration (denoted by $B_d$) satisfies
\begin{align}
    B_d''(X) -\text{Pe} B_d'(X) = 0.
\end{align}
Using the boundary condition at $X =1$, it yields
\begin{align}
    B_d(X) = B_d^0.
\end{align}
Then, using Eq.~\eqref{eq:B_F} gives
\begin{align}
    B_d^0 = \alpha F_{\text{in}}.
\end{align}

\noindent\textbf{Matching of the upstream and the downstream solutions.} We now match the upstream and downstream solutions at $X = X_0$ in order to determine the two unknowns $B_u^0$ and $X_0$:
\begin{align}
    B_u(X_0) &= B_d(X_0) = \alpha F_{\text{in}} ,\label{eq:matching_x0_1}\\
    B_u'(X_0) &= B_d'(X_0) = 0 .\label{eq:matching_x0_2}
\end{align}
\cref{eq:matching_x0_1} yields
\begin{align}
    B_u^0 = \dfrac{\alpha F_{\text{in}} e^{-\text{Pe} X_0/2} }{\dfrac{2{K_w}}{\text{Pe}} \cos(K_w X_0) +\sin(K_wX_0)},
\end{align}
and \cref{eq:matching_x0_2} yields
\begin{align}
    \tan(K_w X_0) = \frac{{K_w} \text{Pe}}{{K_w}^2 - \text{Pe}^2/4}, \label{eq:cond_X0}
\end{align}
which provides the location $X_0$ of the step in the division rate:
\begin{align}
    X_0 = \frac{1}{{K_w}} \left[\arctan\left(\frac{{K_w} \text{Pe}}{{K_w}^2 -\text{Pe}^2/4} \right) + \ell \pi \right], \label{eq:cond_X02}
\end{align}
where $\ell$ is an integer number. At $\text{Pe}=\sqrt{2\lambda_w}$, the argument of the arctangent in Eq.~\eqref{eq:cond_X02} tends to $+\infty$ from below and to $-\infty$ from above. Hence, using the principal branch (between $-\pi/2$ and $\pi/2$), the arctangent jumps from $\pi/2$ to $-\pi/2$ at that point. One must thus choose the integer $\ell$ on each side of $\text{Pe}=\sqrt{2\lambda_w}$ to ensure that $X_0$ remains continuous. 
In the limit $\text{Pe} \rightarrow 0$, $X_0\rightarrow 0$, which entails $\ell =0$ for $\text{Pe} < \sqrt{2\lambda_w}$. The continuity of $X_0$ at $\text{Pe} =  \sqrt{2\lambda_w}$ then imposes that 
\begin{align}
    X_0\left(\text{Pe} \rightarrow  \sqrt{2\lambda_w}^{-}\right) &= X_0\left(\text{Pe} \rightarrow  \sqrt{2\lambda_w}^{+} \right), \\
    \text{i.e.}\;\;\frac{1}{K_w}\frac{\pi}{2} &= \frac{1}{K_w} \left( - \frac{\pi}{2} + \ell \pi \right).
\end{align}
Therefore $\ell = 1$ for $\text{Pe} > \sqrt{2\lambda_w} $. Finally, the solution for $B$ is
\begin{align}
    B(X) &= \alpha F_{\text{in}} e^{\text{Pe} (X-X_0)/2} \frac{2{K_w}\, \cos({K_w}X) + \text{Pe} \,\sin({K_w}X)}{ 2{K_w}\,\cos({K_w} X_0) + \text{Pe}\,\sin({K_w}X_0)} , & \textrm{for}\;X<X_0, \label{B_explicit1}\\
    B(X)& = \alpha F_{\text{in}}, & \textrm{for}\;X>X_0, \label{B_explicit2}
\end{align}

with
\begin{equation}
    X_0 = \left\{ 
\begin{array}{ll}
    \dfrac{1}{{K_w}} \arctan\left(\dfrac{{K_w} \text{Pe}}{{K_w}^2 -\text{Pe}^2/4} \right), & \ \text{if} \;\; \text{Pe} < \sqrt{2\lambda_w}, \\[3mm]
    \dfrac{1}{{K_w}} \left[\arctan\left(\dfrac{{K_w} \text{Pe}}{{K_w}^2 -\text{Pe}^2/4} \right) + \pi \right] ,& \ \text{if} \;\;  \sqrt{2\lambda_w}<\text{Pe}<\text{Pe}_\text{max}. \label{X0expl}
\end{array} 
    \right.
\end{equation}
The position $X_0$ of the step is thus determined self-consistently, from the coupled system. Note that, while Eq.~\eqref{X0expl} is written for all $\text{Pe}<\text{Pe}_0$, the step-profile requires $X_0\leq 1$, which in fact corresponds to the more restrictive condition $\text{Pe}<\text{Pe}_\text{max}$, see Section~\ref{sec:washout} below. 

Our explicit solution for the profile of $B(X)$ in the case of a step-function division rate profile coincides with that of Ref.~\cite{ghosh2022emergent} (Eqs.~S140–S145) in the limit $L\to\infty$. However, here, we fully address the case of a finite $L$, and we fix the amplitude of the profile using food conservation.

\begin{figure}[htbp]
\centering
\includegraphics[width=0.6\textwidth]{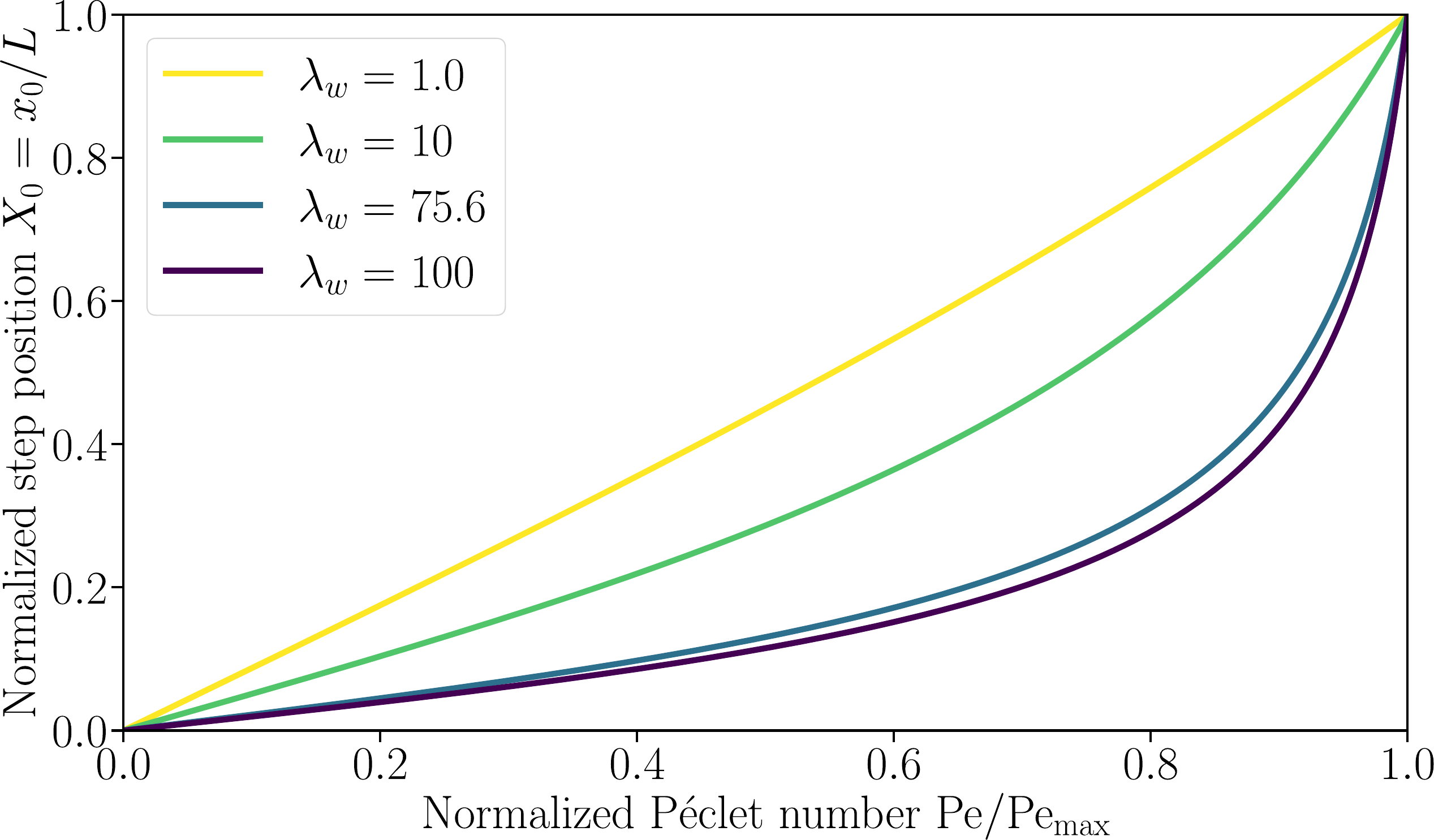}
\caption{Relative location $X_0=x_0/L$ within the gut of the step of the division rate (when $\rho$ is approximated by a step function) versus the normalized Péclet number $\text{Pe}/\text{Pe}_{\text{max}}$ for different values of $\lambda_w = r_w L^2 / D$ (recall that our usual value is $\lambda_w = 75.6$). The position $X_0$ is plotted using Eq.~\eqref{X0expl}. 
\label{Fig_X0}}
\end{figure}

Fig.~\ref{Fig_X0} shows how the self-consistently determined step position $X_0$ from Eq.~\eqref{X0expl} increases with the Péclet number for various values of $\lambda_w= r_w L^2 / D$. We observe that this increase arises for larger Pe values when $\lambda_w$ is larger. This is because quickly-dividing bacteria deplete food faster, resulting in a smaller region where division is non-negligible. 

\begin{figure}[htbp]
\centering
\includegraphics[width=0.55\textwidth]{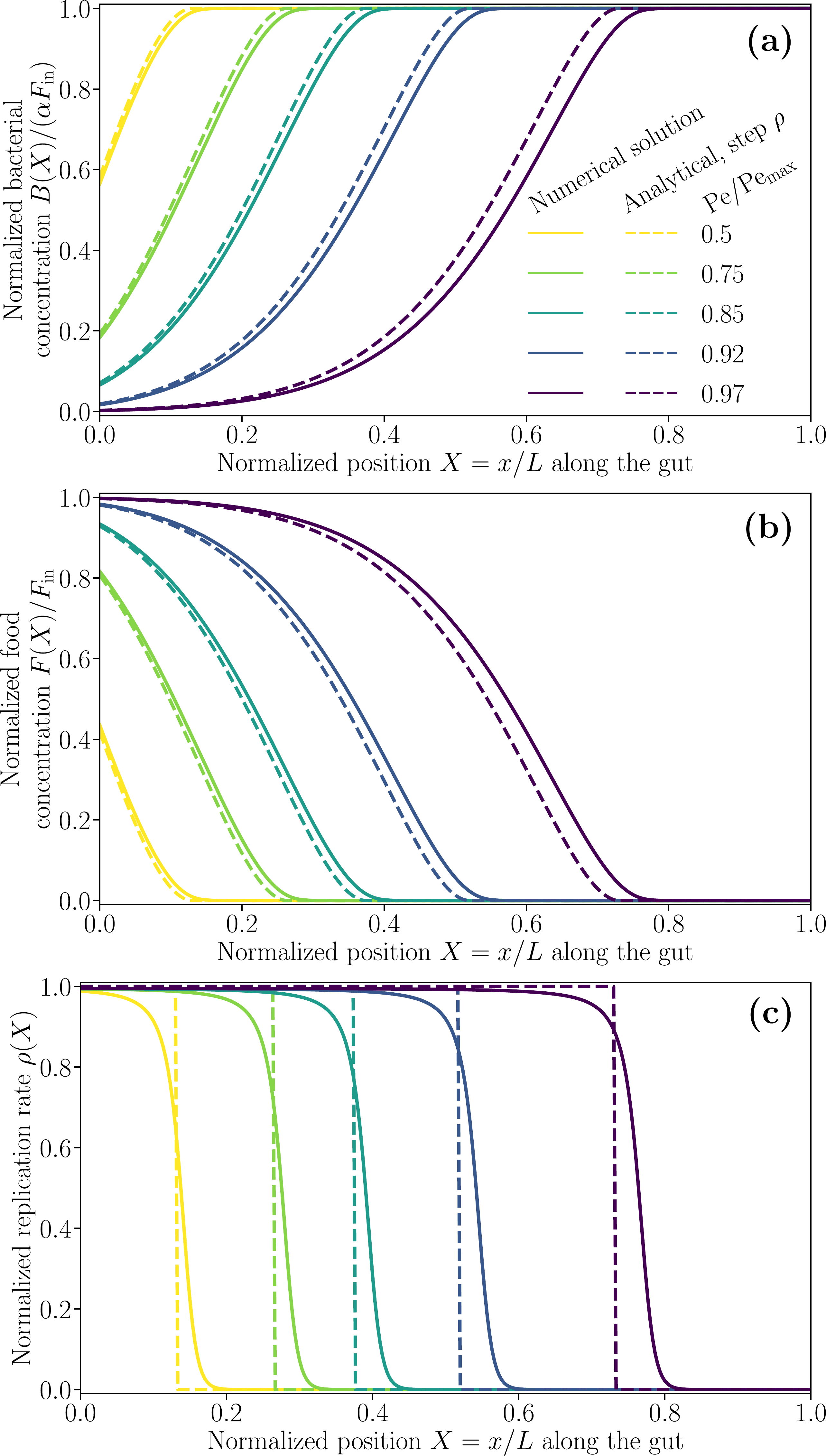}
\caption{Comparison of the exact system at steady state with its step-function approximation. Steady-state profiles of wild-type bacterial concentration $B$ (a), food concentration $F$ (b), and division rate $\rho$ (c) are obtained from numerical solutions of Eqs.~\eqref{eq:eq_F},~\eqref{eq:eq_B} at steady state. They are compared to those obtained with a step-function $\rho$, using the analytical solution for $B$ in Eqs.~\eqref{B_explicit1}, \eqref{B_explicit2} and \eqref{X0expl}. Parameter values: as in main text figures 1, 2 and 3, $\lambda_w = r_w L^2 / D= 75.6$ (yielding $\mathrm{Pe}_\mathrm{max}=16.6$), $k/ F_{\text{in}} =5 \times 10^{-3}$, $\alpha = 200$.
\label{Fig_step_profiles}}
\end{figure}

Fig.~\ref{Fig_step_profiles} further shows examples of profiles of wild-type bacterial concentration $B$, food concentration $F$ and division rate $\rho$ for different values of Pe, and compares the exact profiles obtained numerically to the analytical approximations using a step-function division rate profile. We observe that the step function approximation is quite good for this value of $k/ F_{\text{in}}$, which is small, ($k/ F_{\text{in}} =5 \times 10^{-3}$). Note that this value is realistic, see Section~\ref{humanparams}. 

\clearpage
\subsection{Washout condition} \label{sec:washout}

Beyond a certain flow velocity, washout occurs in the system, meaning that all bacteria get washed out of the system~\cite{cremer2016effect}. Mathematically, the washout condition
corresponds to a bifurcation point where the trivial steady state solution $F(x) = F_\text{in}$
and $B(x) = 0$ becomes stable~\cite{labavic2022hydrodynamic}. In previous works~\cite{cremer2016effect,Cremer2017,labavic2022hydrodynamic}, the washout condition was determined by comparing key length
scales and time scales of the system. Two conditions yielding two branches were thus identified. Assuming that the division rate profile is a step function allows us to perform an exact determination of the washout condition within that approximation. We further show that the two known branches are found as asymptotic limits of this exact condition.

Within our step-function approximation, washout is obtained when $X_0 =1$, meaning that the step occurs at the downstream boundary of the system. Thus, we aim to determine the value $\text{Pe}_{\text{max}} = v_{\text{max}} L / D$ of the Péclet number $\text{Pe}$ for which $X_0 =1$. Eq.~\eqref{eq:cond_X0} then gives the condition
\begin{align}
    \tan({K_{\text{max}}}) = \frac{{K_{\text{max}}}\text{Pe}_{\text{max}}}{{K_{\text{max}}}^2 - \text{Pe}_{\text{max}}^2/4}, \label{tanKmax}
\end{align}
where we introduced
\begin{align}
     {K_{\text{max}}} = \frac{\text{Pe}_0}{2}\sqrt{1-{(\text{Pe}_{\text{max}}/ \text{Pe}_0)}^2}, \label{defKmax}
\end{align}
and used $\text{Pe}_0=2\sqrt{\lambda_w}$ and $K_w=\sqrt{\text{Pe}_0^2 -\text{Pe}^2}/2$ (see Section~\ref{dimlessvars}). Note that, as in the previous section, we assume $\text{Pe} < \text{Pe}_0$, and in particular $\text{Pe}_{\text{max}} < \text{Pe}_0$.
Eq.~\eqref{tanKmax} can be reformulated as
\begin{align}
    \tan({K_{\text{max}}}) &=  \frac{2(\text{Pe}_{\text{max}}/ \text{Pe}_0) \sqrt{1- {(\text{Pe}_{\text{max}}/ \text{Pe}_0)}^2}}{1-2 {(\text{Pe}_{\text{max}}/ \text{Pe}_0)}^2} .
\end{align}
Since $0\leq\text{Pe}_{\text{max}}/ \text{Pe}_0<1$, let us introduce $\theta=\arcsin(\text{Pe}_{\text{max}}/ \text{Pe}_0)\in [0,\pi/2)$. This leads to
\begin{align}
    \tan({K_{\text{max}}}) =  \frac{2 \sin(\theta) \cos (\theta)}{1-2 \sin^2(\theta)} =  \tan(2\theta) .
\end{align}
Combining Eq.~\eqref{eq:cond_X02} (written for $X_0=1$) with Eq.~\eqref{tanKmax} provides $K_{\text{max}} = \arctan[\tan(2\theta)] + \ell\pi$. 
Recall that $0\leq 2\theta<\pi$. For $0\leq 2\theta<\pi/2$, the principal branch of the arctangent gives $\arctan[\tan(2\theta)] = 2\theta$, and we determined above that $\ell=0$, which gives $K_{\text{max}} = 2\theta$. For $\pi/2<2\theta<\pi$, the principal branch of the arctangent gives $\arctan[\tan(2\theta)] = 2\theta-\pi$,
and we determined above that $\ell=1$, which also gives $K_{\text{max}} = 2\theta$. Thus, we have 
\begin{equation}
    K_{\mathrm{max}} = 2\theta=2\arcsin(\text{Pe}_{\text{max}}/ \text{Pe}_0).
\end{equation}
Using Eq.~\eqref{defKmax}, this yields the washout condition
\begin{align}
    \text{Pe}_0 = \frac{4\arcsin{(\text{Pe}_{\text{max}}/ \text{Pe}_0)} }{\sqrt{1-(\text{Pe}_{\text{max}}/ \text{Pe}_0)^2}}. \label{eq:eq_Pe_max}
\end{align}

\begin{figure}[htbp]
\centering
\includegraphics[width=0.6\textwidth]{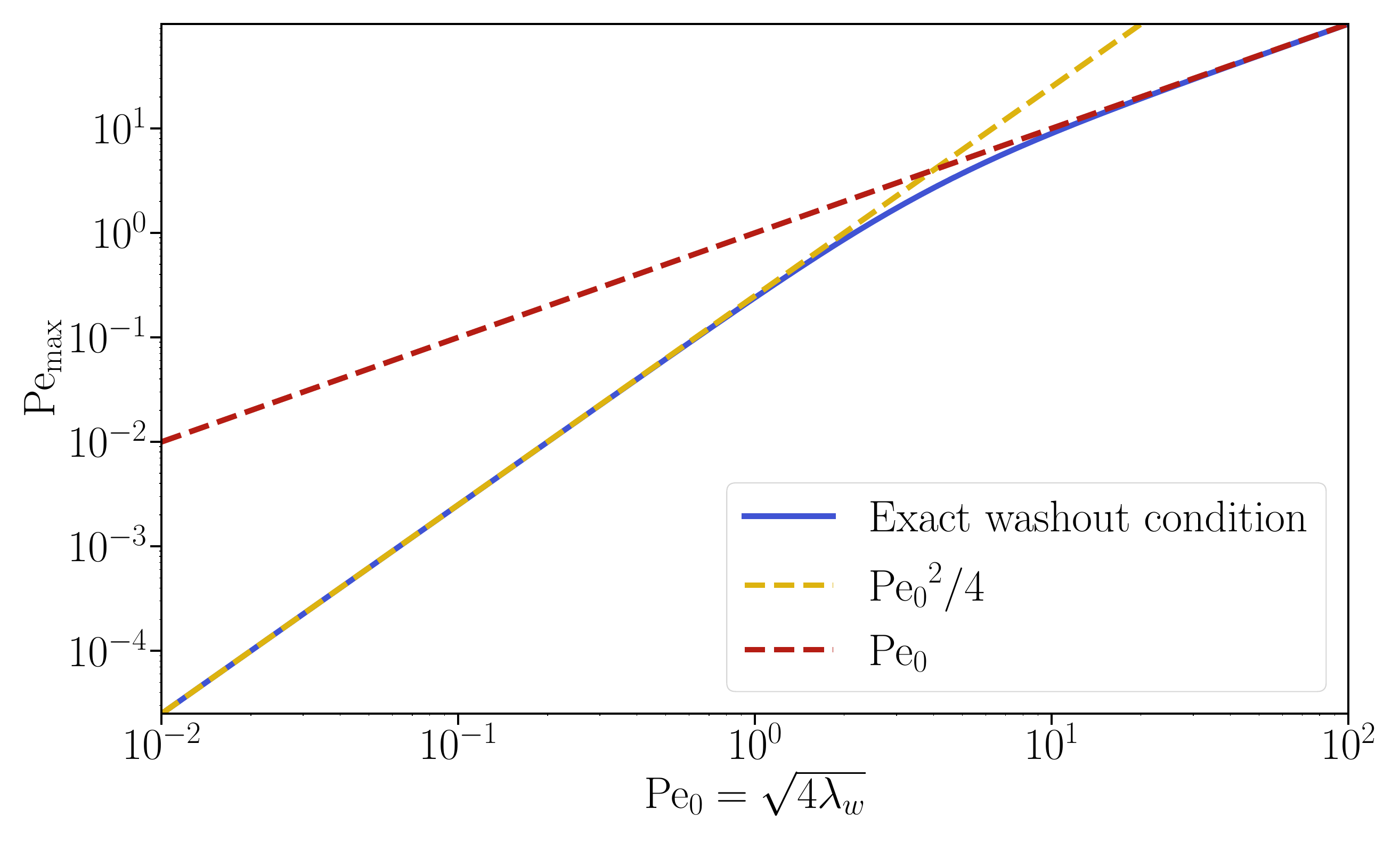}
\caption{Exact washout condition in the step-function division-rate case, and asymptotic regimes. The Péclet number above which washout occurs is plotted as a function of $\text{Pe}_0$. Solid blue line: Péclet number $\text{Pe}_{\text{max}}$ given by Eq.~\eqref{eq:eq_Pe_max}. Dashed yellow line: small $\text{Pe}_0$ asymptotic regime in Eq.~\eqref{wo_asympt_1}: $\text{Pe}_{\text{max}} = {\text{Pe}_0}^2 / 4$. Dashed red line: large $\text{Pe}_0$ asymptotic regime  in Eq.~\eqref{wo_asympt_2}: $\text{Pe}_{\text{max}} = {\text{Pe}_0}$.
\label{FigWashout}}
\end{figure}

\noindent\textbf{Asymptotic regimes.} First, for $\text{Pe}_{\text{max}} \ll \text{Pe}_0\ll 1$, Eq.~\eqref{eq:eq_Pe_max} gives
\begin{align}
    \text{Pe}_{\text{max}} \approx \frac{\text{Pe}_0^2}{4}= \lambda_w = \frac{r_w L^2}{D} , \label{wo_asympt_1}
\end{align}
and thus
\begin{align}
    v_{\text{max}} \approx r_w L.
\end{align}
Second, in the opposite regime $\text{Pe}_0 \gg 1$, Eq.~\eqref{eq:eq_Pe_max} gives $\text{Pe}_{\text{max}}\to \text{Pe}_0$, and we have 
\begin{align}
    \text{Pe}_{\text{max}} \approx \text{Pe}_0= 2\sqrt{\lambda_w} = 2\sqrt{\frac{r_w L^2}{D}} ,  \label{wo_asympt_2}
\end{align}    
and hence
\begin{align}
    v_{\text{max}} \approx 2\sqrt{r_wD}.
\end{align}
These two asymptotic regimes are in agreement with those found in Refs. \cite{labavic2022hydrodynamic, cremer2016effect}. Fig.~\ref{FigWashout} illustrates the exact washout condition found here and these two asymptotic regimes.

\subsection{Fixation probability}

\subsubsection{Moderate selection regime}
\label{step_fix_mod}

In Section~\ref{phi-expansion}, we performed a perturbative calculation of the fixation probability in the moderate selection regime where $[1+c_v^2(f)]/N_\mathrm{active}\ll s\ll 1$ (see Section~\ref{sec:neutral}), and we obtained the following result (see Eqs.~\eqref{phi_order1} and~\eqref{deff}, written using the dimensionless variables from Section~\ref{dimlessvars}): 
\begin{equation}
    \phi(X)=s\frac{\left<f\right>}{\left<f^2\right>}f(X)+O(s^2), \label{phi_order1-step}
\end{equation}
with
\begin{equation}
f(X)=B(X) e^{-\text{Pe}X}, \label{deff-step}
\end{equation}
where $B(X)$ is the steady-state wild-type bacteria concentration profile in the absence of mutants. In the case of the step-function division rate profile, $B(X)$ has an explicit analytical expression, given by Eqs.~\eqref{B_explicit1},~\eqref{B_explicit2} and~\eqref{X0expl}. Furthermore, since $\rho(X)=\Theta(X_0-X)$, the division-event-weighted averages involved in Eq.~\eqref{phi_order1-step} can be expressed as
\begin{align}
     \left<f\right>  = \frac{\int_0^{X_0} \dd X \,B(X)f(X) }{\int_0^{X_0} \dd X\, B(X)} \;\;\text{and}\;\;\left<f^2\right>  = \frac{\int_0^{X_0} \dd X \,B(X)f^2(X) }{\int_0^{X_0} \dd X\, B(X)}.\label{rwaf}
\end{align}
Hence, with the step-function division rate profile, we obtain an explicit analytical representation of the first-order spatial fixation profile from Eqs.~\eqref{phi_order1-step},~\eqref{B_explicit1},~\eqref{B_explicit2} and~\eqref{X0expl}. Consequently, Eq.~\eqref{phibar_order1} gives an explicit analytical representation of the division-event-weighted average of $\phi$ as well. Computing the integrals yields the explicit formula
\begin{align}
    \left<\phi\right>  = s \,\frac{4 \lambda_w (9 \lambda_w-2 \mathrm{Pe}^2) (\mathrm{Pe}+\lambda_w X_0)^2}{(1+8 e^{\mathrm{Pe} X_0/2}) (\mathrm{Pe}^3-4 \lambda_w \mathrm{Pe})^2}+O(s^2).
\end{align}
Introducing $\beta=\arcsin(\mathrm{Pe}/\mathrm{Pe}_0)$ yields
\begin{align}
    \left<\phi\right>  = s \,\frac{(9 -8 \sin^2\beta) (\beta+\sin\beta \cos\beta)^2}{4(1+8\, e^{2\beta\tan\beta}) \sin^2\beta \cos^6\beta}+O(s^2), \label{phimoy_beta}
\end{align}
which shows that to first order in $s$, $\left<\phi\right>$ only depends on $\mathrm{Pe}/\mathrm{Pe}_0=v/v_F$, where $v_F=2\sqrt{r_wD}$ is the Fisher speed.

Note that the differential equation for the fixation probability (Eq.~\eqref{eq:phi}) linearized around $\phi=0$ in the step-function division rate profile case has a nontrivial solution if and only if
\begin{align}
    \frac{\tan(K_m X_0)}{K_m \text{Pe}}\left[ {K_m}^2 - \frac{\text{Pe}^2 }{4} \right]& = 1 ,
\end{align}
where we assume $\cos(K_mX_0)\neq0$. 
For $K_m=K_w$, or equivalently $\lambda_m=\lambda_w$ or $r_m=r_w$, this is the same condition as for the stationary wild-type profile $B(X)$, namely Eq.~\eqref{eq:cond_X0}. This provides a consistency check that $s=0$ is the critical point, in agreement with the general branching-process result above, see Section~\ref{critpt}.

\subsubsection{Impact of the step-function division rate approximation}

\begin{figure}[hb]
    \centering
    \includegraphics[width=0.6\linewidth]{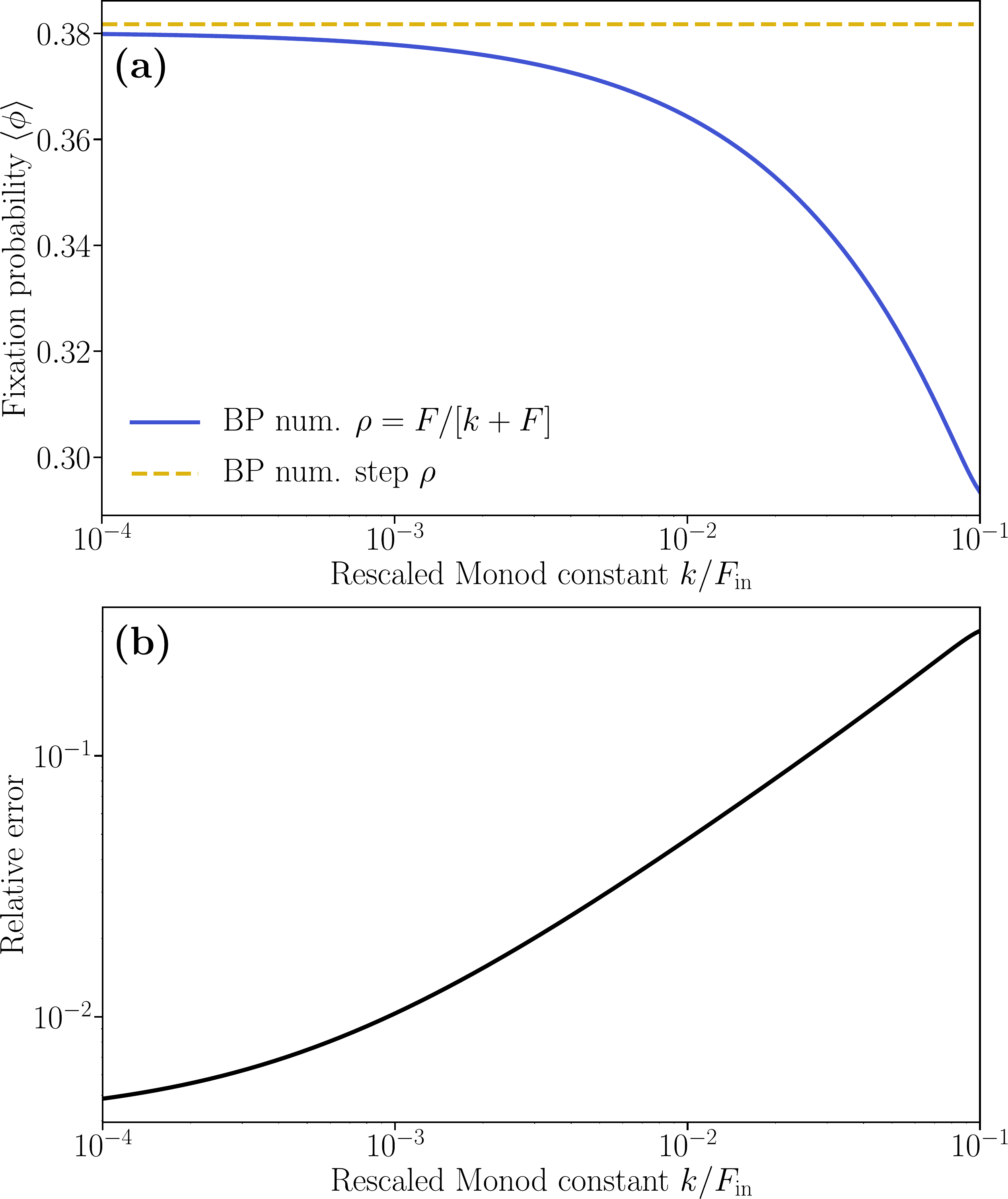}
    \caption{Comparison of the mutant fixation probability in the exact system and in its step-function approximation. (a) The division-event-weighted mutant fixation probability $\langle\phi\rangle$ is plotted versus the rescaled Monod constant $k/F_\mathrm{in}$, for a mutant relative fitness advantage $s=1$. Numerical solutions of the branching process Eq.~\eqref{eq:phi1} use either the steady-state mutant-free profile of the division rate $\rho$ (``BP num. $\rho=F/[k+F]$'') or its step approximation (``BP num. step $\rho$''). Note that the Moran model for large populations, which is the well-mixed reference, gives $\langle\phi\rangle=s/(1+s)=0.5$ for $s=1$. (b) Relative error between the two curves plotted in (a). Parameters: same $\lambda_w$ (and $\mathrm{Pe}_\mathrm{max}$), $\alpha$, and $\text{Pe} =v L / D = 15.3=0.92 \,\mathrm{Pe}_\mathrm{max}$ (with $\mathrm{Pe}_\mathrm{max}=16.6$ corresponding to washout) as in main text Figs.~1 and 2. 
    \label{step_approx}}
\end{figure}

The validity of the step-function division rate approximation is expected to depend on the rescaled Monod constant $k/F_\mathrm{in}$: if it is very small, the real profile of $\rho$ is close to a step function, while if it is large, it is substantially smoother. Fig.~\ref{step_approx} reports the division-event-weighted mutant fixation probability $\langle\phi\rangle$ versus the rescaled Monod constant $k/F_\mathrm{in}$. It shows that the step-function division rate approximation leads to an error that grows with $k/F_\mathrm{in}$, but remains smaller than ten percent for $k/F_\mathrm{in}<10^{-2}$. This is in the realistic range for the human colon, see Section~\ref{humanparams}.

\subsection{Mean number of division events and mean exit time}

Here, we focus on the mean number $R_1(x)$ of division events of a tracked mutant parental branch starting at $x$ before it exits the system (see Section~\ref{sec-Rn}), first in the case of constant division rate throughout the system, and then in the case of a step-profile division rate. 
By definition, $R_1(x)$ (see Eq.~\eqref{eq:R1def0}) is the expectation value over all stochastic trajectories $X_t$ such that $X_0=x$ of the number of division events undergone along the tracked parental branch. Hence,
\begin{equation}
    R_1(x)=\mathbb{E}_x\left[\int_0^{t_\text{exit}} r_m\rho(X_t)\,\dd t\right], \label{R1again}
\end{equation}
where $\mathbb{E}_x$ denotes the expectation value over all stochastic trajectories $X_t$ starting at $x$, and $t_\text{exit}$ denotes the time at which the tracked parental branch exits the system. Thus, we investigate how $R_1(x)$ is related to the mean exit time $T(x)$ of a bacterium.\\

\noindent\textbf{Mean exit time.} Let us first consider the mean time $T(x)$ until the tracked branch, starting at $x$, exits the system through the downstream boundary $x=L$. This mean exit time satisfies 
\begin{align}
    DT''(x) + vT'(x) = -1, 
    \label{eqT}
\end{align}
and the backward boundary conditions Eqs.~\eqref{bc_bwd1_mt} and \eqref{bc_bwd2_mt}. 
We find the explicit solution
\begin{align}
    T(x) = \frac{D}{v^2} \left(1 - e^{-vx/D} \right) + \frac{L-x}{v} .\label{solT}
\end{align}
Note that this result holds for any division rate profile $\rho(x)$. Indeed, the time spent by one bacterium in the system depends on transport and not on division. This equation entails in particular that $T(0)=L/v$, and that $T(L) = \frac{D}{v^2} \left(1 - e^{-vL/D} \right)>0$: a branch starting at the downstream boundary stays a finite mean time in the system, consistent with the boundary conditions employed.\\

\noindent\textbf{$R_1$ for a constant division rate.} $R_1(x)$, formally defined in Eq.~\eqref{eq:R1def0}, satisfies the differential equation Eq.~\eqref{eq:eq_R1} and the boundary conditions Eqs.~\eqref{bc_bwd1_mt} and \eqref{bc_bwd2_mt}. In the specific case where the division rate is constant throughout the system, $r_m\rho(x) = r_m$ (i.e.\ $\rho(x) = 1$), Eq.~\eqref{eq:eq_R1} becomes
\begin{equation}
    DR_1''(x) + vR_1'(x) =- r_m. \label{eqR1cst}
\end{equation}
We notice that $R_1/r_m$ satisfies the same differential equation as $T$ (see Eq.~\eqref{eqT}), and the same boundary conditions. Hence, in the case of a constant division rate, we have $R_1(x)=r_m T(x)$, with $T(x)$ given in Eq.~\eqref{solT}: the expected number of division events is the division rate times the expected residence time. More formally, Eq.~\eqref{R1again} simplifies when the division rate is constant, as it can be taken out of the integral and of the expectation value, yielding the simple product $R_1(x)=r_m T(x)$.\\

\noindent\textbf{$R_1$ for a step division rate.} Let us now return to the case where $\rho(X) = \Theta(X_0-X)$, and let us also return to dimensionless variables, as elsewhere in this Section. In this case, Eq.~\eqref{eq:eq_R1} becomes
\begin{equation}
    R_1''(X) + \text{Pe}\,R_1'(X) =- \lambda_m \Theta(X_0-X),
\end{equation}
and we find the following analytical expression for $R_1$:
\begin{equation}
    R_1(X) =
    \left\{ \begin{array}{cl}
        \dfrac{\lambda_m}{\text{Pe}} \left( \dfrac{1- e^{-\text{Pe}X}}{\text{Pe}} +X_0 - X\right) & ,\ \text{if} \  X< X_0, \\
        \dfrac{\lambda_m}{\text{Pe}^2} e^{-\text{Pe} X} \left(  e^{\text{Pe}X_0} - 1\right)  &, \ \text{else}.
    \end{array}  \right. \label{R1expl}
\end{equation}
The second line shows that for $X>X_0$, we have $R_1(X)=R_1(X_0)e^{-\text{Pe}(X-X_0)}$: a tracked parental branch starting there cannot divide locally, and its nonzero $R_1$ arises entirely from the possibility of diffusing back into the region where division is possible before exiting the system. Note also that, from the above explicit formula for $R_1$, we can check that the division-event-weighted average is $\left< R_1\right>= 1 + s$, consistent with our general result in Eq.~\eqref{Rnbar}. 

Above, we saw that in the case of a constant division rate, $R_1(x)=r_mT(x)$, i.e.\ $R_1(X)=\lambda_mT(X)$ (with the dimensional $T(x)$ related to the dimensionless one $T(X)$ via $T(x)=T(X)L ^2/D$). To see how this is modified for a heterogeneous division rate, let us consider the initial-position division rate multiplied by the total residence time, namely $\lambda_m\rho(X) T(X)$, and compare it to $R_1(X)$. 
Using Eq.~\eqref{solT} with dimensionless variables and $\rho(X) = \Theta(X_0-X)$ provides
\begin{align}
    \lambda_m\rho(X) T(X)  =\dfrac{\lambda_m}{\text{Pe}} \Theta(X_0- X) \left( 1- X + \frac{1 - e^{-\text{Pe} X}}{ \text{Pe}} \right).
\end{align}
Comparing this result to Eq.~\eqref{R1expl}, we notice that
\begin{align}
    \lambda_m T(X) = R_1(X) + \frac{\lambda_m}{\text{Pe}} (1-X_0)\;\;\;\;  \text{for}\; X<X_0.
\end{align}
Taking the division-event-weighted average, where only $X<X_0$ contributes, thus yields
\begin{align}
    \left< \lambda_m\rho T\right> = \left<R_1\right> + \frac{\lambda_m}{\text{Pe}} (1-X_0).
\end{align}
This result reduces to $\left<R_1\right> = \left<\lambda_m\rho T\right>$ in the limit $X_0 \rightarrow 1$, consistent with our constant division rate result above. By contrast, for $X_0\in(0,1)$, we have $\left<R_1\right> < \left<\lambda_m\rho T\right>$. The extra positive term in $\left<\lambda_m\rho T\right>$ quantifies the overestimate of $\left<R_1\right>$ obtained by assuming that division is possible during the whole time the tracked parental branch spends in the system, while division only occurs in the region $X<X_0$.

\subsection{Deterministic growth of the mutant concentration}

Here, we apply the method described in Section~\ref{detgrowth-general} to the case of a step-function growth profile. 
Rewriting the differential equation for the mutant concentration $M$, namely Eq.~\eqref{eq:eq_M}, in terms of dimensionless variables (see Section~\ref{dimlessvars}), and looking for a separable solution of the form $M(X, \tau) = g(X) h(\tau) $, gives
\begin{align}
    \frac{1}{h(\tau)} \frac{\dd h }{\dd \tau }(\tau) = \frac{g''(X)}{g(X)}  - \text{Pe} \,\frac{g'(X)}{g(X)}  + \lambda_m  \Theta(X_0-X) \ = \tilde{\gamma},
\end{align}
where $\tilde{\gamma}$ is an eigenvalue of the problem, and can be interpreted as a dimensionless deterministic growth exponent. Indeed, solving the differential equation for $h$ yields $h(\tau) = h_0 e^{\tilde{\gamma} \tau}$, where $h_0$ is a constant. The dimensionless $\tilde{\gamma}$ is related to $\gamma$ from Section~\ref{detgrowth-general} via $\tilde{\gamma}=L^2\gamma/D$. As in Section~\ref{detgrowth-general}, our goal is to find the largest eigenvalue $\tilde{\gamma}_{\max}$, which determines the mutant invasion growth rate. 
Let us consider the differential equation for $g$:
\begin{align}
    g''(X) - \text{Pe}\, g'(X) + \left[\lambda_m  \Theta(X_0-X) - \tilde{\gamma}\right] g(X) = 0, \label{eqg}
\end{align}
with boundary conditions given by the dimensionless forms of Eqs.~\eqref{bc_fwd1} and~\eqref{bc_fwd2}, namely $g'(0)-\text{Pe}g(0)=0$ and $g'(1)=0$. Note that Eq.~\eqref{eqg} corresponds to Eq.~S211 in Ref.~\cite{ghosh2022emergent}, but that the limit $L\to\infty$ was taken there.

On the downstream side $X> X_0$, Eq.~\eqref{eqg} reduces to
\begin{align}
    g_d''(X) - \text{Pe}\, g_d'(X) - \tilde{\gamma} g_d(X) = 0.
\end{align}
The solution is
\begin{align}
    g_d(X) = A_d e^{r_+ X} + C_d e^{r_- X} ,
\end{align}
where $A_d$ and $C_d$ are constants and
\begin{align}
    r_{\pm} = \frac{\text{Pe}\, \pm \sqrt{\text{Pe}^2+ 4 \tilde{\gamma}}}{2}.
\end{align}
The downstream boundary condition $g_d'(1) = 0$ yields
\begin{align}
    g_d(X) = A_d \left( e^{r_+ X} - \frac{r_+}{r_-} e^{r_- X + r_+- r_-}  \right).
\end{align}
Since we are interested in the dominant eigenvalue in the case of a beneficial mutant, which is expected to satisfy $\tilde{\gamma}_{\max}>0$ (see Section~\ref{critpt} and the first-order result in Section~\ref{detgrowth-general}), we assume that $\tilde{\gamma}>0$. This ensures that $\text{Pe}^2+ 4 \tilde{\gamma}>0$ and $r_-<0$.

On the upstream side $X< X_0$, Eq.~\eqref{eqg} reduces to
\begin{align}
    g_u''(X) - \text{Pe}\, g_u'(X) + (\lambda_m - \tilde{\gamma}) g_u(X) = 0,
\end{align}
with the upstream boundary condition $g_u'(0)- \text{Pe}\, g_u(0) = 0$. 

There are now \textit{a priori} two cases to consider, depending on the sign of the discriminant $\text{Pe}^2-4(\lambda_m - \tilde{\gamma})$. 
Let us make the transformation to a Sturm-Liouville problem introduced in Section~\ref{detgrowth-general}. Introducing $y(X)=e^{-\text{Pe}X/2}g(X)$, Eq.~\eqref{eqg} becomes
\begin{equation}
    y''(X)=\left[\tilde{\gamma}+\frac{\text{Pe}^2}{4}-\lambda_m\Theta(X_0-X)\right]y(X).
\end{equation}
When $\tilde{\gamma}>\lambda_m-\text{Pe}^2/4$, the coefficient on the right-hand side is positive everywhere. The principal eigenfunction of this Sturm-Liouville problem does not change sign and can be chosen positive~\cite{teschl_book}. With $y(X)>0$ for $X\in[0,1]$, we also have $y''(X)>0$ for $X\in[0,1]$, and thus, $y'$ is strictly increasing. However, the boundary conditions give
\begin{equation}
    y'(0)=\frac{\text{Pe}}{2}y(0)>0,
    \qquad
    y'(1)=-\frac{\text{Pe}}{2}y(1)<0,
\end{equation}
which is incompatible with $y'(X)$ being strictly increasing. Hence $\tilde{\gamma}_{\max}\leq\lambda_m-\text{Pe}^2/4$, so all eigenvalues $\tilde{\gamma}$ satisfy $\tilde{\gamma}\leq\lambda_m-\text{Pe}^2/4$.

Let us thus focus on the case where $\tilde{\gamma}<\lambda_m-\text{Pe}^2/4$. 
In this case, we have
\begin{align}
    g_u(X) =  e^{\text{Pe} X/2} \left[ A_u\cos(Q_mX)+ C_u \sin(Q_mX) \right],
\end{align}
with 
\begin{align}
     Q_m = \frac{\sqrt{4\lambda_m - 4 \tilde{\gamma} - \text{Pe}^2}}{2}.
\end{align}
The upstream boundary condition yields
\begin{align}
    C_u = \frac{\text{Pe}}{2Q_m} A_u.
\end{align}
We now determine $\tilde{\gamma}$ by matching the upstream and downstream solutions so that
\begin{align}
    g_u(X_0) &= g_d(X_0) ,\label{eq:glgr} \\
    g_u'(X_0) &= g_d'(X_0) .\label{eq:glprim_grprim} 
\end{align}
This yields the system of equations
\begin{equation} \left\{
    \begin{aligned}
   A_u e^{\text{Pe} X_0/2} \left[\cos(Q_mX_0)+ \frac{\text{Pe}}{2Q_m }\sin(Q_mX_0) \right] &= A_d \left( e^{r_+ X_0} - \frac{r_+}{r_-} e^{r_- X_0 + r_+- r_-}  \right), \\
 A_u e^{\text{Pe} X_0/2} \left[ \text{Pe} \cos(Q_mX_0)+ \left(\frac{\text{Pe}^2}{4Q_m } - Q_m\right)\sin(Q_mX_0) \right] & = A_d \left( r_+e^{r_+ X_0} - r_+ e^{r_- X_0 + r_+- r_-}  \right) .
    \end{aligned} \right.
\end{equation}
This system admits nontrivial solutions if the associated determinant vanishes, which gives 
\begin{align}
     &r_+ \left[ e^{(r_+-r_-) (X_0-1)} - 1 \right] \left[1+ \frac{\text{Pe}}{2Q_m }\tan(Q_mX_0) \right] - \left[ e^{(r_+-r_-) (X_0-1)} - \frac{r_+}{r_-}  \right]  \left[ \text{Pe} + \left(\frac{\text{Pe}^2}{4Q_m } - Q_m\right)\tan(Q_mX_0) \right] = 0 \label{eq:eq_reff}.
\end{align} 
Note that this form assumes $\cos(Q_mX_0)\neq0$. Note also that the special case where $\tilde{\gamma}=\lambda_m-\text{Pe}^2/4$ corresponds to $Q_m=0$ and can be obtained by taking the $Q_m\to0$ limit of the expressions above. 
The largest solution $\tilde{\gamma}_{\text{max}}$ of Eq.~\eqref{eq:eq_reff} yields the effective mutant invasion growth rate: 
\begin{align}
    \gamma_{\text{max}} = \frac{D\, \tilde{\gamma}_{\text{max}}}{L^2}.
\end{align} 

Overall, compared to the general calculation in Section~\ref{detgrowth-general}, the step-function division rate profile allowed us to obtain an explicit equation to determine $\tilde{\gamma}_{\text{max}}$, namely Eq.~\eqref{eq:eq_reff}. This calculation did not require expanding in small $s$, and is thus not restricted to the weak-selection regime of Eq.~\eqref{gammamax-pert}. 

In addition, in the case of a step-function division rate profile, Eq.~\eqref{gammamax-pert} reduces to 
\begin{equation}
    \gamma_\mathrm{max}=s \,r_w\frac{\int_0^{X_0}B^2(X)e^{-\text{Pe}X}\,\dd X}{\int_0^1 B^2(X)e^{-\text{Pe}X}\,\dd X}+O(s^2), \label{gammamax-pert-step}
\end{equation}
which can be evaluated using the analytical expressions of $B(X)$ in Eqs.~\eqref{B_explicit1} and~\eqref{B_explicit2} and of $X_0$ in Eq.~\eqref{X0expl}. We find
\begin{equation}
\gamma_\mathrm{max}= s\, r_w \frac{2  \text{Pe} (\text{Pe} + \lambda_w X_0)}{\text{Pe}^2 + 2 \lambda_w (2 + \text{Pe} X_0)+ e^{-\text{Pe}(1- X_0)} (\text{Pe}^2-4 \lambda_w) }+O(s^2).
\end{equation}
Introducing $\beta=\arcsin(\mathrm{Pe}/\mathrm{Pe}_0)$ as in Eq.~\eqref{phimoy_beta} gives
\begin{equation}
    \gamma_\mathrm{max}= s\,r_w\,\frac{2\tan\beta\,\left(\beta+\sin\beta\cos\beta\right)}{1+\sin^2\beta+2\beta\tan\beta-\cos^2\beta\;e^{-\left(\mathrm{Pe}_0\sin\beta-4\beta\tan\beta\right)}}+O(s^2).
    \label{gammamax_beta}
\end{equation}

For reference, we also calculate $\gamma_\mathrm{WM}$ given in Eq.~\eqref{gammaWM} in the case of the step-function division rate profile:
\begin{align}
    \gamma_{\text{WM}} & = s r_w \frac{\text{Pe}}{\text{Pe} + \lambda_w (1- X_0)} = s r_w \frac{\text{Pe}}{\text{Pe} +  (1- X_0) {\text{Pe}_0}^2/4} \nonumber\\
    & = s\,r_w\,\frac{2\sin 2\beta}{2\sin 2\beta+\mathrm{Pe}_0\cos\beta-4\beta}\,,
\end{align}
where the expression in the second line is written in terms of $\beta=\arcsin(\mathrm{Pe}/\mathrm{Pe}_0)$ and $\mathrm{Pe}_0$.

\begin{figure}[htbp]
    \centering
    \includegraphics[width=0.6\linewidth]{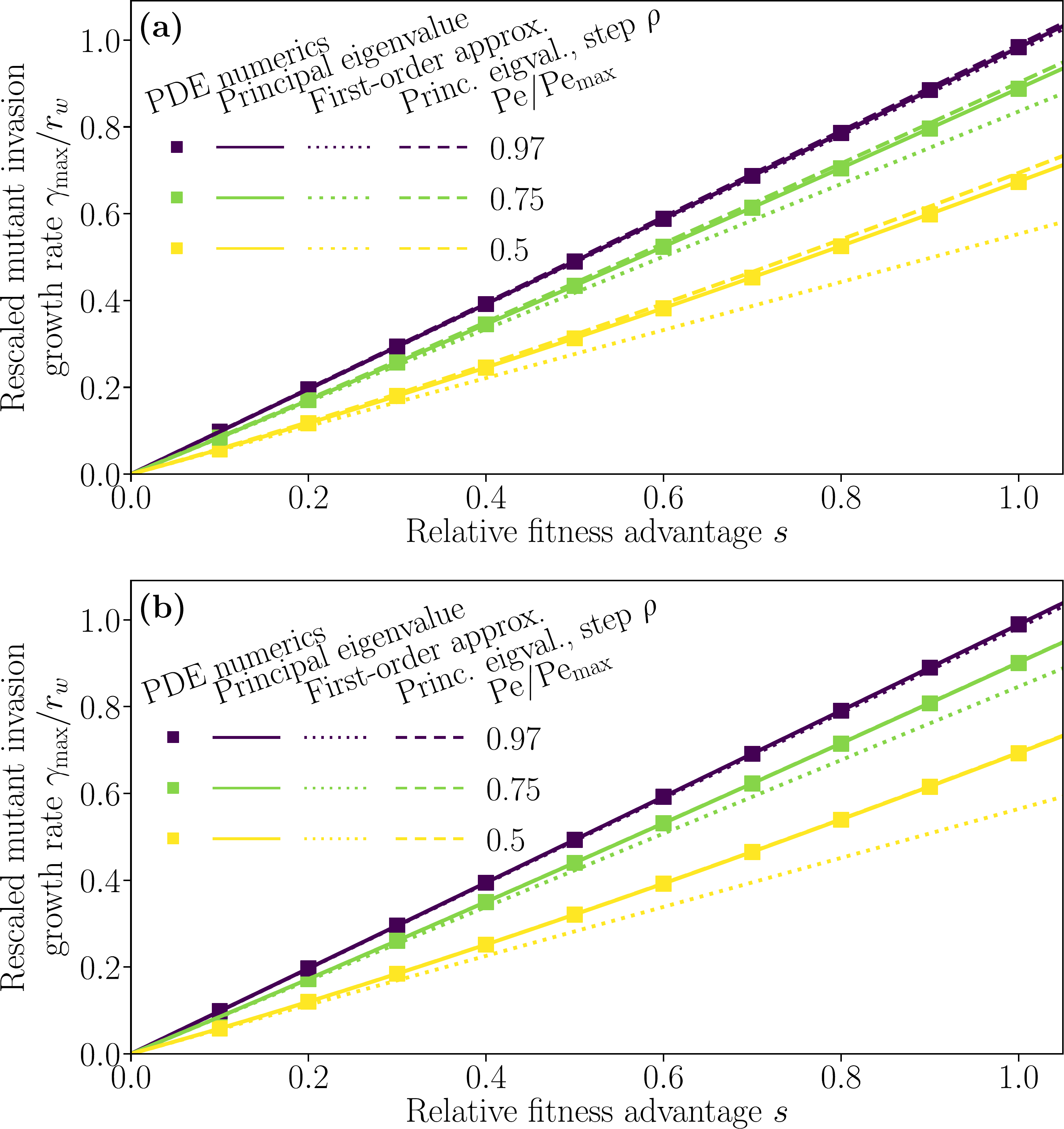}
    \caption{Mutant invasion growth rate $\gamma_\mathrm{max}$ versus mutant relative fitness advantage $s$, for different values of the Péclet number Pe. The mutant invasion growth rate $\gamma_\mathrm{max}$, scaled by the wild-type division rate $r_w$, is reported from different approaches. First, numerical integrations of Eqs.~\eqref{eq:eq_F},~\eqref{eq:eq_B},~\eqref{eq:eq_M} were performed with an initial small concentration of mutants, and the initial growth rate of mutant number was determined by a linear fit in a semilogarithmic plot (``PDE numerics''). Second, the principal eigenvalue $\gamma_\mathrm{max}$ of the operator $\mathcal{A}_s^\dagger$ was determined numerically (``Principal eigenvalue''). Its explicit first-order approximation is also shown (``First-order approx.''). Finally, the principal eigenvalue was also determined  using Eq.~\eqref{eq:eq_reff} in the case where $\rho$ is approximated by a step function (``Princ. eigval., step $\rho$''). (a) Parameter values: same $\lambda_w$ (and hence $\mathrm{Pe}_\mathrm{max}$), $k/F_\mathrm{in}$ and $\alpha$ as in main text Figs.~1, 2 and~3. (b) Same parameter values, except that $k/F_\mathrm{in}=10^{-4}$. In this case, the solid and dashed lines are superposed.
    }
    \label{fig:r_eff}
\end{figure}

Fig.~\ref{fig:r_eff} shows that the principal eigenvalue $\gamma_\mathrm{max}$ of $\mathcal{A}_s^\dagger$ indeed describes well the mutant invasion growth rate. We also note that $\gamma_\mathrm{max}$ is convex beyond the first-order approximation given in Eq.~\eqref{gammamax-pert}. Furthermore, we observe that $\gamma_\mathrm{max}$ grows with the Péclet number, and that, as $\mathrm{Pe}\to \mathrm{Pe}_\mathrm{max}$, $\gamma_\mathrm{max}/r_w\to s$, which is the well-mixed expectation for $k\ll F_\mathrm{in}$. Finally, Fig.~\ref{fig:r_eff}(b) shows that the expression in Eq.~\eqref{eq:eq_reff} within the step-function growth rate approximation provides an excellent approximation of the exact mutant invasion growth rate for $k/F_\mathrm{in}=10^{-4}$.

\newpage

\subsection{Fixation probability of a neutral mutant and active population size}

In Section~\ref{sec:neutral}, we showed that the fixation probability of a neutral mutant can be related to the active population size, see Eq.~\eqref{eq:Nactive}. In the case where the division rate profile $\rho$ is a step function, we obtain the explicit expression 
\begin{equation}
    \frac{N_\mathrm{active}}{N}= \frac{4 \lambda_w + \text{Pe}^2+ 
 2 \lambda_w \text{Pe} X_0 + e^{-\text{Pe}(1 - X_0)} ( \text{Pe}^2 - 4 \lambda_w ) }{2 (\lambda_w + \text{Pe} - \lambda_w X_0)(\text{Pe} + \lambda_w X_0)}\,, \label{NactiveStep}
\end{equation}
which can be rewritten in terms of $\beta=\arcsin(\mathrm{Pe}/\mathrm{Pe}_0)$ and $\mathrm{Pe}_0$ as
\begin{equation}
\frac{N_\mathrm{active}}{N}
=\frac{4\cos^2\beta\left[1+\sin^2\beta+2\beta\tan\beta-\cos^2\beta\,e^{-(\mathrm{Pe}_0\sin\beta-4\beta\tan\beta)}\right]}
{\left(2\beta+\sin2\beta\right)\left(\mathrm{Pe}_0\cos\beta-4\beta+2\sin2\beta\right)}.
\end{equation}

\newpage

\section{Comparison with Ref. \cite{ghosh2022emergent}} \label{comp_GG}

In Ref.~\cite{ghosh2022emergent}, the dynamics of a selected mutant lineage in the gut was also studied in the regime where mutants are rare. However, there are several differences from our analysis, which we describe here. 

First, here, we focus on the case where $D$ and $v$ are constant, and we neglect bacterial death. Ref.~\cite{ghosh2022emergent} formally addressed the more general case where $D$ and $v$ may depend on $x$ and included bacterial death. However, their subsequent analytical work also mainly focused on constant $D$ and $v$, and no bacterial death. Here, we restrict our comparison to this case.

Second, formally, in Ref.~\cite{ghosh2022emergent}, the change in local mutant number density over one short time interval was separated into deterministic and stochastic parts. The deterministic change contains the mean effects of birth, death, advection, and diffusion. Meanwhile, fluctuations around these means in the stochastic increment are approximated using independent, zero-mean, unit-variance Gaussian variables associated with the different microscopic event types. Here, we did not employ such an approximation.

Third, derivations in Ref.~\cite{ghosh2022emergent} are based on the Laplace generating functional of the fluctuating spatial mutant number density field. 
Combining the generating-functional representation in Eq.~S28 of Ref.~\cite{ghosh2022emergent} with the nonzero long-time fixed point in Eq.~S101 shows that the extinction probability, starting from a single initial mutant at position $x$, is $\tilde u(x)=e^{-\tilde w(x)}$. Eq.~13 of Ref.~\cite{ghosh2022emergent} is the stationary equation satisfied by the logarithmic field $\tilde w(x)=-\log[\tilde{u}(x)]$. 
Note that we denote their logarithmic field by $\tilde w$, while henceforth, $w$ will denote the corresponding exact log-extinction field in our branching process. These two fields are \textit{a priori} not identical, because $\tilde w$ follows from the Gaussian approximation of Ref.~\cite{ghosh2022emergent}, while $w$ follows from our exact branching process.

Starting from our Eq.~\eqref{eq:u} on the mutant extinction probability $u(x)$ in the exact fixed-environment branching process, and introducing $w(x)=-\log \left[u(x)\right]$, we obtain
\begin{equation}
0=D w''(x)+v w'(x)-D[w'(x)]^2+r_m\rho(x)\left[1-e^{-w(x)}\right].
\label{eq:exactwstationary}
\end{equation}
Note that a gradient-square term is generated by the logarithmic change of variables.
Performing a Taylor expansion of the exponential term for small $w$ leads to
\begin{equation}
	0= Dw''(x)+vw'(x)-D[w'(x)]^2+r_m\rho(x)w(x)-\frac{r_m\rho(x)}{2}w^2(x)+O\left(w^3(x)\right).
	\label{eq:exactwexpanded}
\end{equation}
Neglecting terms whose leading contribution is $O(w^3)$, this is the same bulk differential equation as Eq.~13 in Ref.~\cite{ghosh2022emergent}, specialized to constant $D$ and $v$, zero mutant death rate ($d_m=0$ in their notations), and with mutant division rate $b_m=r_m\rho(x)$. 
Thus, up to second order, the equation satisfied by our $w$ has the same form as the equation satisfied by their $\tilde w$. 
Indeed, the Gaussian approximation reproduces the first and second cumulants of demographic terms. Note that in our branching process, higher demographic cumulants generate cubic and higher-order terms absent from Eq.~13 of Ref.~\cite{ghosh2022emergent}. 

Despite this agreement up to second order in the logarithmic field $w$, it is important to note that
\begin{equation}
	\phi(x)\equiv 1-u(x)=1-e^{-w(x)}=w(x)-\frac{w^2(x)}{2}+O\left(w^3(x)\right).
	\label{eq:westapprox}
\end{equation}
Hence, the mutant log-extinction field $w$ and the mutant fixation probability $\phi$ only match to first order. Ref.~\cite{ghosh2022emergent} interprets its stationary logarithmic field as a fixation probability. Overall, beyond leading order, the two approaches differ both in the stochastic description and in the relation between the logarithmic field and fixation probability.

Finally, another difference is that although Ref.~\cite{ghosh2022emergent} specifies finite-$L$ boundary conditions for the bacterial density, their explicit minimal-model establishment calculation takes the downstream part of the colon to be infinite, with $\widetilde w(x)\to 0$ as $x\to\infty$. Here, we retain the finite colon model throughout.

\begin{figure}[htbp]
    \centering\includegraphics[width=0.55\linewidth]{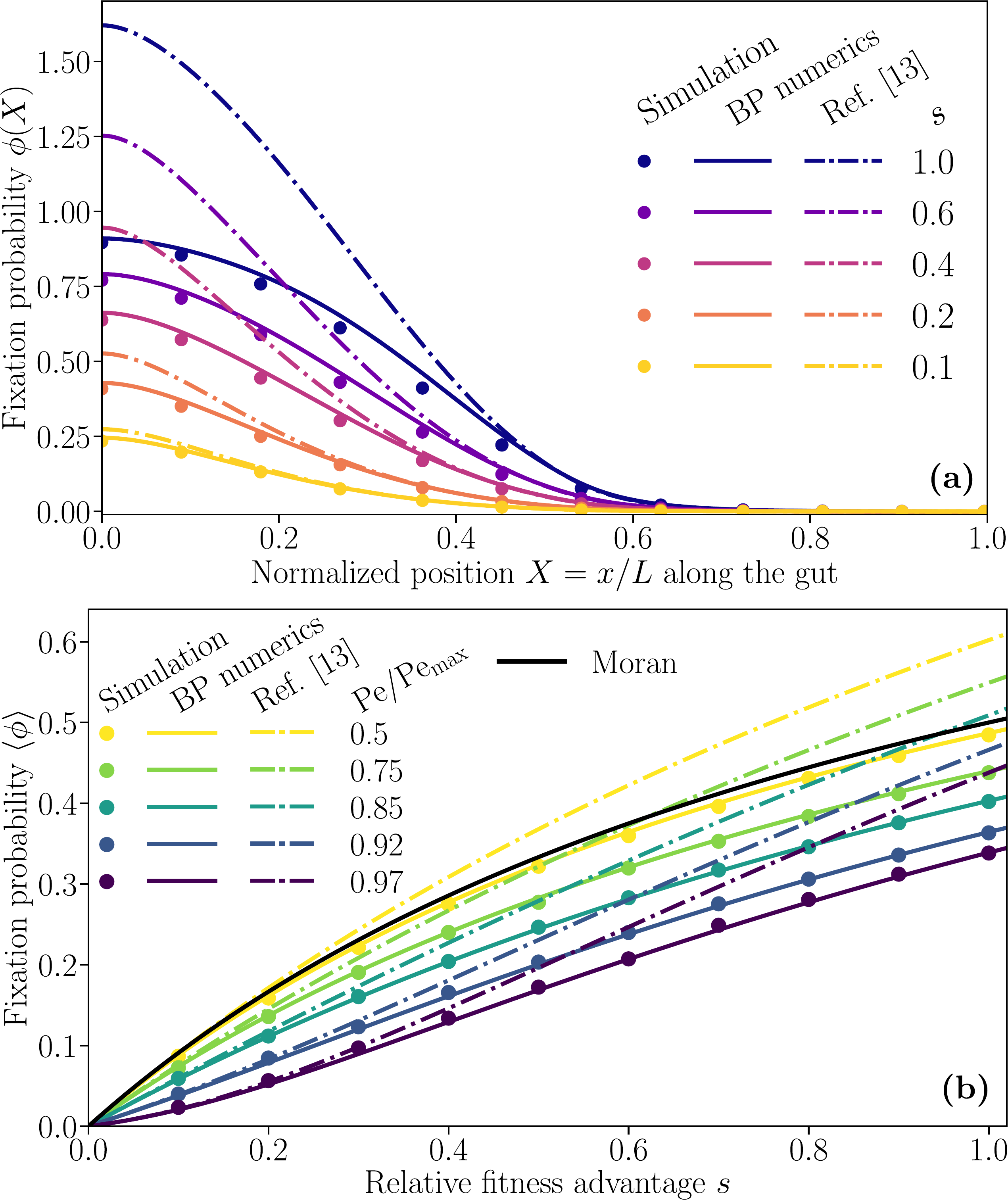}
    \caption{Comparison between the present results and those of Ref.~\cite{ghosh2022emergent} for the mutant fixation probability. \protect\\ (a) Mutant fixation probability $\phi(X)$ versus normalized initial mutant position $X=x/L$ along the gut, for different values of the mutant relative fitness advantage $s$. Stochastic simulations are described in Section~\ref{sec:stochsim}. Numerical solutions of the branching process Eq.~\eqref{eq:phi1} use the steady-state mutant-free profile of the division rate $\rho$ (``BP numerics''). Numerical solutions of Eq.~\eqref{eq:exactwexpanded}, matching Eq.~13 in Ref.~\cite{ghosh2022emergent}, use the same steady-state mutant-free profile of the division rate $\rho$ (``Ref.~\cite{ghosh2022emergent}''). Parameter values: same as in main text Figs.~1 and 2. \protect\\(b) Division-event-weighted mutant fixation probability $\langle\phi\rangle$ versus mutant relative fitness advantage $s$, for different values of the Péclet number Pe. Stochastic simulations are described in Section~\ref{sec:stochsim}. Numerical solutions of the branching process Eq.~\eqref{eq:phi1} use the steady-state mutant-free profile of the division rate $\rho$ (``BP numerics''). Numerical solutions of Eq.~\eqref{eq:exactwexpanded}, matching Eq.~13 in Ref.~\cite{ghosh2022emergent}, use the same steady-state mutant-free profile of the division rate $\rho$ (``Ref.~\cite{ghosh2022emergent}''). The Moran model for large populations, $\langle\phi\rangle=s/(1+s)$, gives the well-mixed reference. Parameters: same $\lambda_w$ (and $\mathrm{Pe}_\mathrm{max}$), $k/F_\mathrm{in}$ and $\alpha$ as in main text Figs.~1, 2 and 3. }
    \label{fig:Comp_Good}
\end{figure}

Fig.~\ref{fig:Comp_Good} compares our results from Eq.~\eqref{eq:phi1} and from our stochastic simulations to the results obtained using Eq.~\eqref{eq:exactwexpanded}, matching Eq.~13 in Ref.~\cite{ghosh2022emergent}. We observe that all results agree in the limit of small relative mutant fitness advantage (consistent with the fact that the first-order results coincide), but that our results are in better agreement with those of stochastic simulations for larger $s$. These conclusions hold both for the fixation probability $\phi(x)$ of a mutant introduced at a given location $x$ along the gut (Fig.~\ref{fig:Comp_Good}(a)) and for the division-event-weighted mutant fixation probability $\langle\phi\rangle$ (Fig.~\ref{fig:Comp_Good}(b)).

\section{Numerical methods}

\subsection{Stochastic simulations}
\label{sec:stochsim}

Our stochastic simulation method extends that of Ref.~\cite{labavic2022hydrodynamic}, which considered the neutral mutant case. Space and time are discretized, and we denote the corresponding discretization steps by $\delta x$ and $\delta t$, respectively. Specifically, as in Sections~\ref{discrete1} and~\ref{sec:discrete}, the system is discretized into $K$ intervals of length $\delta x = L/ K$, resulting into $K+1$ discrete sites, located at $x_i=i\delta x$, with $0\leq i\leq K$ (see Fig.~\ref{fig:fig1SI}). Boundary conditions are also implemented as specified there: there is no upstream jump from $x_0$, and there are only advective downstream jumps from $x_K$. 

We denote by $n_w(x_i,t)$ and $n_m(x_i,t)$ the numbers of wild-type and mutant bacteria at site $x_i$, respectively, and by $\mathcal{F}(x_i,t)=F(x_i,t)\delta x$ the amount of food at this site. Bacteria are discrete, while food is treated as a continuous quantity. The stochastic model is constructed to recover the continuous equations Eqs.~\eqref{eq:eq_F},~\eqref{eq:eq_B},~\eqref{eq:eq_M} in the continuum limit (where $\delta x$ and $\delta t$ tend to zero and bacterial number to infinity, and fluctuations become negligible).\\

\noindent\textbf{Transport and division.} During each time step $\delta t$, bacteria undergo transport and division, while food is consumed and transported by advection-diffusion. The corresponding rules are the following:

\begin{itemize}

\item Each bacterium has a probability
\begin{equation}
    p_-=\frac{D\delta t}{(\delta x)^2}
\end{equation}
to move to its upstream neighboring site, and a probability
\begin{equation}
    p_+=\frac{D\delta t}{(\delta x)^2}
    +\frac{v\delta t}{\delta x}
\end{equation}
to move to its downstream neighboring site. The probability to remain at the same site is
$1-p_--p_+$.

\item Each wild-type bacterium at site $x_i$ divides with probability
\begin{equation}
    p_w(x_i,t)
    =r_w\delta t
    \frac{\mathcal{F}(x_i,t)}{\tilde{k}+\mathcal{F}(x_i,t)},
\end{equation}
where $\tilde{k}=k\delta x$, while each mutant bacterium divides with probability
\begin{equation}
    p_m(x_i,t)
    =r_m\delta t
    \frac{\mathcal{F}(x_i,t)}{\tilde{k}+\mathcal{F}(x_i,t)}.
\end{equation}
Here, $r_w$ and $r_m$ denote the maximal division rates of wild-type and mutant bacteria, respectively.

\item Each division event produces one additional bacterium of the same type at the same site and consumes an amount $1/\alpha$ of food. Consequently, if $n_{\mathrm{rep}}(x_i,t)$ division events occur at site $x_i$ during $\delta t$, the local amount of food is decreased by
\begin{equation}
    \Delta \mathcal{F}(x_i,t)
    =-\frac{n_{\mathrm{rep}}(x_i,t)}{\alpha}.
\end{equation}
As in Ref.~\cite{labavic2022hydrodynamic}, this discretization may lead to transient negative local amounts of food. Whenever this occurs, the division probability is set to zero.

\item Food transport is implemented using a finite-difference discretization of the advection-diffusion equation. For interior sites, the food amount is updated according to
\begin{equation}
\begin{aligned}
\mathcal{F}(x_i,t+\delta t) = \mathcal{F}(x_i,t)
&+\frac{D\delta t}{(\delta x)^2}
\left[\mathcal{F}(x_{i+1},t)-2\mathcal{F}(x_i,t)+\mathcal{F}(x_{i-1},t)\right] -\frac{v\delta t}{2\delta x}
\left[\mathcal{F}(x_{i+1},t)-\mathcal{F}(x_{i-1},t)\right].
\end{aligned}
\end{equation}
The boundary sites are updated separately to account for the prescribed inlet conditions and the downstream boundary conditions.
Note that this implementation of food transport slightly differs from Ref.~\cite{labavic2022hydrodynamic}. However, the relative difference between the associated effective diffusion coefficients is $v\delta x/(2D)=\mathrm{Pe}/(2K)$, where $K$ is the number of intervals between discretized sites. Hence, the two schemes match for large enough $K$. In practice, we used $K=300$.\\

\end{itemize}

\textbf{Simulation procedure.} Simulations are performed as follows:

\begin{itemize}

\item \textit{Initialization.}
We first determine the stationary wild-type bacterial and food profiles by numerically solving the deterministic reaction-advection-diffusion equations. The stationary solution is obtained by iterating the discretized equations until the total absolute change in the bacterial and food profiles between two consecutive iterations falls below a prescribed convergence threshold.

The initial total number of wild-type bacteria is obtained from the stationary bacterial profile, rounded with the floor function. Individual bacteria are then randomly distributed among the sites, with probabilities proportional to the local stationary bacterial density. The initial food distribution is taken directly from the deterministic stationary profile.

\item \textit{Stochastic equilibration.}
Starting from this initial configuration, the wild-type population is left to relax to its stochastic steady-state by simulating its dynamics for a duration $t_{\mathrm{init}}$.

At each time step, the following operations are performed:
\begin{enumerate}
    \item Bacterial transport is simulated by drawing, at each site, the numbers of bacteria moving upstream, moving downstream, or remaining at the same site from a multinomial distribution. Boundary sites are treated separately to account for the boundary conditions.

    \item The number of dividing bacteria at each site is drawn from a binomial distribution, with the division probability determined by the local food amount. The bacterial population and food amount are updated accordingly.

    \item Food is transported deterministically according to the discretized advection-diffusion equation.
\end{enumerate}

The resulting bacterial and food distributions are used as the initial configuration for the subsequent fixation simulations.

\item \textit{Mutation.}
Before introducing a mutant, the wild-type population is left to relax for a duration $t_{\mathrm{inter}}$. A mutant is introduced using either of two possible procedures:
\begin{enumerate}
\item A mutant is added at a given site $x_i=i\delta x$. No other modification is performed (no food is removed, no wild-type is removed), but we checked that such modifications of the protocol do not impact conclusions. This protocol allows us to estimate the fixation probability $\phi(x)$ of a mutant appearing at position $x$ along the gut (see main text Fig.~2).
\item A mutant is introduced during the next division event. Specifically, if several bacteria reproduce during the same time step, one of the newly produced bacteria is chosen uniformly at random and assigned the mutant type. Thus, the spatial position of the introduced mutant is sampled according to the distribution of division events, rather than uniformly along the system. This protocol allows us to estimate the division-event-weighted fixation probability $\langle\phi\rangle$ of a mutant appearing in the gut (see main text Fig.~3). It models mutants appearing with a fixed probability at each bacterial division.
\end{enumerate}

\item \textit{Dynamics until fixation or extinction.}
Following mutant introduction, both wild-type and mutant populations are transported in the same way, but they divide with their respective maximal division rates $r_w$ and $r_m$.

The simulation continues until one of the two populations disappears. Three outcomes are distinguished: extinction of the mutant, fixation of the mutant, and complete washout of both populations. Specifically, a mutant is considered fixed when no wild-type bacteria remain while at least one mutant bacterium is still present. Conversely, mutant extinction occurs when all mutant bacteria disappear while wild-type bacteria remain. Complete washout corresponds to the disappearance of both populations. Although this outcome is possible in principle, it was never observed in our simulations, which were conducted below the washout threshold.

\end{itemize}

The stochastic procedure is repeated over $N_{\mathrm{sim}}$ realizations. As mentioned above, between successive realizations, the wild-type population is left to relax for a duration $t_{\mathrm{inter}}$, starting from the configuration used to initialize the preceding realization. This procedure allows us to sample different configurations of the wild-type population in the stochastic steady state, rather than systematically starting from the same initial configuration, thereby reducing potential biases in the estimation of the fixation probability. 

The fixation probability is estimated from the fraction of realizations resulting in mutant fixation:
\begin{equation}
    \hat{\phi}
    =\frac{N_{\mathrm{fix}}}
    {N_{\mathrm{sim}}},
\end{equation}
where $N_{\mathrm{fix}}$ denotes the number of fixation events.\\

\noindent\textbf{Parameter values. }
The parameter values employed in our stochastic simulations are listed below:
\begin{center}
\begin{tabular}{l|l}
Parameter & Value \\
\hline
Maximal wild-type division rate $r_w$ & $0.42$~h$^{-1}$~\cite{labavic2022hydrodynamic} \\
Length $L$ of the system & $6$~cm~\cite{labavic2022hydrodynamic} \\
Effective diffusion coefficient $D$ & $0.2$~cm$^2$/h~\cite{labavic2022hydrodynamic} \\
Cross-section area $S$ of the gut & $1$~cm$^2$~\cite{labavic2022hydrodynamic} \\
Conversion factor $\alpha$ & $200$~bacteria/mM \\
Incoming food concentration $F_{\mathrm{in}}$ & $2$~mM \\
Monod constant $k$ & $0.01$~mM 
\end{tabular}
\end{center}
Note that the linear concentrations used throughout this work are related to volume ones used in Ref.~\cite{labavic2022hydrodynamic} via $F_{\mathrm{linear}}=S F_{\mathrm{volume}}$. 
These values give the dimensionless parameter values $\lambda_w=75.6$ and $k/F_{\mathrm{in}}=5\times10^{-3}$. 
In addition, $v$ is varied from $0.28$~cm/h to $0.54$~cm/h while keeping $F_{\mathrm{in}}$ fixed, giving Péclet numbers from $0.5\,\mathrm{Pe}_\mathrm{max}$ to $0.97\,\mathrm{Pe}_\mathrm{max}$, and total population sizes in the $1200-2400$ range, with active population sizes around $500$. 

Our discretization employs $K=300$ (corresponding to $\delta x=L/K=0.02$~cm) and $\delta t=2\times 10^{-4}$~h, as in Ref.~\cite{labavic2022hydrodynamic}. To obtain each marker in our figures, $N_\mathrm{sim} = 10^5$ simulation replicates were performed. The values of $t_\mathrm{init}$ and $t_\mathrm{inter}$ used in the stochastic simulations reported in our figures are respectively $10^7$ and $10^4$ time steps, i.e.\ 2000 hours and 2 hours.

\subsection{Numerical solutions}
\label{numerics}

We performed numerical integration of the coupled partial differential equations in Eqs.~\eqref{eq:eq_F}, \eqref{eq:eq_B} and \eqref{eq:eq_M} in Python, using an explicit Euler scheme in time and finite differences in space. Our discretization uses time step $\delta t$ and spatial step $\delta x$. 
The values of $\delta t$ and $\delta x$ we employed in our figures are as follows: 
\begin{itemize}
\item Main text Figs.~1 and~4, and Figs.~\ref{fig:Rn_X}, \ref{fig:phi_gamma}, \ref{Fig_step_profiles}: $D\delta t / L^2 = 1.4\times 10^{-7}$ and $\delta x/ L = 1.7\times 10^{-3}$. 
\item Main text Figs.~2 and~3, and Figs.~\ref{fig:phi_Pe},  \ref{fig:phi_2profiles}, \ref{Nactive_fig}, \ref{phimoy_smalls}, \ref{fig:r_eff}, \ref{fig:Comp_Good}: $D\delta t / L^2 = 5.6\times 10^{-7}$ and $\delta x/ L = 3.3\times 10^{-3}$. 
\item Fig.~\ref{step_approx}: $D\delta t / L^2 = 2\times 10^{-7}$ and $\delta x/ L = 2\times 10^{-3}$.
\end{itemize}
Stationary solutions were obtained by using the first time point where the absolute difference between two successive time points was smaller than $10^{-10}$.

Numerical solutions of ordinary differential equations such as Eq.~\eqref{eq:phi1} were obtained using the function \texttt{solve\_bvp} of the Python \texttt{scipy.integrate} library with a tolerance of $10^{-9}$ and a maximum number of mesh nodes of $2\times 10^5$.

\section{Realistic parameter values in the human gut}
\label{humanparams}

Let us now discuss the realistic range of values of the dimensionless parameters in the human colon. \\

\noindent\textbf{Relevant values of $k/F_\mathrm{in}$.} Let us start by considering $k/F_\mathrm{in}$, and for this, let us first estimate $F_\mathrm{in}$. Bacteria in the large intestine consume a mix of different nutrients that have not (or not completely) been absorbed in the small intestine. It is estimated that about 20~g of fibers per day and about the same weight of undigested starch reach the colon for the standard British diet (and potentially more for non-Western diets richer in fiber) \cite{arnoldini2025quantifying}.
Given that the inflow of digesta in the colon is about 2 liters per day \cite{debongnie1978capacity}, the order of magnitude of the incoming food concentration $F_\mathrm{in}$ in the colon is of the order of 20 g/L. The Monod constant $k$ depends on many factors, including the substrate and the bacterial strain. Even restricting to \textit{Escherichia coli} growing on glucose, $k$ can range from 0.03 to 5 mg/L depending on adaptation~\cite{kovarova1998growth}, and it is typically higher for other substrates (e.g.\ 5 to 900 mg/L for different substrates~\cite{lawrence1969kinetics}). Realistic values for $k/F_\mathrm{in}$ thus span a wide range, but in all cases, they are much smaller than 1, at most of order 0.05. In practice, as long as $k$ is small relative to $F_\mathrm{in}$, we obtain similar results, and the analytical results in the limit of a step function, which is the limit of a very small $k/F_\mathrm{in}$, are a good approximation of the full system (see Fig.~\ref{step_approx}). \\

\noindent\textbf{Relevant values of $\lambda_w$.} The effective diffusion coefficient $D$ models mixing of the digesta~\cite{cremer2016effect}, and is thus challenging to measure directly. However, based on analysis of experiments from Ref.~\cite{hammer1993fluid} of the transport of markers through the gut, Ref.~\cite{Cremer2017} estimated $D$ to be in the range of $(5-20) \times 10^5\, \mu \mathrm{m^2 /s}$.

A lower bound for the bacterial replication time is given by the minimal doubling time of fast replicating bacteria such as \textit{Escherichia coli} in good conditions, which can be as low as 20 minutes \cite{gibson2018distribution}, yielding a growth rate around $6\times 10^{-4}\mathrm{/s}$. Besides, an upper bound for the bacterial replication time in the gut is what is needed to survive turnover. Feces weigh about 130 g/day and contain $25-50$\% of bacteria in mass \cite{rose2015characterization}. The typical mass of a bacterium being 1 pg, this means that about $(3-6)\times 10^{13}$ bacteria per day are lost in feces, to be compared with about $4\times10^{13}$ bacteria in the colon \cite{sender2016}, leading to about one renewal per day. However, this upper bound assumes that bacteria replicate at their maximal rate everywhere in the colon, so it is very likely that the bacteria maximal division rate is significantly higher. The growth rate of \textit{Bacteroides thetaiotaomicron} in the mouse digestive tract has been measured to be of about 0.4~/h~\cite{hoces2023fitness}, i.e.\ about $10^{-4}\mathrm{/s}$.  This probably better reflects the expected range for commensal bacteria. Overall, it is reasonable to take $r_w$ between $0.6\times 10^{-4}\mathrm{/s}$ and $6\times 10^{-4}\mathrm{/s}$.

The total length of the colon is $190\pm 26$~cm, with $23\pm7$~cm for the ascending colon \cite{khashab2009colorectal}, where most of the microbial activity takes place \cite{Cremer2017}. So the length $L$ of the system of interest is in the $20-200$~cm range, depending on whether the focus is on the ascending colon only, or on the whole colon (with reduced downstream bacterial activity due to nutrient depletion).

Taken together, these estimates lead to $\lambda_w=r_w L^2/D \approx [(0.6-6)\times 10^{-4}]\times[(0.2 - 2) \times 10^6]^2/[(5-20) \times 10^5] \approx 1 - 5000$. The values used in our figures are in this range. Note that $\mathrm{Pe}_0=2\sqrt{\lambda_w}$ is thus of the order of $2-140$.\\

\noindent\textbf{Relevant values of $\mathrm{Pe}$.} 
It is empirically well demonstrated that there is a strong gradient of bacterial concentration along the colon \cite{donaldson2016}, and that most nutrients that could be used by bacteria are consumed by the end of the colon \cite{arnoldini2025quantifying}. The washout condition (Section \ref{sec:washout}) is $\mathrm{Pe}_\mathrm{max} \approx \mathrm{Pe}_0^2/4$ for $\mathrm{Pe}_0<1$, and $\mathrm{Pe}_\mathrm{max} \approx \mathrm{Pe}_0$ for $\mathrm{Pe}_0>10$. When $D$ increases, the system becomes more mixed, so for small $\mathrm{Pe}$, gradients will be weaker. Thus, realistic Péclet numbers are smaller than $\mathrm{Pe}_\mathrm{max}$, since the bacterial population persists, but of the same order of magnitude as $\mathrm{Pe}_\mathrm{max}$, since observed gradients are strong. As $\mathrm{Pe}_0>1$, the regime where $\mathrm{Pe}_\mathrm{max} \approx \mathrm{Pe}_0$ is most relevant, and we thus expect $\mathrm{Pe}$ to be in the $1-140$ range.

A second way to estimate $\mathrm{Pe}$ is to directly use its definition: $\mathrm{Pe}=vL/D$. As mentioned above, $D$ is in the range of $(5-20) \times 10^5\, \mu \mathrm{m^2 /s}$, and $L$ in the $(2-20) \times 10^{5}\, \mu \mathrm{m}$ range. The average diameter of the ascending colon is approximately $46$~mm~\cite{mark2021colorectal}. As mentioned above, the volume entering the colon per day is about 2 liters. Taken together, this gives an estimate of $v \approx 14 \, \mu \mathrm{m/s}$. We thus obtain a range of $1.4-56$ for Pe, which is compatible with our first estimate, while using different arguments. 

Note also that $\mathrm{Pe}/\mathrm{Pe}_0=v/(2\sqrt{r_w D})$. Using the previously mentioned value for $v$, and ranges for $r_w$ and $D$, we obtain $\mathrm{Pe}/\mathrm{Pe}_0$ in the range $0.2-1.3$. If $\mathrm{Pe}_\mathrm{max} \approx \mathrm{Pe}_0$, this entails that $\mathrm{Pe}/\mathrm{Pe}_\mathrm{max}$ is around washout, with a substantial fraction of the range below washout and a smaller fraction above it. The range considered in our figures is below washout but close to it.\\

\noindent\textbf{Relevant values of $s$.} 
In \textit{Escherichia coli} bacteria well-adapted to their environment, most non-lethal mutations have small fitness effects: an average fitness cost of 0.3\% was reported in Ref.~\cite{Robert2018}, and for beneficial mutants, fitness advantages $s<0.1$ were reported in Ref.~\cite{Couce2024}. However, large fitness differences (up to $s=0.4$) between evolved lineages have been observed in \textit{Escherichia coli} during adaptation to the gut as a new environment~\cite{Barroso2014}. This illustrates the importance of making predictions both for small fitness advantages and for larger ones.

\end{document}